%% file: Article.tex
\documentclass[conference]{IEEEtran}

\input{Format}

\input{Macros}

\input{Figures}

\input{Tables}

\hypersetup
  {
  pdftitle  = {Quantifying over Trees in Monadic Second-Order Logic},
  pdfauthor = {M. Benerecetti, L. Bozzelli, F. Mogavero, A. Peron.}
  }

\begin{document}

  \title{\LARGE Quantifying over Trees in Monadic Second-Order Logic}

  \author{%
    \IEEEauthorblockN{Massimo Benerecetti, Laura Bozzelli, Fabio Mogavero}%
    \IEEEauthorblockA{Universit\`a degli Studi di Napoli Federico II \\
    emails: \{massimo.benerecetti,laura.bozzelli,fabio.mogavero\}@unina.it}
  \and
    \IEEEauthorblockN{Adriano Peron}
    \IEEEauthorblockA{Universit\`a degli Studi di Trieste\\
    email: adriano.peron@units.it} }


  \maketitle

\input{Abstract}

\input{Introduction}

\input{Background}

\input{SectionI}

\input{SectionII}







\input{Discussion}



  \bibliographystyle{IEEEtranS}
  \bibliography{References}

  \onecolumn

  \newpage
  \appendix

\input{AppendixA}

\input{AppendixB}

\input{AppendixC}


\end{document}

%% file: Format.tex
\usepackage[latin1]{inputenc}
\usepackage[T1]{fontenc}

\usepackage[english]{babel}

\usepackage{graphicx}

\usepackage{multicol}

\usepackage{caption}

\usepackage[fig]{fmocdmac}

\makeatletter
\def\orcidID#1{\smash{\href{http://orcid.org/#1}{\protect\raisebox{-1.25pt}{
\protect\includegraphics{orcid.eps}}}}}
\makeatother

\AtEndPreamble
  {

  }


%% file: Macros.tex
\newtheorem{definition}{Definition}

\newtheorem{proposition}{Proposition}
\newtheorem{lemma}{Lemma}

\newtheorem{theorem}{Theorem}
\newtheorem{corollary}{Corollary}

\usrmth{Cnt}{}{sym}[\odot]

\DeclareRobustCommandx{\OSMC}[3][1=, 2=, 3=]
  {\MC[#1][#2][1s\arglef{,}{#3}]}
\DeclareRobustCommandx{\OSAFMC}[3][1=, 2=, 3=]
  {\AFMC[#1][#2][1s\arglef{,}{#3}]}
\DeclareRobustCommandx{\OSGMC}[3][1=, 2=, 3=]
  {\GMC[#1][#2][1s\arglef{,}{#3}]}
\DeclareRobustCommandx{\OSAFGMC}[3][1=, 2=, 3=]
  {\AFGMC[#1][#2][1s\arglef{,}{#3}]}

\cmdtxtoparname{STLS}[STL*]

\cmdtxtoparname{CCTLS}[CCTL*]
\cmdtxtoparname{WCCTLS}[CCTL*]

\cmdtxtoparname{GCTL}
\cmdtxtoparname{GCTLS}[GCTL*]

\newcommand{\UU}
  {\newmth[mathbb]{U}}
\newcommand{\RR}
  {\newmth[mathbb]{R}}
\renewcommand{\SS}
  {\newmth[mathbb]{S}}
\newcommand{\BB}
  {\newmth[mathbb]{B}}

\newcommand{\details}[1]{{}}
\newcommand{\DefinedAs}{\defeq}
\newcommand{\tpl}[1]{(#1)}
\newcommand{\Nat}{{\SetN}}

\cmdtxtoparname{Class}[C]

\newcommand{\Prop}{\APSet}
\newcommand{\Val}{\mathcal{V}}
\newcommand{\Logic}{\mathit{L}}
\newcommand{\Lab}{\mathit{Lab}}
\newcommand{\Var}{\VarSet}
\newcommand{\Child}{{\mthsym{child}}}
\newcommand{\Path}{{\mthsym{path}}}
\newcommand{\Tree}{{\mthsym{tree}}}
\newcommand{\Fin}{{\mthsym{fin}}}
\newcommand{\Inf}{{\mthsym{inf}}}

\cmdtxtoparname{SD}
\newcommand{\LT}{\mathcal{T}}
\newcommand{\ND}{\mathit{ND}}
\newcommand{\D}{\mathit{D}}
\newcommand{\NAcc}{\mathit{NA}}
\newcommand{\Acc}{\mathit{A}}
\newcommand{\N}{\mathit{N}}
\newcommand{\Rel}{\mathit{R}}
\newcommand{\last}{\mathit{lst}}
\newcommand{\FloorL}[1]{\lfloor #1\rfloor}

\cmdtxtoparname{SO}

\DeclareRobustCommand{\CoWMSO}
  {\coWMSO}

\DeclareRobustCommand{\CoWMPL}
  {\coWMPL}

\DeclareRobustCommand{\CoWMTL}
  {\coWMTL}

\DeclareRobustCommand{\CTLStar}
  {\CTLS}

\DeclareRobustCommand{\CCTLStar}
  {{\txtname{C}}\CTLS}

  \DeclareRobustCommand{\WCCTLStar}
  {{\txtname{WC}}\CTLS}

    \DeclareRobustCommand{\CoWCCTLStar}
  {{\txtname{coWC}}\CTLS}

\DeclareRobustCommand{\CCTLS}
  {{\txtname{C}}\CTLS}
\DeclareRobustCommand{\WCCTLS}
  {{\txtname{W}}\CCTLS}

\newcommand{\Until}{{\U}}
\newcommand{\Next}{{\X}}

\newcommand{\Eventually}{{\F}}
\newcommand{\EQ}{{\E}}
\newcommand{\AQ}{{\A}}
\newcommand{\DC}{{\DSym}}

\def\M{{\mathcal{M}}}

\cmdmthargfun{tr}
\usrmth{post}{}{argfun}
\usrmth{root}{}{lbop}[\downarrow]
\usrmth{child}{Mac}{argsym}
\usrmth{pnf}{Mac}{argsym}
\usrmth{maxsubtree}{Mac}{argsym}[mst]
\usrmth{down}{}{lbop}[\ssearrow]
\usrmth{up}{}{lbop}[\nnearrow\!]
\usrmth{nonblocking}{Mac}{argsym}[nb]

\newcommand{\kmodels}[1]{%
\:\parbox[b][1.2em]{1.5em}
  {$\models$\hspace{-0.6em}\raisebox{0.50em}{\tiny$#1$}}%
}


%% file: Figures.tex
\tikzstyle{tnodea} =
	[regular polygon, regular polygon sides = 3, draw = black, fill = gray!25,
	inner sep = 0.15em]
\tikzstyle{tnodeb} =
	[regular polygon, regular polygon sides = 3, draw = black, fill = gray!75,
	inner sep = 0.20em]

 \tikzstyle{recnode} =
	[rectangle, minimum width = 0.5 cm, minimum height = 1cm, fill = gray!75,
	inner sep = 0.20em]

 \tikzstyle{bullnode} =
	[circle, fill = gray!75,
	inner sep = 0.20em]

\AfterEndPreamble{

\newcommand{\figLogFinBrn}{
\begin{tikzpicture}[node distance = 5em, bend angle = 45]

  \matrix (L) [matrix of nodes, ampersand replacement = \&,
    row sep = 1em, column sep = 7.5em] {
    \coWMSO \& \MSO \&       \\
            \&      \& \WMSO \\
    \coWMTL \& \MTL \&       \\
            \&      \& \WMTL \\
    \coWMPL \& \MPL \&       \\
            \&      \& \WMPL \\
  };

  \path[<-]
    (L-1-1) edge [] node[text=blue] {}
              (L-3-1)
    (L-1-2) edge  [] node[nodraw, text=blue]
            {\text{\scriptsize \cite{Rab70}}}
              (L-2-3)
            edge [] node[nodraw, text=blue]
            {\text{\scriptsize Cor~\ref{cor:MTLversusMSO}}}
              (L-3-2)
    (L-2-3) edge [] node[nodraw, text=blue]
            {\text{\scriptsize Thm~\ref{thm:WMTLversusWMSO}}}
              (L-4-3)
    (L-3-1) edge [] node[text=blue] {}
              (L-5-1)
    (L-3-2) edge [] node[nodraw, text=blue]
            {\text{\scriptsize Thm~\ref{thm:PowerCoWMTL}}}
              (L-4-3)
            edge [] node[nodraw, text=blue]
            {\text{\scriptsize Cor~\ref{cor:HierarchyFullQuant}}}
              (L-5-2)
    (L-4-3) edge [] node[nodraw, text=blue]
            {\text{\scriptsize Cor~\ref{cor:WeakVariantsHierarchy}}}
              (L-6-3)
    (L-5-2) edge[dashed, red] node[nodraw] {\color{red} ?}
              (L-4-3)
            edge [] node[nodraw, text=blue]
            {\text{\scriptsize Thm~\ref{thm:InexpressivenessWeakMPL}}}
              (L-6-3)
  ;

  \path[<->]
    (L-1-1) edge [] node[nodraw, text=blue]
            {\text{\scriptsize Thm~\ref{thm:PowerCoWMSO}}}
              (L-1-2)
    (L-3-1) edge [] node[nodraw, text=blue]
            {\text{\scriptsize Thm~\ref{thm:PowerCoWMTL}}}
              (L-3-2)
    (L-5-1) edge [] node[nodraw, text=blue]
            {\text{\scriptsize Thm~\ref{thm:InexpressivenessWeakMPL}}}
              (L-5-2)
  ;

\end{tikzpicture}
}

\newcommand{\figLogArbBrn}{
\begin{tikzpicture}[node distance = 5em, bend angle = 22.5]
  \matrix (L) [matrix of nodes, ampersand replacement = \&,
    row sep = 2.5em, column sep = 7.5em] {
          \& \coWMSO  \&        \\
    \MSO  \&          \& \WMSO \\
          \& \coWMTL  \&        \\
    \MTL  \&          \& \WMTL  \\
    \MPL  \& \coWMPL  \&        \\
          \&          \& \WMPL  \\
  };
  \path[<-]
    (L-1-2) edge [] node[nodraw, text=blue]
            {\text{\scriptsize Thm~\ref{thm:PowerCoWMSO}}}
              (L-2-1)
            edge [] node[nodraw, text=blue]
            {\text{\scriptsize Thm~\ref{thm:PowerCoWMSO}}}
              (L-2-3)
            edge [] node[nodraw, text=blue]
            {\text{\scriptsize Thm~\ref{thm:CoWeakVariantsHierarchy}}}
              (L-3-2)
    (L-2-1) edge [] node[nodraw, text=blue]
            {\text{\scriptsize Cor~\ref{cor:MTLversusMSO}}}
              (L-4-1)
    (L-2-3) edge [] node[nodraw, text=blue]
            {\text{\scriptsize Thm~\ref{thm:WMTLversusWMSO}}}
              (L-4-3)
    (L-3-2) edge [] node[nodraw, text=blue]
            {\text{\scriptsize Thm~\ref{thm:PowerCoWMTL}}}
              (L-4-1)
            edge [] node[nodraw, text=blue]
            {\text{\scriptsize Thm~\ref{thm:PowerCoWMTL}}}
              (L-4-3)
            edge [] node[nodraw, text=blue]
            {\text{\scriptsize Thm~\ref{thm:CoWeakVariantsHierarchy}}}
              (L-5-2)
    (L-4-1) edge [] node[nodraw, text=blue]
            {\text{\scriptsize Cor~\ref{cor:HierarchyFullQuant}}}
              (L-5-1)
    (L-4-3) edge [] node[nodraw, text=blue]
            {\text{\scriptsize Cor~\ref{cor:WeakVariantsHierarchy}}}
              (L-6-3)
    (L-5-2) edge [] node[nodraw, text=blue]
            {\text{\scriptsize Thm~\ref{thm:InexpressivenessWeakMPL}}}
              (L-6-3)
  ;
  \path[<->]
    (L-5-1) edge [] node[nodraw, text=blue]
            {\text{\scriptsize Thm~\ref{thm:InexpressivenessWeakMPL}}}
              (L-5-2)
  ;

\end{tikzpicture}
}

\newcommand{\figND}
  {
  \begin{figure}[htbp]
    \vspace{-1.00em}
    \begin{center}
      \footnotesize
      \scalebox{0.95}[1.00]
        {
        \begin{tikzpicture}[node distance = 5.25em, bend angle = 45]

          \node [nodraw]
                (X)
                []
                {};

          \node [nodraw]
                (A)
                [node distance = 8.5em, left of = X]
                {$\ND_2$};
          \node [bullnode]
                (AE)
                [node distance = 1.5em, below of = A]
                {};
          \node [recnode]
                (A1)
                [below of = AE]
                {$\ND_1$};
          \node [recnode]
                (A0)
                [node distance = 3.0em, left of = A1]
                {$\ND_1$};
          \node [tnodea]
                (A2)
                [node distance = 4.2em, inner sep = -0.20em, right of = A1]
                {$\ND_2$};

          \node [nodraw]
                (B)
                [node distance = 8.5em, right of = X]
                {$\D_2$};
          \node [bullnode]
                (BE)
                [node distance = 1.5em, below of = B]
                {};
          \node [tnodea]
                (B2)
                [below of = BE]
                {$\D_2$};
          \node [recnode]
                (B1)
                [node distance = 4.2em,left of = B2]
                {$ND_1$};
          \node [recnode]
                (B0)
                [node distance = 3.2em,left of = B1]
                {$ND_1$};
            \node [tnodea]
                (B3)
                [right of = B2]
                {$\D_2$};

          \path[->]
            (AE)	edge	[bend right]
                        node [] {}
                        (A0.north)
                  edge	[]
                        node [] {}
                        (A1.north)
                  edge	[bend left]
                        node [] {}
                        (A2.north)
            (BE)	edge	[bend right]
                        node [] {}
                        (B0.north)
                  edge	[bend right]
                        node [] {}
                        (B1)
                  edge	[]
                        node [] {}
                        (B2.north)
                        edge	[bend left]
                        node [] {}
                        (B3.north)
            ;

        \end{tikzpicture}
        }
      \caption{\label{fig:recND} \small The classes of  trees  $\ND_n$ and $\D_n$ for $n=2$.
      }
    \end{center}
    \vspace{-1em}
  \end{figure}
  }

\newcommand{\figAN}
  {
  \begin{figure}[htbp]
    \vspace{-1.00em}
    \begin{center}
      \footnotesize
      \scalebox{0.95}[1.00]
        {
        \begin{tikzpicture}[node distance = 5.25em, bend angle = 45]

          \node [nodraw]
                (X)
                []
                {};

          \node [nodraw]
                (A)
                [node distance = 8.5em, left of = X]
                {$\Acc_3$};
          \node [bullnode]
                (AE)
                [node distance = 1.5em, below of = A]
                {};
          \node [tnodea]
                (A1)
                [inner sep = -0.20em, below of = AE]
                {$\Acc_2$};
          \node [tnodea]
                (A0)
                [inner sep = -0.20em, node distance = 4.0em, left of = A1]
                {$\Acc_2$};
          \node [tnodea]
                (A2)
                [node distance = 4.0em, inner sep = -0.20em, right of = A1]
                {$\Acc_2$};
          \node [nodraw]
                (B)
                [node distance = 8.5em, right of = X]
                {$\NAcc_3$};
          \node [bullnode]
                (BE)
                [node distance = 1.5em, below of = B]
                {};
                \node [nodraw]
                (BX)
                [below of = BE]
                {};
          \node [tnodea]
                (B2)
                [inner sep = -0.20em,node distance = 2.0em, left of = BX]
                {$\Acc_3$};
          \node [tnodea]
                (B1)
                [inner sep = -0.20em,node distance = 4.5em,left of = B2]
                {$\NAcc_3$};
          \node [tnodea]
                (B0)
                [inner sep = -0.20em, node distance = 2.0em, right of = BX]
                {$\Acc_3$};
            \node [tnodea]
                (B3)
                [inner sep = -0.20em, node distance = 4.5em, right of = B0]
                {$\Acc_3$};

          \path[->]
            (AE)	edge	[bend right]
                        node [] {}
                        (A0.north)
                  edge	[]
                        node [] {}
                        (A1.north)
                  edge	[bend left]
                        node [] {}
                        (A2.north)
            (BE)	edge	[bend left]
                        node [] {}
                        (B0.north)
                  edge	[bend left]
                        node [] {}
                        (B3.north)
                  edge	[bend right]
                        node [] {}
                        (B2.north)
                        edge	[bend right]
                        node [] {}
                        (B1.north)
            ;

        \end{tikzpicture}
        }
      \caption{\label{fig:recNA} \small
      The classes of Kripke trees $\Acc_n$ and $\NAcc_n$
      for $n=3$.
      }
    \end{center}
    \vspace{-1em}
  \end{figure}
  }

}


%% file: Tables.tex
\AfterEndPreamble{

\newcommand{\tabonestpgmc}{
\setlength{\columnsep}{-3.5em}
\begin{multicols}{3}
\begin{itemize}
\item
  $\trFun[\varElm]{\Ff} \defeq \Ff$;
\item
  $\trFun[\varElm]{\Tt} \defeq \Tt$;
\item
  $\trFun[\varElm]{\apElm} \defeq \PapElm[\apElm](\varElm)$;
\item
  $\trFun[\varElm]{\XvarElm} \defeq \varElm \in \XvarElm$;
\item
  $\trFun[\varElm]{\neg \varphiFrm} \defeq \neg \trFun[\varElm]{\varphiFrm}$;
\item
  $\trFun[\varElm]{\varphiFrm[1] \Cnt\, \varphiFrm[2]} \defeq
  \trFun[\varElm]{\varphiFrm[1]} \Cnt\, \trFun[\varElm]{\varphiFrm[2]}$, $\Cnt
  \in \{ \wedge,\! \vee \}$;
\end{itemize}
\end{multicols}
\vspace{-1em}
\begin{itemize}
\item
  $\trFun[\varElm]{\DMod[\geq k]\, \varphiFrm} \defeq \exists \yvarElm[1],
  \ldots, \yvarElm[k] \ldotp (\bigwedge_{i \neq j} (\yvarElm[i] \neq
  \yvarElm[j]) \!\wedge\! \bigwedge_{i = 1}^{k} \childMac{\varElm, \yvarElm[i]})
  \wedge (\bigwedge_{i = 1}^{k} \trFun[ {\yvarElm[i]} ]{\varphiFrm})$;
\item
  $\trFun[\varElm]{\BMod[< k]\, \varphiFrm} \defeq \forall \yvarElm[1], \ldots,
  \yvarElm[k] \ldotp (\bigwedge_{i \neq j} (\yvarElm[i] \neq
  \yvarElm[j]) \!\wedge\! \bigwedge_{i = 1}^{k} \childMac{\varElm, \yvarElm[i]})
  \!\rightarrow\! (\bigvee_{i = 1}^{k} \trFun[ {\yvarElm[i]} ]{\varphiFrm})$;
\end{itemize}
\begin{itemize}
\item
  $\trFun[\varElm]{\mu \XvarElm \ldotp \varthetaFrm} \defeq \neg
  \trFun[\varElm]{\nu \XvarElm \ldotp \pnfMac{\neg {\varthetaFrm}[\XvarElm /
  \neg \XvarElm]}}$;
\item
  $\trFun[\varElm]{\nu \XvarElm \ldotp \varthetaFrm} \defeq \exists^{\TSym}
  \XvarElm \ldotp \varElm \in \XvarElm \wedge \forall \varElm \in \XvarElm
  \ldotp \trFun[\varElm]{\varthetaFrm}$.
\end{itemize}
}

\newcommand{\tabaltfreonestpgmc}{
\begin{itemize}
\item
  $\trFun[\varElm]{\nu \XvarElm \ldotp \varthetaFrm} \defeq
  \begin{cases}
    {\neg \trFun[\varElm]{\mu \XvarElm \ldotp \pnf(\neg ({\varthetaFrm}[\XvarElm
    / \neg \XvarElm][\YvarElm / \neg \YvarElm]))}},
  & \text{if } {\varthetaFrm \in \PhiSet[\ZSet', \OSet'][\ASym\FSym]};
  \\
    {\trFun[\varElm]{{\varthetaFrm}[\XvarElm / \YvarElm]}},
  & \text{otherwise};
  \end{cases}$
\item
  $\trFun[\varElm]{\mu \XvarElm \ldotp \varthetaFrm} \defeq
  \begin{cases}
    {\exists^{\TSym} \XvarElm \ldotp \varElm \in \XvarElm \wedge \forall
    \varElm \in \XvarElm \ldotp \exists^{\TSym} \YvarElm \ldotp
    \maxsubtreeMac{\YvarElm, \XvarElm, \varElm} \wedge
    \trFun[\varElm]{{\varphiFrm}[\XvarElm / \YvarElm] \down[\YvarElm]}},
  & \text{if } {\varthetaFrm \in \PhiSet[\ZSet', \OSet'][\ASym\FSym]};
  \\
    {\trFun[\varElm]{{\varthetaFrm}[\XvarElm / \YvarElm]}},
  & \text{otherwise};
  \end{cases}$ \\
\end{itemize}
where $\ZSet' \defeq \ZSet \cup \{ \YvarElm, \XvarElm \}$, $\OSet' \defeq \OSet
\cup \{ \YvarElm, \XvarElm \}$, and $\maxsubtreeMac{\YvarElm, \XvarElm, \varElm}
\defeq \forall \yvarElm \ldotp (\yvarElm \in \YvarElm) \leftrightarrow (\varElm
\leq \yvarElm \wedge \forall \zvarElm \ldotp (\varElm \leq \zvarElm \leq
\yvarElm) \rightarrow \zvarElm \in \XvarElm)$, for some fresh fixpoint variable
$\YvarElm \in \VarSet[2] \setminus \free{\varthetaFrm}$, where the two
second-order existential quantifiers range over finite trees.
}

\newcommand{\tabstl}{
\begin{itemize}
\item
  $\trFun[\XvarElm, \varElm]{\varphiFrm[1] \UU[\phiFrm] \varphiFrm[2]} \defeq
  \exists^{\TSym} \XvarElm' \ldotp (\nonblockingMac{\XvarElm'} \wedge \XvarElm'
  \sqsubset_{\phiFrm}^{\XvarElm} \XvarElm) \wedge (\trFun[\XvarElm',
  \varElm]{\varphiFrm[2]} \wedge \forall^{\TSym} \XvarElm'' \ldotp
  (\nonblockingMac{\XvarElm''} \wedge \XvarElm' \sqsubset_{\phiFrm}^{\XvarElm}
  \XvarElm'' \sqsubset_{\phiFrm}^{\XvarElm} \XvarElm) \rightarrow
  \trFun[\XvarElm'', \varElm]{\varphiFrm[1]})$;
\item
  $\trFun[\XvarElm, \varElm]{\varphiFrm[1] \RR[\phiFrm] \varphiFrm[2]} \defeq
  \forall^{\TSym} \XvarElm' \ldotp (\nonblockingMac{\XvarElm'} \wedge \XvarElm'
  \sqsubset_{\phiFrm}^{\XvarElm} \XvarElm) \rightarrow (\trFun[\XvarElm',
  \varElm]{\varphiFrm[2]} \vee \exists^{\TSym} \XvarElm'' \ldotp
  (\nonblockingMac{\XvarElm''} \wedge \XvarElm' \sqsubset_{\phiFrm}^{\XvarElm}
  \XvarElm'' \sqsubset_{\phiFrm}^{\XvarElm} \XvarElm) \wedge \trFun[\XvarElm'',
  \varElm]{\varphiFrm[1]})$;
\item
  $\trFun[\XvarElm, \varElm]{\varphiFrm[1] \SS[\phiFrm] \varphiFrm[2]} \defeq
  \exists^{\TSym} \XvarElm' \ldotp (\nonblockingMac{\XvarElm'} \wedge \XvarElm
  \sqsubset_{\phiFrm}^{\XvarElm} \XvarElm') \wedge (\trFun[\XvarElm',
  \varElm]{\varphiFrm[2]} \wedge \forall^{\TSym} \XvarElm'' \ldotp
  (\nonblockingMac{\XvarElm''} \wedge \XvarElm \sqsubset_{\phiFrm}^{\XvarElm}
  \XvarElm'' \sqsubset_{\phiFrm}^{\XvarElm} \XvarElm') \rightarrow
  \trFun[\XvarElm'', \varElm]{\varphiFrm[1]})$;
\item
  $\trFun[\XvarElm, \varElm]{\varphiFrm[1] \BB[\phiFrm] \varphiFrm[2]} \defeq
  \forall^{\TSym} \XvarElm' \ldotp (\nonblockingMac{\XvarElm'} \wedge \XvarElm
  \sqsubset_{\phiFrm}^{\XvarElm} \XvarElm') \rightarrow (\trFun[\XvarElm',
  \varElm]{\varphiFrm[2]} \vee \exists^{\TSym} \XvarElm'' \ldotp
  (\nonblockingMac{\XvarElm''} \wedge \XvarElm \sqsubset_{\phiFrm}^{\XvarElm}
  \XvarElm'' \sqsubset_{\phiFrm}^{\XvarElm} \XvarElm') \wedge \trFun[\XvarElm'',
  \varElm]{\varphiFrm[1]})$;
\end{itemize}
where $\nonblockingMac{\XvarElm} \defeq \forall \varElm \!\in\! \XvarElm \ldotp
\exists \yvarElm \!\in\! \XvarElm \ldotp \childMac{\xvarElm, \yvarElm}$ and
$\YvarElm \!\sqsubset_{\phiFrm}^{\XvarElm}\! \ZvarElm \defeq (\YvarElm
\!\subset\! \ZvarElm) \wedge (\forall \yvarElm \!\in\! \YvarElm \ldotp \forall
\zvarElm \!\in\! \ZvarElm \ldotp (\trFun[\XvarElm, \yvarElm]{\phiFrm} \wedge
\childMac{\yvarElm, \zvarElm}) \!\rightarrow\! \zvarElm \!\in\! \YvarElm)$.
}

}


%% file: Abstract.tex


\begin{abstract}

\emph{Monadic Second-Order Logic} (\MSO) extends \emph{First-Order Logic} (\FO)
with variables ranging over sets and quantifications over those variables.
We introduce and study \emph{Monadic Tree Logic} (\MTL), a fragment of \MSO
interpreted on infinite-tree models, where the sets over which the variables
range are arbitrary subtrees of the original model.
We analyse the expressiveness of \MTL compared with variants of \MSO and \MPL,
namely \MSO with quantifications over paths.
We also discuss the connections with temporal logics, by providing non-trivial
fragments of the \emph{Graded \MC} that can be embedded into \MTL and by showing
that \MTL is enough to encode temporal logics for reasoning about strategies
with \FO-definable goals.

\end{abstract}


%% file: Introduction.tex


\section{Introduction}

The study of \emph{Monadic Second-Order Logic} (\MSO), namely \emph{First-Order
Logic} (\FO) extended with variables and quantifications ranging over sets, has
attracted a lot of attention over the years, mainly because of its high
expressive power and nice computational properties, particularly when
interpreted over words and trees.
This, in turn, makes it a good fit as a formal framework for reasoning about
\emph{regular languages} and \emph{computational systems} in general, whose set
of possible dynamic evolutions is often represented as a tree structure.

A seminal result in the field was originally provided by
B\"uchi~\cite{Buc60}, who proved the equivalence between the monadic
second-order logic of one successor with variables ranging over finite sets and
\emph{finite-state automata on finite words}~\cite{RS59}, which he exploited to
devise a decision procedure for that logic.
The result was then extended to the case of variables ranging over arbitrary
sets and \emph{finite-state automata on infinite words} in~\cite{Buc62,Buc66}.
Rabin~\cite{Rab69} later proved the decidability of \MSO interpreted over binary
trees, by means of an automata-theoretic characterisation of the expressive
power of logic on these structures.
This result was then extended by Walukiewicz~\cite{Wal96,Wal01}, who provided a
general framework for investigating \MSO by means of a class of automata that
captures the expressive power of \MSO on trees with arbitrary (finite and
infinite) branching degree.
By exploiting Rabin's result, Muller and Schupp~\cite{MS81,MS85} have shown that
\MSO is decidable for graphs with bounded tree-width, while
Courcelle~\cite{Cou90,Cou89,Cou92} has conducted a quite extensive study of \MSO
on graphs in  connection with both graph theory and complexity theory.

Variants of \MSO over tree models have also been studied.
\emph{Weak Monadic Second-Order Logic} (\WMSO) is one such variant, where the
second-order variables can only range over finite sets.
An automata-theoretic characterisation of \WMSO on binary trees has been
proposed by Rabin~\cite{Rab70} to show that \WMSO is strictly less expressive
than \MSO on this class of structures.
Automata for \WMSO on these trees have been proposed in~\cite{MSS86,MSS92},
where \emph{weak alternating tree automata} have been introduced.
It has been shown, moreover, that \WMSO is equivalent to the alternation-free
fragment of \emph{Modal \MC}~\cite{Koz83}, when direct access to the left and
right children of a node is available~\cite{AN92}.
A deeper analysis of this connection, when bisimulation invariant fragments are
considered, has been recently carried out in~\cite{FVZ13,CFVZ14,CFVZ20}.

Another noteworthy variant is \emph{Monadic Path Logic}~\cite{GS85,HT87} (\MPL),
where second-order variables are restricted to range over paths.
This restriction makes it strictly less expressive than full \MSO over trees.
The interest attracted by \MPL resides in the fact that many temporal logics,
most notably \CTLS, can be embedded into \MPL~\cite{HT87,Tho87,MR99,MR03}.

One needs to jump, however, directly to \MSO in order to be able to capture the
full \emph{Modal \MC}~\cite{JW96} and its \emph{graded extension}~\cite{KSV02}.
One reason for this is that one can express path properties in \MC that are not
expressible in \FOL and may be witnessed by \emph{non-connected} (\ie,
\emph{non-convex}) sets of nodes.
One such property is the one true at a node $\tElm$ if all the nodes at even
positions from $\tElm$ onward along some path satisfy a given atomic proposition
$\apElm$.
A \MC formula that collects all those witnesses is, for instance, $\nu \XvarElm
\ldotp (\apElm \wedge \DMod \DMod \XvarElm)$, where the modal formula $\DMod
\varphi$ holds at a node if one of its successors satisfies $\varphi$.
The nodes of the tree which occur at even positions along some path are not
connected to one another and 
they are all witnesses of the property.
More importantly, those non-connected witnesses depend on one another, in the
sense that, for a node $\tElm$ to witness the property, some node two steps
further ahead must also witness it and removing $\apElm$ from it would prevent
$\tElm$ from becoming a witness.
The possible non-connectedness of the witnesses is what makes the \MSO ability
to quantify over sets of, possibly non-connected, nodes an essential feature.
This contrasts, for instance, with the property true at a node $\tElm$ when
there exists a path from it and a prefix $\pthElm$ of that path where $\qapElm$
holds at the last node and $\apElm$ holds at all the previous ones.
This can be expressed in \CTLS by the formula $\E (\apElm \U \qapElm)$.
In this case, indeed, if a node $\tElm$ satisfies the property, all the nodes
along the witnessing prefix up to the node witnessing $\qapElm$ satisfy it as
well.
Non-connected witnesses may exist for this property too, however, they are all
independent from each other, in the sense that removing the property from a
prefix $\pthElm$ (\eg, by removing $\apElm$ and $\qapElm$ from the labelling of
the connected witnesses forming $\pthElm$) would bear no consequences for the
other witnesses outside $\pthElm$.
All properties expressible in \CTLS are of this kind and quantifying over paths,
which are connected sets of nodes, suffices to capture them all in \MPL.

It turns out that a similar connectedness property holds true even for more
expressive logics than \CTLS, such as languages for reasoning about strategies
and games, like \emph{Alternating-Time Temporal Logic}
(\ATLS)~\cite{AHK97,AHK02}, \emph{Strategy Logic}
(\SL)~\cite{CHP07,CHP10,MMV10a,MMPV14,MMPV17}, and \emph{Substructure Temporal
Logic} (\STLS)~\cite{BMM13,BMM15}.
For instance, it has been shown that \STLS, an extension of \CTLS that
implicitly allows for quantification over subtrees, is strictly more expressive
than \CTLS, since the latter is bisimulation invariant, while the former is not.
By means of subtree quantifications, \STLS is able to model strategies and,
therefore, to encode games with \FO-definable goals.

Guided by these observations, it appears natural to consider the seemingly
missing fragment of \MSO in which quantifications range over subtrees.
In this work, we study precisely this semantic restriction, that we call
\emph{Monadic Tree Logic}, \MTL for short, interpreted over non-blocking trees.
We provide a full picture of the expressive power of the logic, together with
some variants that restrict the range of the second-order variables to finite
trees only, giving rise to Weak \MTL (\WMTL), and to infinite ones only, leading
to Co-Weak \MTL (\coWMTL), and compare them to the corresponding variants of
\MSO and \MPL.
Each variant turns out to be strictly less expressive than the corresponding
\MSO variant and strictly more expressive than the analogous \MPL variant.
We also show that \MTL (\resp, \MSO, \MPL) is equivalent to \coWMTL (\resp,
\coWMSO, \coWMPL) and more expressive than \WMTL (\resp, \WMSO, \WMPL), when
interpreted over finitely-branching trees.
Interestingly enough, though, the situation changes drastically when we
interpret \MSO and \MTL on arbitrary trees that allow for infinitely-branching
degrees.
In this case, the co-weak variants of each logic strictly contain both the
corresponding full and weak versions, while the latter two become incomparable.

The second part of the article analyses the connections between variants of \MTL
and temporal logics.
More specifically, we identify a non-trivial fragment of the Graded \MC, called
\emph{One-Step Graded \MC} (\OSGMC), that can be captured by \MTL and can
express properties of trees that cannot be expressed in \MPL.
We also show that the alternation-free fragment of \OSGMC (\OSAFGMC) can be
captured by \WMTL and not by \WMPL.
Finally, we provide an encoding of \STLS into \MTL, showing that quantification
over trees is powerful enough to reason about games with \FO-definable goals.


%% file: Background.tex


\section{Background}
\label{sec:background}

Let  $\Nat$ be the set of natural numbers.
For a  finite or infinite word $w$ over some alphabet, $|w|$ denotes the length of $w$
($|w|=\infty$ if $w$ is infinite) and for all $0\leq i<|w|$, $w(i)$ is the
$(i+1)$-th letter of $w$.\vspace{0.2cm}

\noindent \textbf{Trees.} A tree $T$ is a subset of $\Nat^{*}$ such that there
is an element $\tau_0\in T$, called the \emph{root of $T$}, so that:
\begin{itemize}
  \item for each $\tau\in T$, $\tau$ is of the form $\tau_0\cdot\tau' $ for some
    $\tau'\in \Nat^{*}$;
  \item for all $\tau,\tau' \in \Nat^{*}$, if $\tau_0\cdot \tau\in T$ and $\tau'$ is a prefix of $\tau$, then
  $\tau_0\cdot \tau'\in T$.
\end{itemize}
Elements of $T$ are called \emph{nodes}. For $\tau\in T$, a \emph{child} of
$\tau$ in $T$ is a node in $T$ of the form $\tau\cdot n$ for some $n\in \Nat$. A
\emph{descendant}  of $\tau$ in $T$ is a node in $T$ of the form $\tau\cdot
\tau'$, for some $\tau'\in \Nat^{*}$. 
  A \emph{subtree of $T$} is a subset of
$T$ which is a tree.
The \emph{subtree of $T$ rooted at a node $\tau\in T$} is the tree
consisting of all the descendants of $\tau$ in $T$. A \emph{forest of $T$} is a
union of subtrees of $T$.  A \emph{path} of $T$ is a subtree $\pi$ of $T$ that
is totally ordered by the child-relation (i.e., each node of $\pi$ has at most
one child in $\pi$).  In the following, a path $\pi$ of $T$ is also viewed as a
word over $T$, in accordance with the total ordering in $\pi$ induced by the
child relation.
A tree $T$ is said to be:
\begin{itemize}
\item \emph{finitely-branching} if each node in $T$ has finitely many children
  in $T$ (and \emph{infinitely-branching}, otherwise);
\item \emph{blocking} if some node in $T$ has no children in $T$ (and
  \emph{non-blocking}, otherwise);
\item a \emph{chain} if it has a unique maximal path from the root (each node
  has at most one child). Note that a path of a tree corresponds to a chain.  
\item a \emph{complete binary tree} if each node has exactly two children;
\item \emph{dense} if it contains a subtree $T'$ such that each node of $T'$ has
  a descendant in $T'$ having at least two distinct children in $T'$.
\end{itemize}
Note that a dense tree has an uncountable number of infinite paths or,
equivalently, contains a complete binary tree as minor.
Dense trees correspond  to thick trees in~\cite{BIS13,ISB16}.
\vspace{0.1cm}

\noindent \textbf{Labelled trees and Kripke trees.}  For an alphabet $\Sigma$, a
$\Sigma$-labelled tree is a pair $\LT =\tpl{T, \Lab}$ consisting of a tree and a
labelling $\Lab:T \mapsto \Sigma$ assigning to each node in $T$ a symbol in
$\Sigma$.
Note that the dynamic behaviour of a system starting from an initial state can
be modelled by a $2^{\Prop}$-labelled tree, where $\Prop$ is a finite set of
atomic propositions.
A node in the tree $T$ represents a state of the system and the root corresponds
to the initial state. The maximal paths in the tree starting from the root
correspond to the possible executions of the system from the initial
state. Moreover, a node of a tree is labelled by elements in $\Prop$,
representing the atomic predicates that hold at the given state of the
computation.
Since we consider labelled trees modelling the dynamic behaviour of
\emph{reactive} systems and for these systems the executions are in general
infinite, we will restrict the interpretation of the considered logics to
labelled trees which are non-blocking. A non-blocking tree $T$ is infinite, and
maximal paths in $T$ are infinite as well.

Given a finite  set $\Prop$ of atomic propositions, a \emph{Kripke tree} over $\Prop$ is
a  non-blocking $2^{\Prop}$-labelled tree.\vspace{0.1cm}

\noindent \textbf{Relative Expressiveness.}
Let $\M$ be a set of models,
and $\Logic$ and $\Logic'$ be two logical languages interpreted over models in
$\M$.  Given two formulas $\varphi\in\Logic$ and $\varphi'\in\Logic'$, we say
that $\varphi$ and $\varphi'$ are \emph{equivalent} if for each model $M\in\M$,
$M$ satisfies $\varphi$ \iff $M$ satisfies $\varphi'$.  The language
$\Logic$ \emph{is subsumed by} $\Logic'$, denoted $\Logic\leq \Logic'$, if each
formula in $\Logic$ has an equivalent formula in $\Logic'$.
The language $\Logic$ \emph{is strictly less expressive than} $\Logic$, written
$\Logic<\Logic'$, if $\Logic\leq \Logic'$ and there is a $\Logic'$-formula which
has no equivalent in $\Logic$.
Two logics $\Logic$ and $\Logic'$ \emph{are expressively incomparable}, denoted
by $\Logic \not \sim \Logic'$, if both $\Logic\not\leq \Logic'$ and
$\Logic'\not\leq \Logic$.  Finally, two logics $\Logic$ and $\Logic'$ are
\emph{expressively equivalent}, denoted $\Logic \equiv \Logic'$, if both $\Logic
\leq \Logic'$ and $\Logic' \leq \Logic$.\vspace{0.2cm}

\noindent \textbf{Counting-$\CTLStar$.} We recall syntax and semantics of
Counting-$\CTLStar$ ($\CCTLStar$ for short~\cite{MR03}), which extends the
classic branching-time temporal logic \CTLS~\cite{EH86} by counting operators.
The syntax of $\CCTLStar$ is given by specifying inductively the set of
\emph{state formulas} $\varphi$ and the set of \emph{path formulas} $\psi$ over
a given finite set $\Prop$ of atomic propositions:
\[
\begin{array}{l}
	\varphi ::= \top \ | \ a \ | \ \neg \varphi \ | \ \varphi \wedge
	\varphi  \ | \ \EQ
	\psi \ | \ \DC^{n}
	\varphi\\
	\psi ::=   \varphi \ | \ \neg \psi \ | \ \psi \wedge
	\psi \ | \ \Next \psi\ | \ \psi \Until \psi
\end{array}
\]
where $a\in \Prop$, $\Next$ and $\Until$ are the standard ``next" and ``until"
temporal modalities, $\EQ$ is the existential path quantifier, and $\DC^{n}$ is
the counting operator with $n\in\Nat$.  The language of $\CCTLStar$ consists of
the state formulas of $\CCTLStar$.  We also use the standard shorthands
$\AQ\varphi \DefinedAs \neg \EQ \neg\varphi$ (``universal path quantifier") and
$\Eventually \psi \DefinedAs \top\Until\psi$ (``eventually"). 

The semantics is given \wrt Kripke trees $\LT = (T,\Lab)$ (over $\Prop$). For a
node $\tau$ of $T$, a path $\pi$ of $T$, and $0\leq i<|\pi|$, the satisfaction
relations $\LT,\tau\models \varphi$, for state formulas $\varphi$ (meaning that
$\varphi$ holds at node $\tau$ of $\LT$), and $\LT,\pi,i\models \psi$, for path
formulas $\psi$ (meaning that $\psi$ holds at position $i$ of the path $\pi$ in
$\LT$), are inductively defined as follows (Boolean connectives are treated as
usual):
\[ \begin{array}{ll}
	\LT,\tau\models a & \Leftrightarrow a\in \Lab(\tau);\\
	\LT,\tau\models \EQ \psi & \Leftrightarrow \LT,\pi,0\models \psi \text{
    for some path $\pi$ of $T$}\\ & \text{starting at node $\tau$}\\
	\LT,\tau\models \DC ^{n} \varphi & \text{$\Leftrightarrow$ there are at least $n$ distinct children}\\ & \text{$\tau'$ of $\tau$ in $T$ such that  }\LT,\tau'\models \varphi\\
	\LT,\pi,i\models \varphi &  \Leftrightarrow \LT,\pi(i)\models \varphi\\
	\LT,\pi,i\models \Next\psi &  \Leftrightarrow i+1<|\pi| \text{ and } \LT,\pi,i+1\models \psi\\
	\LT,\pi,i\models \psi_1\Until \psi_2 & \Leftrightarrow \text{for some $i\leq
    j<|\pi|$: }
	\LT,\pi,j\models\psi_2\\
& \text{ and } \LT,\pi,k\models\psi_1 \text{ for all } i\leq k<j.
\end{array}
\]
%
A Kripke tree $\LT$ satisfies a state formula $\varphi$, written $\LT \models
\varphi$, if $\LT,\tau_0\models \varphi$, where $\tau_0$ is the root of $\LT$.
Given a non-blocking tree $T$, we write $T\models \varphi$ to mean that
$T,\Lab_\emptyset\models \varphi$, where $\Lab_\emptyset(\tau)=\emptyset$ for
all $\tau\in T$.


We also consider two semantic variants of $\CCTLS$, that we call \emph{Weak}
$\CCTLS$ ($\WCCTLS$) and \emph{coWeak} $\CCTLS$ ($\CoWCCTLStar$), where the path
quantifiers $\EQ$ and $\AQ$ range over finite paths and infinite paths,
respectively, starting from the current node.  Standard \CTLS~\cite{EH86} is the
syntactical fragment of $\CoWCCTLStar$ where the counting operators are not
allowed.


%% file: SectionI.tex


\section{Monadic Tree Logic}
\label{sec:mtl}

We start by defining in~Section~\ref{sec:mtl;sub:msol} the three main logics we
shall consider: \MSO, \MTL, and \MPL.
The three languages do not differ at the syntactic level, but only on the range
of quantification  of the second-order variables.
For convenience, though, we provide a unified language (\MSOL), where the
second-order quantifiers are decorated with a symbol $\alpha$ that explicitly
indicates the domain of the quantified variable: $\SSym$ for sets, $\TSym$ for
trees, and $\PSym$ for paths.

\begin{figure}[htbp]
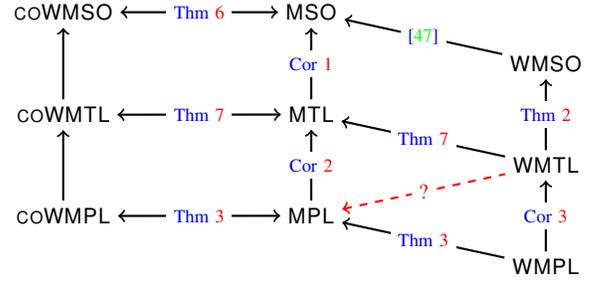

  \footnotesize\centering
  \scalebox{1.00}[1.00]{
    \figLogFinBrn
  }
  \caption{\label{fig:exprfin} Expressiveness results for \MSOL over
    finitely-branching Kripke trees.}
  \vspace{-1.5em}
\end{figure}

This section is mainly devoted to the analysis of the expressiveness of the
various semantic fragments of $\MSOL$ interpreted over non-blocking trees.
In Section~\ref{sec:mtl;sub:fqn}, we compare the expressiveness of $\MPL$,
$\MTL$, and $\MSO$ in the general case, where second-order variables are
interpreted over both finite and infinite paths, subtrees, and sets,
respectively, of the considered model tree.
Then, in Sections~\ref{sec:mtl;sub:wqn} and~\ref{sec:mtl;sub:cqn}, we consider
similar expressiveness issues for the  \emph{weak semantic variant}
(second-order variables range over finite paths, finite subtrees, and finite
sets, respectively) and the \emph{co-weak semantic variant} (second-order
variables range over infinite paths, infinite subtrees, and infinite sets,
respectively).

\begin{figure}[htbp]
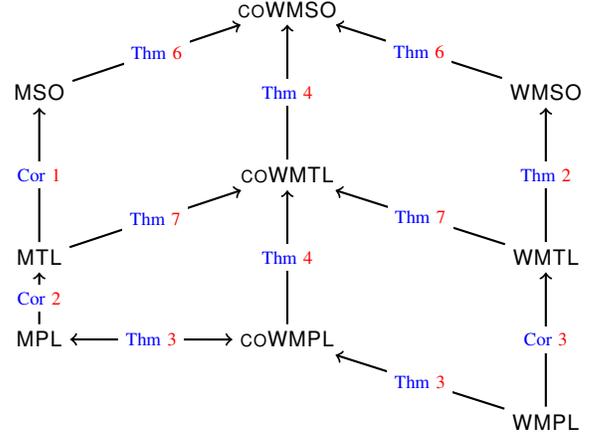

  \vspace{-1em}
  \footnotesize\centering
  \scalebox{1.00}[1.00]{
    \figLogArbBrn
  }
  \caption{\label{fig:exprarb} Expressiveness results for \MSOL over arbitrary
    Kripke trees.}
  \vspace{-.5em}
\end{figure}

Finally, in Section~\ref{sec:mtl;sub:WeakVersusCoWeak} we consider
expressiveness issues of weak semantics versus co-week semantics.
The complete picture of results is summarised in Figure~\ref{fig:exprfin} for
the case of finitely-branching Kripke trees and in Figure~\ref{fig:exprarb} for
the general case.
The two figures have to be interpreted as follows.
An edge connecting two logics has the following meaning: if the edge has a
single arrow, then the target logic is more expressive than the source;
otherwise, the two logics are expressively equivalent.
If there is no edge between two distinct logics and no relation is deducible by
the other edges, then the two logics are expressively incomparable.
A red edge decorated with a question mark indicates a currently open question.

\input{SectionI-A}

\input{SectionI-B}

\input{SectionI-C}

\input{SectionI-D}

\input{SectionI-E}


%% file: SectionI-A.tex


\subsection{Monadic Second-Order Logics}
\label{sec:mtl;sub:msol}

For a given finite set $\Prop$ of atomic propositions, \MSOL is a language
defined over the signature $\{\leq\}\cup \{P_a \mid a\in \Prop\}$, where
second-order quantification is restricted to monadic predicates, $\leq$ is a
binary predicate, and $P_a$ is a monadic predicate for each
$a\in\Prop$.

\begin{definition}[\MSOL Syntax]
  \label{def:syn(msol)}
  Given a finite set $\Prop$ of atomic propositions, a finite set $\Var_1$
  of first-order variables (or \emph{node} variables), and a finite set $\Var_2$
of second-order variables (or \emph{set} variables), the syntax of \emph{Monadic
Second-Order/Tree/Path Logic} (\MSO/\MTL/\MPL, for short) is the set of formulae
built according to the following grammar, where $\apElm \in \Prop$, $\varElm,
\yvarElm \in \Var_1$, and $\XvarElm \in \Var_2$:
\[
\varphiFrm \seteq P_a(\varElm) \mid \varElm \leq \yvarElm \mid \varElm \in
\XvarElm \mid \neg \varphiFrm \mid \varphiFrm \wedge \varphiFrm
\mid \exists \varElm \ldotp \varphiFrm
\mid \exists^{\alpha} \XvarElm \ldotp \varphiFrm,
\]
where $\apElm \in \Prop$, $\varElm, \yvarElm \in \Var_1$, $\XvarElm \in \Var_2$,
and $\alpha$ is $\SSym$ for \MSO, $\TSym$ for \MTL, and $\PSym$ for \MPL.
\end{definition}

Note that \MSO (\resp,  \MTL, \MPL) corresponds to the syntactical fragment of
\MSOL where the second-order existential quantification takes only the form
$\exists^{\SSym}$ (\resp, $\exists^{\TSym}$, $\exists^{\PSym}$).  We also exploit the
standard logical connectives $\vee$ and $\rightarrow$ as abbreviations, the
universal first-order quantifier $\forall x$, defined as $\forall
x.\varphi\DefinedAs \neg\exists x. \neg\varphi$, and the universal second-order
quantifier $\forall^{\alpha} \XvarElm$, defined as $\forall^{\alpha}
X.\varphi\DefinedAs \neg\exists^{\alpha} \XvarElm. \neg\varphi$.  We may also
make use of the shorthands (i) $x=y$ for $x\leq y \wedge y\leq x$, (ii) $x<y$
for $x\leq y \wedge \neg (y\leq x)$; (iii) $\exists x\in \XvarElm.\,\varphi$ for
$\exists x.\,(x\in \XvarElm\wedge \varphi)$, and (iv) $\forall x\in
\XvarElm.\,\varphi$ for $\forall x.\,(x\in \XvarElm \rightarrow \varphi)$.

 As usual, a \emph{free variable} of a formula $\varphi$ is a variable
 occurring in $\varphi$ that is not bound by a quantifier.  A \emph{sentence} is
 a formula with no free variables.  The language of \MSOL consists of its
 sentences.  
 We also consider the first-order fragment
 \FO  of \MSOL, where second-order quantifiers and second-order variables are not allowed.\vspace{0.2cm}

\noindent \textbf{Semantics of $\MSOL$.}
Formulas of $\MSOL$ are interpreted over Kripke trees over $\Prop$.
A Kripke tree $\LT=\tpl{T,\Lab}$ induces the relational structure with domain $T$,
where the binary predicate $\leq$ corresponds to the descendant
relation in $T$, and $P_a$ denotes the set $\{\tau\in T : a\in \Lab(\tau)\}$ of
$a$-labelled nodes.

Let us fix a Kripke tree    $\LT=(T,\Lab)$ over $\Prop$.  A \emph{first-order
valuation for the tree $T$} is a mapping $\Val_1: \Var_1 \mapsto T$ assigning to
each first-order variable a node of $T$. A \emph{second-order valuation for the
tree $T$} is a mapping $\Val_2: \Var_2 \mapsto 2^{T}$ assigning to each
second-order variable a subset of $T$.


\begin{definition}[\MSOL Semantics]
  \label{def:sem(msol)}
 Given a $\MSOL$ formula $\varphi$, a Kripke tree
 $\LT=\tpl{T,\Lab}$ over $\Prop$, a first-order valuation $\Val_1$ for $T$, and a
 second-order valuation $\Val_2$ for $T$, the satisfaction relation
 $\LT,\Val_1,\Val_2\models \varphi$, meaning that $\LT$ satisfies the formula
 $\varphi$ under the valuations $\Val_1$ and $\Val_2$, is
 defined as follows (the treatment of Boolean connectives is
 standard):
\[
 \begin{array}{l@{\hspace{2pt}}l}
\LT,\Val_1,\Val_2\models   P_a(x) &  \Leftrightarrow  a\in \Lab(\Val_1(x));\\
\LT,\Val_1,\Val_2\models x\leq y & \Leftrightarrow
\text{$\Val_1(y)$ is a descendant of}
\text{ $\Val_1(x)$ in $T$};\\
\LT,\Val_1,\Val_2\models x\in X & \Leftrightarrow \Val_1(x)\in \Val_2(X);\\
\LT,\Val_1,\Val_2\models \exists x. \varphi & \Leftrightarrow T,\Lab,\Val_1[x
    \mapsto \tau],\Val_2 \models \varphi \text{ for some }\\
    & \phantom{\Leftrightarrow}\ \tau\in T;
\end{array}
\]
\[
 \begin{array}{l@{\hspace{2pt}}l}
\LT,\Val_1,\Val_2\models \exists^{\SSym} X. \varphi & \Leftrightarrow
  \LT,\Val_1,\Val_2[X \mapsto S]\models \varphi \text{ for some set} \\
  & \phantom{\Leftrightarrow} \text{ of nodes } S \subseteq T;\\
\LT,\Val_1,\Val_2\models \exists^{\TSym} X. \varphi & \Leftrightarrow
  \LT,\Val_1,\Val_2[X \mapsto T']\models \varphi \text{ for some} \\
  & \phantom{\Leftrightarrow} \text{ subtree } T' \text{ of } T;\\
\LT,\Val_1,\Val_2\models \exists^{\PSym} X. \varphi & \Leftrightarrow
  \LT,\Val_1,\Val_2[X \mapsto \pi]\models \varphi \text{ for some path} \\
  & \phantom{\Leftrightarrow} \pi \text{ of } T.
\end{array}
\]
\emph{where for a first-order valuation $\Val_1$ for $T$, a node $\tau\in T$,
and a first-order variable $x\in T$, $\Val_1[x \mapsto \tau]$ denotes the
first-order valuation defined as follows: $\Val_1[x \mapsto \tau](x) = \tau$ and
$\Val_1[x \mapsto \tau](y)=\Val_1(y)$ if $y\not=x$.
The meaning of notation $\Val_2[X \mapsto S]$, for a second-order valuation
$\Val_2$ for $T$, a set $S\subseteq T$, and a second-order variable $X$ is
similar.}
\end{definition}

Note that the satisfaction relation $\LT,\Val_1,\Val_2\models \varphi$, for
fixed $\LT$ and $\varphi$, depends only on the values assigned by $\Val_1$ and
$\Val_2$ to the variables occurring free in $\varphi$.  In particular, if
$\varphi$ is a sentence, we say that $\LT$ \emph{satisfies} $\varphi$, written
$\LT\models \varphi$, if $\LT,\Val_1,\Val_2\models \varphi$ for some valuations
$\Val_1$ and $\Val_2$. In this case, we also say that $\LT$ is a model of
$\varphi$.
A non-blocking tree $T$ satisfies a sentence $\varphi$, written $T\models
\varphi$, if $\tpl{T,\Lab_\emptyset}\models \varphi$, where $\Lab_\emptyset$
assigns to each $T$-node the empty set.\vspace{0.2cm}
%
%

\noindent \textbf{Basic predicates expressible in $\MSOL$.} In the following we
define some useful standard
predicates which can be expressed in $\MSOL$ by using only first-order
quantification:
\begin{itemize}
  \item the second-order binary predicates $X \subset Y$, $X\subseteq Y$, $X=Y$,
    $X\neq Y$ (of expected meaning). For instance, $X\subseteq Y$ can be
    expressed as $\forall x.\, (x\in X \rightarrow x\in Y)$;
  \item the child relation is definable in $\MSOL$ by the binary predicate
    $\Child(x,y) \defeq x<y \wedge \neg \exists z.\,(x<z\wedge z<y) $;
  \item $\Path(X)$ (\resp, $\Path_\infty(X)$, \resp,
    $\Path_f(X)$) capturing the subsets of the given tree which are paths
    (\resp, infinite paths, \resp, finite paths). For example:
      \[
      \begin{array}{c}
      \Path(X) {\ \defeq\ } \forall x\in X.\,\forall y\in X.\,((x \leq y\vee y\leq x)\wedge \\ ((x<y
      \wedge \neg \Child(x,y)) \rightarrow \exists z\in X.\, (x<z\wedge z<y)).
      \end{array}
      \]
  \item the property of being a tree is captured by the second-order predicate:
      \[
      \begin{array}{c}
      \Tree(X) {\ \defeq\ } \exists x.\,\forall y.\, (x\leq y) \wedge \forall x\in X.\,\forall y\in
      X.\, \\((x<y \wedge \neg \Child(x,y)) \rightarrow \exists z\in X.\,
      (x<z\wedge z<y)).
        \end{array}
      \]
      %
      %
      %
\end{itemize}

\noindent \textbf{Weak and coWeak semantic variants.} We also consider the
\emph{Weak semantics variants} of $\MSO$, $\MTL$, and $\MPL$, denoted $\WMSO$,
$\WMPL$ and $\WMTL$, respectively. In $\WMSO$, second-order variables are
interpreted as \emph{finite} sets of the given tree. Similarly, in $\WMTL$ and
$\WMPL$, second-order quantification ranges over \emph{finite} subtrees and
\emph{finite} paths, respectively, of the given tree.
Finally, we consider the \emph{coWeak semantics variants} of $\MSO$, $\MTL$, and
$\MPL$ (written $\CoWMSO$, $\CoWMPL$ and $\CoWMTL$, respectively) where
second-order variables are interpreted as infinite sets of the given tree
(infinite paths and infinite subtrees in the case of $\CoWMPL$ and $\CoWMTL$,
respectively).


%% file: SectionI-B.tex


\subsection{Expressiveness under Full Quantifications}
\label{sec:mtl;sub:fqn}

In this section, we compare the expressiveness of the logics $\MPL$, $\MTL$, and
$\MSO$. We prove that $\MTL$ strictly lies between $\MPL$ and $\MSO$ even in the
case of \emph{finitely-branching setting} (\ie, over the class of
finitely-branching Kripke trees).  First, we show that $\MTL$ is strictly less
expressive than $\MSO$. In fact, the result already holds over the class of
$2^{\Prop}$-labelled infinite chains. For this subclass of Kripke trees,
quantification over trees reduces to quantification over paths, which in turn
can be simulated by first-order quantification~\cite{Buc66}.
Thus, since $\FO <\MSO$ even over infinite chains, we obtain the following
result (for details, see Appendix~\ref{app:mtl;sub:fqn}).

\begin{proposition}[name =, restate=propMTLChains]
  \label{prop:MTLChains}
  Over $2^{\Prop}$-labelled infinite chains, it holds that:
  \begin{itemize}
  \item $\MTL\equiv \WMTL \equiv \CoWMTL \equiv \FO <\MSO$, and
  \item by~\cite{Buc66}, $\MSO\equiv \WMSO \equiv \CoWMSO$.
\end{itemize}
\end{proposition}

Since the class of chains can be trivially captured in $\FO$, by
Proposition~\ref{prop:MTLChains}, the following hold.

\begin{proposition}\label{prop:ChainResultsCorollary}
  For all $\Logic\in \{\MSO,\WMSO,\CoWMSO\}$ and $\Logic'\in
  \{\MTL,\WMTL,\CoWMTL\}$, $\Logic \not \leq \Logic'$ even in the
  finitely-branching setting.
\end{proposition}

Clearly, $\MTL$ is subsumed by $\MSO$ (the predicate $\Tree(X)$ can be expressed
in $\MSO$). Therefore, by Proposition~\ref{prop:ChainResultsCorollary}, we
obtain the desired result.

\begin{corollary}[restate = thm:MTLversusMSO]
  \label{cor:MTLversusMSO}
  $\MTL < \MSO$ even in the finitely-branching setting.
\end{corollary}

Next, we show that $\MPL$ is strictly less expressive than $\MTL$.  Evidently,
$\MPL$ is subsumed by $\MTL$ (path quantification can be simulated by tree
quantification and first-order quantification). In order to show that $\MTL$ is
not subsumed by $\MPL$, we prove that the density property (characterising the
class of dense non-blocking trees) is definable in $\MTL$ but not in $\MPL$.
The density property can be  expressed in $\MTL$ as:
\[
\begin{array}{l}
\exists^{\TSym} X.\,\forall x\in X.\,\exists x_1\in X.\,\exists x_2\in X.\, \\ (
x<x_1 \wedge x<x_2 \wedge \neg\, x_1 \leq x_2 \wedge \neg\, x_2 \leq x_1).
\end{array}
\]
To prove that the density property cannot be expressed in $\MPL$, we need a
preliminary result that generalises the known expressiveness equivalence between
$\CoWMPL$ and $\CoWCCTLStar$~\cite{MR03}.  The easy translation of
$\CoWCCTLStar$ into $\CoWMPL$ can be trivially adapted to show that every
$\CCTLStar$ (\resp, $\WCCTLStar$) formula can be translated in linear time into
an equivalent $\MPL$ (\resp, $\WMPL$) formula. By adapting the compositional
argument in~\cite{MR03} for showing that $\CoWMPL$ is subsumed by
$\CoWCCTLStar$, we obtain the following result.

\begin{proposition}\label{prop:CountingCTLStarVersusMPL}
$\MPL \equiv \CCTLStar$ and $\WMPL \equiv \WCCTLStar$.
\end{proposition}

For each $n\geq 1$, we define two non-empty classes $\ND_n$ and $\D_n$ of
non-blocking finitely-branching trees such that the following holds
\underline{for each $n>1$}:
\begin{itemize}
  \item $\ND_n$ contains only isomorphic trees which does \emph{not} satisfy the
    density property;
  \item $\D_n$ contains only isomorphic trees which  satisfy the density property;
  \item no state formula $\varphi$ in $\CCTLStar$ with size smaller than $n$
    distinguishes the classes $\ND_n$ and $\D_n$, \ie, for all $T\in\D_n$ and
    $T'\in \ND_n$, $T\models \varphi$ \iff $T'\models \varphi$.
\end{itemize}
Thus, by Proposition~\ref{prop:CountingCTLStarVersusMPL}, it follows that the
logic $\MPL$ cannot capture the density property.

In the following, the size $|\varphi|$ of a $\CCTLStar$ formula $\varphi$ is
defined as the length of the string for representing $\varphi$, where we assume
that the natural numbers $k$ in the counting operators $\DC^{k}$ are encoded in
unary. In particular, $|\DC^{k}\varphi|=k+1+|\varphi|$.

\noindent The classes $\ND_n$ and $\D_n$ are defined by induction on $n\geq 1$:
\begin{itemize}
  \item $\ND_1$ and $\D_1$ coincide and consist of the infinite chains;
  \item for each $n>1$, $\ND_n$ is the smallest set of non-blocking trees $T$
    satisfying the following conditions:
   \begin{itemize}
     \item the root of $T$, called \emph{$\ND_n$-node}, has exactly $n\cdot
       (n-1)+1$ distinct children\vspace{0.05cm}

     $\hspace{1cm}
     \tau_{1,1},\ldots,\tau_{1,n},\ldots,
     \tau_{n-1,1},\ldots,\tau_{n-1,n},\tau_n ;
     $ \vspace{0.1cm}

     \item for all $\ell\in [1,n-1]$, the subtrees rooted at the children
       $\tau_{\ell,1},\ldots,\tau_{\ell,n}$ are in $\ND_{\ell}$;
     \item the subtree rooted at $\tau_n$ is in $\ND_{n}$.
   \end{itemize}
  \item for each $n>1$, $\D_n$ is the smallest set of non-blocking trees $T$
    satisfying the following conditions:
   \begin{itemize}
     \item the root of $T$, called \emph{$\D_n$-node}, has exactly $n\cdot
       (n-1)+2$ distinct children\vspace{0.05cm}

     $\hspace{1cm}
       \tau_{1,1},\ldots,\tau_{1,n},\ldots,\tau_{n-1,1},\ldots,\tau_{n-1,n},\tau_n,\tau'_n;
       $\vspace{0.1cm}

   \item for all $\ell\in [1,n-1]$, the subtrees rooted at the children
     $\tau_{\ell,1},\ldots,\tau_{\ell,n}$ are in $\ND_{\ell}$ (note $\ND_\ell$
     and not $\D_\ell$);
     \item the subtrees rooted at $\tau_n$ and $\tau'_n$ are in $\D_{n}$.
   \end{itemize}
\end{itemize}

\figND

By construction, the following holds.

\begin{lemma}[name =, restate=lmmClassesDandND]
\label{lmm:ClassesDandND} For all $n>1$, the trees in $\D_n$ are dense, while those in $\ND_n$ are not.
 \end{lemma}
\begin{proof}
Let $n>1$, $T\in \D_n$, and $S$ be the subset of $T$ consisting only of
$\D_n$-nodes. By construction, $S$ is a complete binary tree. Hence, $T$ is
dense.  Next, we show that for all $k\geq 1$ and $T\in\ND_k$, $T$ is not dense.
Hence, the result follows.  The proof is by induction on $k$. The case $k=1$ is
obvious since, in this case, $T$ is an infinite chain.  Now, let $k>1$. We
assume that $T$ is dense, and derive a contradiction. Hence, there is a subtree
$S$ of $T$ satisfying the following condition:
\begin{itemize}
  \item \emph{reachability
invariance}; every node of $S$ has a descendant  in $S$ having at least two distinct children in $S$.
\end{itemize}

By construction, one of the
following two conditions occurs:
\begin{itemize}
  \item there is $1\leq h<k$ and $T'\in \ND_h$ such that $S\subseteq T'$. By the
    induction hypothesis, $T'$ is not dense, and we derive a contradiction;
  \item for some node $\tau$ of $S$, the subtree $S'$ of $S$ rooted at node
    $\tau$ is a subset of a tree $T'$ in $\ND_{h}$ for some $1\leq h<k$.  Note
    that like $S$, $S'$ satisfies the reachability invariance
    condition. Conversely, by the induction hypothesis, $T'$ is not dense,
    reaching a contradiction.\qedhere
 \end{itemize}
\end{proof}

We can show that any $\CCTLStar$ state formula $\varphi$ does not distinguishes
the classes $\D_n$ and $\ND_m$ for all $n,m\geq |\varphi|$.  The proof is by structural induction on the size of $\varphi$. In the evaluation of the temporal and counting modalities, we need  to compare representatives of the classes  $\D_i$ and $\ND_j$ for (possibly distinct) indexes $i$ and $j$ satisfying the invariant $i,j\geq |\psi|$, where $\psi$ is the currently processed subformula of $\varphi$.

\begin{lemma}[name =, restate=lmmInexpressivenessCCTLStarNDandD] \label{lmm:InexpressivenessCCTLStarNDandD}
 Let $\varphi$ be a $\CCTLStar$ state formula. Then, for all $m,n>1$ such that
 $\min(m,n)\geq |\varphi|$ and for all $T,T'\in \D_n\cup \D_m \cup \ND_n \cup
 \ND_m$, the following holds: $T\models \varphi \Leftrightarrow T'\models
 \varphi$.
\end{lemma}

Thus, by
Lemmata~\ref{lmm:ClassesDandND}--\ref{lmm:InexpressivenessCCTLStarNDandD} and
the equivalence $\MPL\equiv\CCTLStar$, the following holds.

\begin{theorem}[name =, restate = thminexpressivenessDensityMPL]
  \label{thm:inexpressivenessDensityMPL}
 The density property is not expressible in $\MPL$ even in the
 finitely-branching setting.
\end{theorem}

Since the density property can be expressed in $\MTL$, and $\MPL$ is trivially
subsumed by $\MTL$, by Corollary~\ref{cor:MTLversusMSO} and
Theorem~\ref{thm:inexpressivenessDensityMPL}, we obtain the following
expressiveness hierarchy for the logics $\MPL$, $\MTL$, and $\MSO$.

\begin{corollary}
  \label{cor:HierarchyFullQuant}
  $\MPL < \MTL < \MSO$ even in the finitely-branching setting.
\end{corollary}


%% file: SectionI-C.tex


\subsection{Expressiveness under Weak Quantifications}
\label{sec:mtl;sub:wqn}

In this section, we compare the expressiveness of the weak semantic variants of
$\MSO$, $\MTL$, and $\MPL$. As a main result, we establish that $\WMTL$ strictly
lies between $\WMPL$ and $\WMSO$ even in the finitely-branching setting.
%
Clearly, $\WMTL$ is subsumed by $\WMSO$ (the requirement that a second-order
variable ranging over finite sets captures only finite subtrees of the given
tree can be expressed by using only first-order quantification). Thus, by
Proposition~\ref{prop:ChainResultsCorollary}, the following holds.

\begin{theorem}[name =, restate=thmWMTLversusWMSO]
  \label{thm:WMTLversusWMSO}
$\WMTL < \WMSO$ even in the finitely-branching setting.
\end{theorem}

Next, we show that $\WMPL<\WMTL$.
First, we observe that since quantification over finite paths can be expressed
in $\FO$, $\WMPL$ and $\FO$ are expressively equivalent and $\WMPL\leq \WMTL$.
In order to show that $\WMTL$ is more expressive than $\WMPL$ in the general
case, we consider the \emph{infinitely-branching property} requiring that a tree
is infinitely-branching.  This property can be expressed in $\MTL$ under the
weak semantics as $\neg \forall x.\, \exists^{\TSym} X.\,\forall
  y.\,(\Child(x,y) \rightarrow y\in X)$. On the other hand, it is known
by~\cite{CF11} that every satisfiable $\MSO$ formula is satisfied by a
finitely-branching Kripke tree. Thus, since $\WMPL$, $\MPL$ and $\MTL$ are
subsumed by $\MSO$, the following holds.

\begin{proposition}[name =, restate=propinexpressiveInfBranchFullQuantification]
  \label{prop:inexpressiveInfBranchMPL}
  The infinitely-branching property is not definable in the logics $\MSO$,
  $\MTL$, $\MPL$, and $\WMPL$.
 \end{proposition}

In order to show that $\WMTL$ is not subsumed by $\WMPL$ even in the
finitely-branching setting, we show that for each atomic proposition $a$, the
property (called $\emph{$a$-acceptance}$) expressed by the $\CTLStar$ formula
$\AQ\Eventually a$ is not definable in $\WMPL$. Note that the $a$-acceptance
property captures the Kripke trees such that each infinite path from the root
visits a node labeled by $a$.  In the finitely-branching setting, this property
can be expressed in $\WMTL$ by requiring that there is a finite
\emph{tree-prefix} $T_a$ of the given tree $T$ such that (i) each leaf of $T_a$ is
labeled by proposition $a$, and (ii) for each non-leaf node $\tau\in T_a$, each
child of $\tau$ in $T$ is child of $\tau$ in $T_a$ as well.

%
%

In the following, we show that the $a$-acceptance property is not expressible in
$\WMPL$. By Proposition~\ref{prop:CountingCTLStarVersusMPL}, it suffices to
prove that $a$-acceptance is not definable in $\WCCTLStar$.
show that the $\CTLStar$ formula $\AQ\Eventually a$ (called


\noindent For a $\WCCTLStar$ formula $\psi$, we say that $\psi$ is
\emph{balanced} if:
\begin{itemize}
  \item for each subformula $\psi_1\Until \psi_2$ of $\psi$, it holds that
    $|\psi_1|=|\psi_2|$;
  \item for each subformula $\EQ \theta $ of $\psi$, $\theta$ is of the form
    $\theta_1\wedge \theta_2$ with $|\theta_1|=|\theta_2|$.
\end{itemize}

Proving the inexpressiveness result of $a$-acceptance for balanced $\WCCTLStar$
state formulas allows us to state it for any $\WCCTLStar$ state formula, since
(by using conjunctions of $\top$) a $\WCCTLStar$ state formula can be trivially
converted into an equivalent balanced $\WCCTLStar$ state formula.

\figAN

Let $\Prop=\{a\}$. For each $n\geq 1$, we define two non-empty classes $\Acc_n$
and $\NAcc_n$ of \emph{finitely-branching} Kripke trees over $\Prop$ such that
the following holds for each $n\geq 1$:
\begin{itemize}
\item $\Acc_n$ contains only isomorphic Kripke trees which satisfy the
  $a$-acceptance property;
  \item $\NAcc_n$ contains only isomorphic Kripke trees which does \emph{not}
    satisfy the $a$-acceptance property;
  \item no balanced state formula $\varphi$ in $\WCCTLStar$ with size smaller
    than $n$ distinguishes the classes $\NAcc_n$ and $\Acc_n$, \ie, for all
    $(T,\Lab)\in\Acc_n$ and $(T',\Lab')\in \NAcc_n$, $(T,\Lab)\models \varphi$
    iff $(T',\Lab')\models \varphi$.
\end{itemize}

\noindent The classes $\Acc_n$ and $\NAcc_n$ are defined by induction on $n\geq
1$:
\begin{itemize}
  \item $\Acc_1$ consist of the labeled infinite chains where each node is
    labeled by proposition $a$ ($a$-node or $\Acc_1$-node);
  \item for each $n>1$, $\Acc_n$ is the smallest set of Kripke trees
    $\tpl{T,\Lab}$ satisfying the following conditions:
   \begin{itemize}
     \item the root of $T$, called \emph{$\Acc_n$-node}, has empty label and
       exactly $n$ distinct children $\tau_1,\ldots,\tau_n$;
     \item for all $\ell\in [1,n]$, the Kripke subtree rooted at the child
       $\tau_i$ is in $\Acc_{n-1}$.
   \end{itemize}
  \item for each $n\geq 1$, $\NAcc_n$ is the smallest set of Kripke trees
    $\tpl{T,\Lab}$ satisfying the following conditions:
   \begin{itemize}
     \item the root of $T$, called \emph{$\NAcc_n$-node}, has empty label and
       exactly $n+1$ distinct children $\tau_0,\tau_1,\ldots,\tau_n$;
   \item for all $\ell\in [1,n]$, the Kripke subtree rooted at the child
     $\tau_i$ is in $\Acc_{n}$;
     \item the Kripke subtree rooted at $\tau_0$ is in $\NAcc_{n}$.
   \end{itemize}
\end{itemize}

Note that since Kripke trees are infinite, it makes sense that a Kripke tree $\tpl{T,\Lab}$ and its subtree rooted at a child  of the root  are isomorphic. Hence, the class $\NAcc_n$ is well defined.
 
Let $n\geq 1$. By construction, for each Kripke tree in $\Acc_n$, each infinite
path from the root has a suffix visiting only $a$-nodes.  On the other hand, for
each Kripke tree in $\NAcc_n$, there is an infinite path from the root visiting
only nodes with empty label (in particular, $\NAcc_n$-nodes).  Hence, the
following holds.

\begin{lemma}\label{lemma:ClassesAandNA}
For all $n\geq 1$, the Kripke trees in $\Acc_n$ satisfy the $a$-acceptance
property, while the Kripke trees in $\NAcc_n$ not.
 \end{lemma}

We can show that any balanced $\WCCTLStar$ state formula $\varphi$ does not
distinguish the classes $\Acc_n$ and $\NAcc_n$ for all $n\geq |\varphi|$.

 \begin{lemma}[name =, restate=lmmhCompatibility]
 \label{lmm:hCompatibility} Let $\varphi$ be a balanced $\WCCTLStar$ state
formula. Then for all $n>|\varphi|$,
 $\tpl{T,\Lab}\in \NAcc_n$ and $\tpl{T',\Lab'}\in \Acc_n$, it holds that
 $\tpl{T,\Lab}\models \varphi$ if and only if $\tpl{T',\Lab'}\models \varphi$.
 \end{lemma}

Thus, by Lemmata~\ref{lemma:ClassesAandNA} and~\ref{lmm:hCompatibility} and
Proposition~\ref{prop:CountingCTLStarVersusMPL}, it follows that the
$a$-acceptance property cannot be expressed in $\WMPL$.  Moreover, note that
$\MPL \equiv \CoWMPL$. This is because (i) quantification over finite paths can
be expressed in $\FO$, and (ii) the requirement that a path is infinite can be
defined in $\MPL$ by using only first-order quantifications.  Thus, since
$a$-acceptance can be expressed in $\MPL$, we easily obtain the following
result.

\begin{theorem}[name =, restate=thmInexpressivenessWeakMPL]
  \label{thm:InexpressivenessWeakMPL}
No $\WMPL$ formula can express the $a$-acceptance property. Moreover, it holds
that $\WMPL \equiv \FO$, $\MPL \equiv \CoWMPL$, and $\WMPL<\MPL$, even in the
finitely-branching setting.
\end{theorem}

Since $\WMTL$ can express the infinitely-branching property and, in the
finitely-branching setting, the $a$-acceptance property too, by
Theorem~\ref{thm:WMTLversusWMSO},
Proposition~\ref{prop:inexpressiveInfBranchMPL}, and
Theorem~\ref{thm:InexpressivenessWeakMPL}, we obtain the following
expressiveness hierarchy for weak variants.

\begin{corollary}[name =, restate = corWeakVariantsHierarchy]
  \label{cor:WeakVariantsHierarchy}
  $\WMPL < \WMTL < \WMSO$ even in the finitely-branching setting.
\end{corollary}


%% file: SectionI-D.tex


\subsection{Expressiveness under coWeak Quantifications}
\label{sec:mtl;sub:cqn}

In this section, we establish an expressiveness hierarchy for the coWeak
versions of the considered logics $\MPL$, $\MTL$, and $\MSO$ similar to the one
for the corresponding Weak versions.

\begin{theorem}[name =, restate = thmCoWeakVariantsHierarchy]
  \label{thm:CoWeakVariantsHierarchy}
$\CoWMPL < \CoWMTL < \CoWMSO$ even in the finitely-branching setting.
\end{theorem}
\begin{proof}
Evidently, $\CoWMPL \leq \CoWMTL$ (quantification over infinite paths can be
simulated by quantification over infinite trees and first-order
quantification). Moreover, $\CoWMTL \leq \CoWMSO$ (the requirement that a
second-order variable in $\CoWMSO$ captures only infinite subtrees of the given
tree can be expressed by using only first-order quantification).  Since
$\CoWMPL\equiv \MPL$ (Theorem~\ref{thm:InexpressivenessWeakMPL}), by
Theorem~\ref{thm:inexpressivenessDensityMPL}, $\CoWMPL$ cannot express the
density property even in the finitely-branching setting. On the other hand, the
density property is expressible in $\CoWMTL$. Indeed, the $\MTL$ formula used in
Section~\ref{sec:mtl;sub:fqn} for expressing the density property is equivalent
to its coWeak semantics variant. Moreover,  by
Proposition~\ref{prop:ChainResultsCorollary}, 
$\CoWMSO$ is not subsumed by $\CoWMTL$.
Hence, the result directly follows.
\end{proof}

%% file: SectionI-E.tex
\subsection{Weak Quantifications versus coWeak Quantifications}
\label{sec:mtl;sub:WeakVersusCoWeak}

In this section, we compare the logics $\MTL$ and $\MSO$ with their
corresponding coWeak and Weak semantics variants.

It is known by~\cite{BIS13,ISB16} that the density property cannot be expressed
in $\WMSO$ even in the finitely-branching setting. Thus, being $\WMTL\leq
\WMSO$, the previous inexpressiveness result holds for $\WMTL$ as well.  On the
other hand, we have seen in Section~\ref{sec:mtl;sub:fqn} that the density
property can be instead expressed in $\MTL$ (hence, in $\MSO$ as
well). Moreover, in Section~\ref{sec:mtl;sub:wqn}, we have proved that the
infinitely-branching property can be expressed in $\WMTL$ (hence, in $\WMSO$
too) but not in $\MSO$ and $\MTL$ (see
Proposition~\ref{prop:inexpressiveInfBranchMPL}).  It follows that over
arbitrary Kripke trees, $\WMTL$ and $\MTL$ (\resp, $\WMSO$ and $\MSO$) are
expressively incomparable.

However, in the finitely-branching setting, it is known that $\WMSO$ is subsumed
by $\MSO$. Indeed, in this setting, the predicate $\Fin(X)$ capturing the finite
sets of the given tree can be expressed in $\MSO$ by the formula
\[
\exists^{\SSym} Y.\, \bigl(\Tree(Y)\wedge X\subseteq Y \wedge \neg
\exists^{\SSym} Z.\, (Z\subseteq Y \wedge \Path_{\infty}(Z)).
\]
Moreover, in the finitely-branching setting, assuming that $X$ is interpreted as
a subtree of the given tree, the predicate $\Fin(X)$, can be defined in $\MTL$
as $\neg \exists^{\TSym} Y.\, (Y\subseteq X \wedge \Path_{\infty}(Y))$. Thus, by
Proposition~\ref{prop:ChainResultsCorollary}, we obtain the following result.

\begin{theorem}\label{theo:FullVersusWeak}
  $\MSO \not \sim \WMSO$, $\MTL \not \sim \WMTL$, $\MSO \not \sim \WMTL$, and
  $\MTL \not \sim \WMSO$. In the finitely-branching setting, $\WMSO < \MSO$,
  $\WMTL<\MTL$, and $\MTL \not \sim \WMSO$.
\end{theorem}

Now, we show that second-order quantification over finite sets can be simulated
in $\CoWMSO$ 
by using the following characterisation of the finite
subsets of a non-blocking tree.

\begin{lemma}[name =, restate = lmmCharacterizationFiniteSet]
\label{lmm:CharacterizationFiniteSet}
Let $T$ be a non-blocking tree and $S\subseteq T$. Then, $S$ is finite
\iff the following condition is fulfilled:
\begin{itemize}
\item[(*)] there exist an infinite tree
$T_{\infty}\subseteq T$, an infinite forest $F_{\infty}\subseteq T$, and an
infinite set $Y_\infty\subseteq T$ such that:
\begin{itemize}
	\item $T_\infty$ is finitely-branching;
	\item $F_\infty \subseteq Y_\infty\subseteq T_\infty$;
	\item for each infinite path $\pi$ of $T_\infty$, there is a suffix of $\pi$
    which visits only nodes of $F_\infty$;
	\item $S= T_\infty\setminus Y_\infty$.
\end{itemize}
\end{itemize}
\end{lemma}
\begin{proof}
If Condition~$(*)$ is satisfied, then $S$ is contained in the tree $T_f$
obtained from $T_\infty$ by removing all the nodes of the forest
$F_\infty$. Since $T_\infty$ is finitely-branching and each infinity path of
$T_\infty$ eventually visits only nodes of $F_\infty$, it follows that $T_f$ is
finite. Hence, $S$ is finite as well.

Now, assume that $S$ is finite. Since $T$ is non-blocking, there must be a
finite subtree $T_f$ of $T$ such that $S\subseteq T_f\setminus L$, where $L$ is
the set of leaves of $T_f$. For each $\tau\in L$, let $\pi_\tau$ any infinite
path of $T$ starting at node $\tau$ (since $T$ is not-blocking such a path
$\pi_\tau$ exists). Let $F_\infty$ be the infinite forest given by
$\bigcup_{\tau\in F}\pi_\tau$. Define $T_{\infty}\DefinedAs T_f\cup F_\infty$
and $Y_\infty = (T_f\setminus S) \cup F_\infty$. Evidently, Condition~$(*)$ is
fulfilled.
\end{proof}

We can easily express in $\CoWMSO$ that an infinite subset of a not-blocking
tree $T$ is a tree (\resp, a forest), and that an infinite tree is
finitely-branching. In particular, assuming that a set variable $Z$ is
interpreted as an infinite tree $T_{\infty}$, the property that $T_{\infty}$ is
finitely-branching can be expressed in $\CoWMSO$ by the formula $\neg \exists
x.\, \exists^{\SSym} X.\,[X\subseteq Z\wedge x\in X \wedge \forall y\in X.\,(y=x
  \vee \Child(x,y))]$.  Thus, by Lemma~\ref{lmm:CharacterizationFiniteSet}, we
easily deduce that both $\MSO$ and $\WMSO$ are subsumed by $\CoWMSO$.

Moreover, the class of infinitely-branching trees can be captured in $\CoWMSO$
by the formula $\exists x.\, \exists X^{\SSym}.\,[x\in X \wedge \forall y\in
  X.\,(y=x \vee \Child(x,y))]$. Hence, by
Proposition~\ref{prop:inexpressiveInfBranchMPL}, it follows that $\CoWMSO$ is in
general more expressive than $\MSO$.  However, in the finitely-branching
setting, since the predicate $\Fin(X)$ is definable in $\MSO$, we have
$\MSO\equiv \CoWMSO$.  Thus, being the density property definable both in $\MSO$
and $\CoWMSO$ but not in $\WMSO$, we obtain the following result.

\begin{theorem}
  \label{thm:PowerCoWMSO}
   $\MSO < \CoWMSO$ and $\WMSO < \CoWMSO$. In the finitely-branching setting,
   $\MSO \equiv \CoWMSO$ and $\WMSO < \CoWMSO$.
\end{theorem}

Now, let us consider the coWeak semantics variant of $\MTL$. We first show that
both $\MTL$ and $\WMTL$ are subsumed by $\CoWMTL$. In other terms, second-order
quantification over finite trees can be simulated in $\CoWMTL$. We exploit the
following characterisation of the finite subtrees of a given not-blocking tree.

\begin{lemma}\label{lemma:CharacterizationFiniteTree}
Let $T$ be a not-blocking tree and $T'$ be a subtree of $T$. Then, $T'$ is
finite \iff the following condition holds:  $(*)$  there is a node
$\tau\in T$ and an infinite tree $T_{\infty}\subseteq T$ such that:
      \begin{itemize}
        \item $\tau\in T_\infty$ and $T_\infty$ is finitely-branching;
        \item each infinite path $\pi$ of $T_\infty$ visits some strict
          descendant of $\tau$ in $T$;
        \item $T'= T_\infty\setminus \{\tau'\in T\mid \tau < \tau'\}$.
      \end{itemize}
\end{lemma}
\begin{proof}
Evidently, Condition~$(*)$ entails that $T'$ is finite.  Vice versa, assume that
$T'$ is finite, and let $\tau$ be any leaf node of $T'$. Since $T$ is
non-blocking, there exists an infinite path $\pi_\tau$ of $T$ starting at node
$\tau$. Define $T_\infty \DefinedAs T'\cup \pi_\tau$. It easily follows that
Condition~$(*)$ is fulfilled.
\end{proof}

By Lemma~\ref{lemma:CharacterizationFiniteTree}, we deduce the following result.

\begin{proposition}[name =, restate=propMTLvSubumedByCoWeakMTL]
\label{prop:MTLvSubumedByCoWeakMTL}  $\MTL \leq \CoWMTL$ and $\WMTL\leq \CoWMTL$.
\end{proposition}

The infinitely-branching property can be expressed in $\CoWMTL$ by the formula
$\exists x.\, \exists^{\TSym} X.\,[x\in X \wedge \forall y\in X.\,(y=x \vee
  \Child(x,y))]$. Hence, by Propositions~\ref{prop:inexpressiveInfBranchMPL} and
\ref{prop:MTLvSubumedByCoWeakMTL}, it follows that $\CoWMTL$ is in general more
expressive than $\MTL$.  However, in the finitely-branching setting, assuming
that $X$ is interpreted as a subtree of the given tree, the predicate $\Fin(X)$
can be defined in $\MTL$. Hence, in this setting, $\MTL\equiv \CoWMTL$.  Thus,
being the density property definable both in $\MTL$ (see
Section~\ref{sec:mtl;sub:fqn}) and $\CoWMTL$ (see the proof of
Theorem~\ref{thm:CoWeakVariantsHierarchy}) but not in $\WMTL$ and $\WMSO$, by
Proposition~\ref{prop:ChainResultsCorollary}, we obtain the following result.

\begin{theorem}
  \label{thm:PowerCoWMTL}
  $\MTL < \CoWMTL$, $\WMTL < \CoWMTL$, $\MSO \not \sim \CoWMTL$, and $\WMSO \not
  \sim \CoWMTL$. In the finitely-branching setting, $\MTL \equiv \CoWMTL$,
  $\WMTL < \CoWMTL$, and $\WMSO \not \sim \CoWMTL$.
\end{theorem}

\noindent \textbf{Additional expressiveness results.} By the results established
so far, in order to have a complete picture about the expressiveness comparison
between the considered logics, we have to compare $\MPL$ with $\WMTL$ and
$\WMSO$. It is known that in the finitely-branching setting, $\CTLStar$ is not
subsumed by $\WMSO$.
This follows from~\cite{Eme96,EC80,EL86}, where the authors prove that the
formula $\E\, \G\!\F \apElm$ cannot be expressed in \AFMC, and~\cite{AN92},
where it is shown that \AFMC is equivalent to $\WMSO$.
Thus, being $\CTLStar\leq \MPL$ and
$\WMTL\leq \WMSO$, and since the infinitely-branching property can be expressed
in $\WMTL$ and $\WMSO$ but not in $\MPL$
(Proposition~\ref{prop:inexpressiveInfBranchMPL}), by
Proposition~\ref{prop:ChainResultsCorollary}, the following holds.

\begin{theorem}
  \label{thm:PowerMPL}
  $\MPL \not\sim  \WMTL$ and $\MPL \not\sim  \WMSO$. In the finitely-branching
  setting, $\MPL \not\leq  \WMTL$ and $\MPL \not\sim  \WMSO$.
\end{theorem}

It remains an open question whether in the finitely-branching setting, $\WMTL$
is subsumed by $\MPL$ or not.

%% file: SectionII.tex


\section{Connections with Temporal Logics}
\label{sec:contmplog}

The results in the previous section show that \MTL is a non-trivial fragment of
\MSO that strictly contains \MPL.
So the question of its relationship with temporal logics becomes worthy of
investigation.
To this end, we first identify a new fragment of the \emph{Graded \MC}
(\GMC)~\cite{KSV02} and its \emph{alternation-free variant} (\AFGMC), whose
semantics can be encoded in \MTL and \WMTL, respectively.
To the best of our knowledge, this is the first non-trivial example (\ie,
not subsumed by other temporal formalisms) of modal fixpoint logics, whose
semantics does not require the full power of set quantifications.
We then present a translation of \emph{Substructure Temporal Logic}
(\STLS)~\cite{BMM13,BMM15} into \coWMTL, which shows that the latter is powerful
enough to reason about games with \FO-definable goals and to encode several
verification problems, such as \emph{reactive synthesis}~\cite{PR89} and
\emph{module checking}~\cite{KVW01}.

\input{SectionII-A}

\input{SectionII-B}

\input{SectionII-C}


%% file: SectionII-A.tex


\subsection{One-Step Graded \MC}
\label{sec:contmplog;sub:onestpgmc}

\begin{table*}[!tb]
  \small\tabonestpgmc
  \caption{\label{tab:onestpgmc} Translation function $\trFun[\varElm]{} \colon
    \ThetaSet[\ZSet, \OSet] \to \MTL$ from \OSGMC to \MTL.}
  \vspace{-2em}
\end{table*}

As observed by Wolper~\cite{Wol83}, there are simple $\omega$-regular
properties that cannot be expressed in classic temporal logics, while they are
easily expressible in Kozen's Modal \MC~\cite{Koz83}.
One of the simplest examples is the existence of a path in a Kripke tree where a
given atomic proposition $\apElm$ holds true at all even positions along it:
$\nu \XvarElm. (\apElm \wedge \DMod\DMod \XvarElm)$.
As we already observed in the Introduction, this formula is witnessed by
non-connected set of nodes and each witness depends on the one two steps ahead
in the path, due to the double nesting of the modal operator $\DMod$ preceding
the fixpoint variable.
This contrasts with classic temporal logic formulae, whose translation into the
Modal \MC does not require multiple nestings of the modal operators in front of
the fixpoint variables.
For instance, the \CTL formula $\E (\apElm \U \qapElm)$ is equivalent to $\mu
\XvarElm. (\qapElm \vee (\apElm \wedge \DMod \XvarElm))$ and a single modality
separates the fixpoint operator from its variable.

On the basis of these observations, it seems natural to conjecture that
preventing multiple nestings of modalities over fixpoint variables suffices to
write formulae whose dependent sets of witnesses are always connected to each
other, while the non-connected ones are independent from one another.
This ``independence'' property seems crucial for the existence of an encoding
into \MTL, which can only predicate over connected sets of nodes and, therefore,
cannot talk about non-connected sets.
In the following, we prove this conjecture, by first introducing the
\emph{one-step fragment} of Graded \MC~\cite{KSV02}, an extension of the Modal
\MC with graded (\ie, counting) modalities~\cite{Fin72}, and then showing a
direct translation of its semantics into \MTL.

\begin{definition}
  \label{def:onestpgmc(syn)}
  The \emph{One-Step Fragment of \GMC} (\OSGMC) is the set of formulae built
  accordingly to the following context-sensitive grammar, where $\ZSet, \OSet
  \subseteq \VarSet[2]$, $\XvarElm \in \ZSet$, and $\apElm \in \APSet$:
  \begin{eqnarray*}
    {\varphiFrm[\ZSet, \OSet]}
  & {\seteq} &
    {\Ff \mid \Tt \mid \apElm \mid
    \neg \varphiFrm[\ZSet, \OSet] \mid
    \varphiFrm[\ZSet, \OSet] \wedge \varphiFrm[\ZSet, \OSet] \mid
    \varphiFrm[\ZSet, \OSet] \vee \varphiFrm[\ZSet, \OSet]}
  \\
  & &
    {\mid \DMod[\geq k]\, \varphiFrm[\OSet, \emptyset] \mid
    \BMod[< k]\, \varphiFrm[\OSet, \emptyset] \mid
    \XvarElm \mid \varthetaFrm[\emptyset, \emptyset]};
  \\
    {\varthetaFrm[\ZSet, \OSet]}
  & {\seteq} &
    {\mu \XvarElm \ldotp \varthetaFrm[\ZSet \cup \{ \XvarElm \}, \OSet \cup \{
    \XvarElm \}] \mid
    \nu \XvarElm \ldotp \varthetaFrm[\ZSet \cup \{ \XvarElm \}, \OSet \cup \{
    \XvarElm \}] \mid
    \varphiFrm[\ZSet, \OSet]}.
  \end{eqnarray*}
  $\PhiSet[\ZSet, \OSet]$ (\resp, $\ThetaSet[\ZSet, \OSet]$) denotes the set of
  formulae described by the first (\resp, second) rule called base (\resp,
  fixpoint) formulae, where every occurrence of a variable is positive (\ie,
  within the scope of an even number of negations).
  Formulae from $\PhiSet[\emptyset, \emptyset]$ and $\ThetaSet[\emptyset,
  \emptyset]$ are called \emph{sentences}.
\end{definition}

The two sets $\ZSet$ and $\OSet$ of fixpoint variables, called \emph{zero-} and
\emph{one-step variables}, respectively, identify the only free variables that
can occur in a \OSGMC formula.
Specifically, the variables in $\ZSet$ can be used out of the scope of any
modalities, while those in $\OSet$ need to occur inside a single nesting of a
modality.
No nesting of modalities is allowed before reaching a fixpoint variable from the
corresponding fixpoint operator.

Examples of \OSGMC sentences are the encodings $\mu \XvarElm \ldotp (\qapElm
\vee (\apElm \wedge \DMod[\geq 1] \XvarElm))$ and $\nu \XvarElm \ldotp \mu
\YvarElm \ldotp (\apElm \wedge \DMod[\geq 1]\, \XvarElm) \vee (\DMod[\geq 1]\,
\YvarElm)$ of the \CTL and \CTLS state formulae $\E (\apElm \U \qapElm)$ and
$\E\, \G\!\F \apElm$, respectively.
Another example of formula from the set $\ThetaSet[\{ \YvarElm \}, \{ \ZvarElm
\}]$ is $\mu \XvarElm. (\XvarElm \vee \YvarElm) \vee \DMod[\geq 2](\XvarElm \vee
\ZvarElm \wedge \BMod[< 1] \apElm)$.
On the contrary, neither $\DMod[\geq 1] \YvarElm$ nor $\DMod[\geq 1] \BMod[\geq
1] \ZvarElm$ belong to $\ThetaSet[\{ \YvarElm \}, \{ \ZvarElm \}]$.
In the first case, indeed, $\YvarElm$ is not a one-step variable, so it is not
allowed in the scope of a modality.
It is still, however, a \OSGMC formula, since it belongs, \eg, to
$\ThetaSet[\emptyset, \{ \YvarElm \}]$.
The second formula, instead, does not belong at all to the one-step fragment,
since the variable $\ZvarElm$ occurs in the scope of two nested modalities.

The semantics of \OSGMC is completely standard (see, \eg,~\cite{KSV02} for the
full definition). Given a Kripke tree $\TName$ and a set of variables $\VSet$,
let $\AsgSet[\TName](\VSet)$ be the set of assignments mapping each variable in
$\VSet$ to some set of nodes of $\TName$.  For every \OSGMC formula
$\varthetaFrm \in \ThetaSet[\ZSet, \OSet]$, Kripke tree $\TName$, and assignment
$\asgElm \in \AsgSet[\TName](\ZSet \cup \OSet)$ over the free variables from
$\ZSet \cup \OSet$, the denotation $\denot{\varthetaFrm}[\asgElm][\TName]$ is
defined recursively on the structure of $\varthetaFrm$.  Here, we only report
the cases for the counting modalities and fixpoint operators, where
$\post{\wElm}$
denotes, as usual, the set of children of the node $\wElm \in \TSet$:
\begin{itemize}
\item
  $\denot{\DMod[\geq k]\, \varphiFrm}[\asgElm][\TName] \defeq \set{ \wElm \in
  \TSet }{ \card{\post{\wElm} \cap \denot{\varphiFrm}[\asgElm][\TName]} \geq k
  }$;
\item
  $\denot{\BMod[< k]\, \varphiFrm}[\asgElm][\TName] \defeq \set{ \wElm \in \TSet
  }{ \card{\post{\wElm} \setminus \denot{\varphiFrm}[\asgElm][\TName]} < k }$;
\item
  $\denot{\mu \XvarElm \ldotp \varthetaFrm}[\asgElm][\TName] \defeq \bigcap
  \set{ \WSet \subseteq \TSet }{ \denot{\varthetaFrm}[{{\asgElm}[\XvarElm
  \mapsto \WSet]}][\TName] \subseteq \WSet }$;
\item
  $\denot{\nu \XvarElm \ldotp \varthetaFrm}[\asgElm][\TName] \defeq \bigcup
  \set{ \WSet \subseteq \TSet }{ \WSet \subseteq
  \denot{\varthetaFrm}[{{\asgElm}[\XvarElm \mapsto \WSet]}][\TName] }$.
\end{itemize}
The satisfaction relation $\TName \models \varphi$ between a Kripke tree and
sentence holds when the root of $\TName$ belongs to $\denot{\varphi}$.

Let us first observe that \OSGMC is able to express the density property of
trees by means of the sentence
\[
  \varphiFrm[Den]
\defeq
  \nu \XvarElm \ldotp \mu \YvarElm \ldotp (\DMod[\geq 2]\, \XvarElm) \vee
  (\DMod[\geq 1]\, \YvarElm).
\]
Indeed, if $\TName \models \varphiFrm[Den]$, the root of $\TName$ belongs to the
denotation $\DeltaSet \defeq \denot{{\varphiFrm[Den]}}[\emptyfun][\TName]$.
By the semantics of greatest-fixpoint, we have $\denot{{\mu \YvarElm \ldotp
(\DMod[\geq 2]\, \XvarElm) \vee (\DMod[\geq 1]\, \YvarElm)}}[\{ \XvarElm \mapsto
\DeltaSet \}][\TName] = \DeltaSet$, obtained by evaluating the least-fixpoint
subformula $\mu \YvarElm \ldotp (\DMod[\geq 2]\, \XvarElm) \vee (\DMod[\geq 1]\,
\YvarElm)$ on the assignment mapping the variable $\XvarElm$ to the entire
denotation $\DeltaSet \subseteq \TSet$.
This equality implies that every node $\vElm$ in $\DeltaSet$ satisfies one of
the following: \emph{(a)} $\vElm$ has at least two distinct children in
$\DeltaSet$; \emph{(b)} $\vElm$ is able to reach a node in $\DeltaSet$ that
satisfies Property~\emph{(a)}.
Hence, $\DeltaSet$ precisely identifies the set of nodes that form a subtree
each of whose nodes has two distinct strict descendants in $\DeltaSet$.
Therefore, $\TName$ enjoys the density property.
On the other hand, if $\TName$ enjoys the density property, there exists a set
$\MSet\subseteq \TSet$ of nodes corresponding to an infinite binary-tree minor
of $\TName$.
Consider the denotation $\DeltaSet' \defeq \denot{{\mu \YvarElm \ldotp
(\DMod[\geq 2]\, \XvarElm) \vee (\DMod[\geq 1]\, \YvarElm)}}[\{ \XvarElm \mapsto
\MSet \}][\TName]$ of least-fixpoint subformula $\mu \YvarElm \ldotp (\DMod[\geq
2]\, \XvarElm) \vee (\DMod[\geq 1]\, \YvarElm)$ of $\varphiFrm[Den]$ for the
assignment mapping the greatest-fixpoint variable $\XvarElm$ to $\MSet$.
Due to the definition of the set $\MSet$, every node in it can reach at least
two distinct nodes in $\MSet$ as well.
Thus, $\MSet \subseteq \DeltaSet'$, which implies that $\MSet \subseteq
\denot{{\varphiFrm[Den]}}[\emptyfun][\TName]$, by the semantics of
greatest-fixpoint.
Moreover, the root of $\TName$ belongs to $\DeltaSet'$, since it can reach any
node in $\MSet$.
Therefore, $\TName \models \varphiFrm[Den]$ and we have the following result.

\begin{table*}[!tb]
  \small\tabaltfreonestpgmc
  \caption{\label{tab:altfreonestpgmc} Translation function $\trFun[\varElm]{}
    \colon \ThetaSet[\ZSet, \OSet][\ASym\FSym] \to \WMTL$ from \OSAFGMC to
    \WMTL, on finitely-branching trees.}
  \vspace{-2em}
\end{table*}

\begin{theorem}[restate = thmonestpgmcprp]
  \label{thm:onestpgmcprp}
  The density property is expressible in \OSGMC.
\end{theorem}

As an immediate corollary of Theorem~\ref{thm:onestpgmcprp}, jointly with
Theorem~\ref{thm:inexpressivenessDensityMPL} and the observation made
in~\cite{BIS13,ISB16} on the inability of \WMSO to characterise the class of
dense trees, we obtain the following expressiveness relation.

\begin{corollary}[restate = coronestpgmcexp]
  \label{cor:onestpgmcexp}
  $\OSGMC \not\leq \MPL$ and $\OSGMC \not\leq \WMSO$.
\end{corollary}

At this point, we can turn to the encoding of the semantics of \OSGMC formulae
into \MTL.
With this aim in mind, we first identify a one-step simulation property enjoyed
by the modal base underlying \OSGMC.
This property rests on an ordering relation on variable assignments called
\emph{one-step simulation}.
Given two assignments $\asgElm, \asgElm' \in \AsgSet(\ZSet \cup \OSet)$, we
state that $\asgElm$ is \emph{one-step simulated} by $\asgElm'$ \wrt a set of
nodes $\WSet \subseteq \TSet$, in symbols $\asgElm \sqsubseteq_{\WSet}^{\ZSet,
\OSet} \asgElm'$, if
\begin{itemize}
\item
  $\asgElm(\XvarElm) \cap \WSet \subseteq \asgElm'(\XvarElm)$, for all $\XvarElm
  \in \ZSet$, and
\item
  $\asgElm(\XvarElm) \cap \post{\WSet} \subseteq \asgElm'(\XvarElm)$, for all
  $\XvarElm \in \OSet$,
\end{itemize}
where $\post{\WSet}$
denotes, as usual, the set of children in $\TName$ of the nodes in $\WSet$.
Essentially, $\asgElm'$ assigns to any zero-step variable at least as many
elements of the context set $\WSet$ as $\asgElm$ and assigns to any one-step
variable all the children of nodes in $\WSet$ that $\asgElm$ assigns.
The informal reading of $\asgElm \sqsubseteq_{\WSet}^{\ZSet, \OSet} \asgElm'$ is
that $\asgElm'$ contains as much information about $\WSet$ as $\asgElm$, when
the visibility on $\WSet$ is limited to at most one step ahead.

We can now show that the semantics of the base fragment of \OSGMC is monotone
\wrt the one-step simulation relation relativised to the same set of nodes
$\WSet$.
Informally, if an assignment $\asgElm$ is simulated by another assignment
$\asgElm'$ \wrt the set of nodes $\WSet$, then every node from $\WSet$ that
belongs to the denotation of a base formula $\varphiFrm$ \wrt $\asgElm$ also
belongs to the denotation of $\varphiFrm$ \wrt $\asgElm'$.
Obviously, this property, ensured by the syntactic restriction on the nesting of
modal operators, is enjoyed, \eg, by the one-step formulae $\DMod[\geq 1]
\XvarElm$ and $\BMod[< 3] \XvarElm$, but not by the non-one-step formulae
$\DMod[\geq 1]\DMod[\geq 2] \XvarElm$ and $\BMod[< 2]\DMod[\geq 1] \XvarElm$.
The \emph{one-step monotonicity} property is stated as follows and can easily be
proved by induction on the structure of the base \OSGMC formulae.
\begin{lemma}[name =, restate = lmmonestpmodlog]
  \label{lmm:onestpmodlog}
  For every base \OSGMC formula $\varphiFrm \in \PhiSet[\ZSet, \OSet]$, Kripke
  tree $\TName$, set of nodes $\WSet \subseteq \TSet$, and pair of assignments
  $\asgElm, \asgElm' \in \AsgSet[\TName](\ZSet \cup \OSet)$ satisfying $\asgElm
  \sqsubseteq_{\WSet}^{\ZSet, \OSet} \asgElm'$, it holds that
  $\denot{\varphiFrm}[\asgElm][\TName] \cap \WSet \subseteq
  \denot{\varphiFrm}[\asgElm'][\TName]$.
\end{lemma}
The monotonicity property is at the core of the ``independence'' property of the
semantics of all \OSGMC formulae mentioned above.
Indeed, we show that every maximal connected component $\DeltaSet[\wElm]$ of the
denotation $\DeltaSet \defeq \denot{\varthetaFrm}[\asgElm][\TName]$ \wrt an
assignment $\asgElm$ of a formula $\varthetaFrm$ rooted at some node $\wElm \in
\DeltaSet$ (\ie, a maximal subtree rooted at $\wElm$ and fully contained in
$\DeltaSet$) can be also computed by using only the restriction to
$\DeltaSet[\wElm]$ of the interpretation of the fixpoint variables.
Essentially, the fact that a node $\vElm$ of $\DeltaSet[\wElm]$ belongs to
$\DeltaSet$ is independent of whether any other node outside (the one-step
extension of) $\DeltaSet[\wElm]$ belongs to $\DeltaSet$ or not.
In other words, disconnected parts of the denotation cannot affect each other.

The maximal connected component of a given set of nodes $\WSet \subseteq \TSet$
rooted at a node  $\wElm \in \TSet$ can be defined as
\[
  \WSet \!\root[\wElm]
\defeq
  \set{ \vElm \in \WSet }{ \wElm \leq \vElm \land \forall \uElm \in \TSet \ldotp
  (\wElm \leq \uElm < \vElm) \implies \uElm \in \WSet },
\]
while the $(\ZSet, \OSet)$-restriction of an assignment $\asgElm \in
\AsgSet[\TName](\ZSet \cup \OSet)$ to the (one-step extension of) $\WSet
\subseteq \TSet$ is defined as
\[
  (\asgElm \rst[\WSet])(\XvarElm)
\defeq
  \begin{cases}
    {\asgElm(\XvarElm) \cap \WSet},
  & \text{if } {\XvarElm \in \ZSet \setminus \OSet};
  \\
    {\asgElm(\XvarElm) \cap \post{\WSet}},
  & \text{if } {\XvarElm \in \OSet \setminus \ZSet};
  \\
    {\asgElm(\XvarElm) \cap (\WSet \cup \post{\WSet})},
  & \text{otherwise}.
  \end{cases}
\]

The \emph{``independence'' property} can be formalised as follows and proved by
induction on the nesting of fixpoint operators, where the base case is proved by
exploiting Lemma~\ref{lmm:onestpmodlog}.

\begin{lemma}[name =, restate = lmmonestpgmc]
  \label{lmm:onestpgmc}
  For every fixpoint \OSGMC formula $\varthetaFrm \in \ThetaSet[\ZSet, \OSet]$,
  Kripke tree $\TName$, assignment $\asgElm \in \AsgSet[\TName](\ZSet \cup
  \OSet)$, and node $\wElm \in \DeltaSet \defeq
  \denot{\varthetaFrm}[\asgElm][\TName]$, it holds that $\DeltaSet[\wElm] \defeq
  \DeltaSet \root[\wElm] = \denot{\varthetaFrm}[\asgElm'][\TName] \root[\wElm]$,
  where $\asgElm' \defeq \asgElm \rst[{\DeltaSet[\wElm]}]$.
\end{lemma}

Table~\ref{tab:onestpgmc} reports a translation function $\trFun{} \colon
\VarSet[1] \to (\ThetaSet[\ZSet, \OSet] \to \MTL)$ turning each \OSGMC formula
into an equivalent \MTL one.
All cases, but those for the fixpoint operators, are standard (see,
\eg,~\cite{JL04}) and reported here just for completeness.
The real interesting case is the one for the greatest-fixpoint formulae.
The idea here is to exploit the ``independence'' property stated above and
reduce the condition for a node $\wElm$ to belong to the greatest-fixpoint to
the condition that $\wElm$ belongs to a subtree which is also a post-fixpoint.
The translation of least-fixpoint formulae combines the translation for the
greatest-fixpoint with the well-known \MC duality property $\mu \XvarElm \ldotp
\varthetaFrm \equiv \neg \nu \XvarElm \ldotp \neg {\varthetaFrm}[\XvarElm / \neg
\XvarElm]$ connecting the two fixpoint operators, where ${\varthetaFrm}[\XvarElm
/ \neg \XvarElm]$ denotes the formula obtained by uniformly replacing each
occurrence of the variable $\XvarElm$ in $\varthetaFrm$ with its negation $\neg
\XvarElm$.
Note that, since negations between fixpoint operators are not allowed in \OSGMC,
we transform the formula $\neg {\varthetaFrm}[\XvarElm / \neg \XvarElm]$ into an
equivalent one in positive normal form via the auxiliary function $\pnfMac{}
\colon \GMC \to \GMC$.
In this way, we ensure that, if $\varthetaFrm$ is a \OSGMC formula,
$\pnfMac{\neg {\varthetaFrm}[\XvarElm / \neg \XvarElm]}$ is a \OSGMC formula as
well.

At this point, the following result can be obtained via structural induction, by
showing that, for every formula $\varthetaFrm \in \ThetaSet[\ZSet, \OSet]$,
Kripke tree $\TName$, assignment $\asgElm \in \AsgSet[\TName](\ZSet \cup
\OSet)$, and node $\wElm \in \TSet$, it holds that $\wElm \in
\denot{\varthetaFrm}[\asgElm][\TName]$ \iff $\TName, \{ \varElm \mapsto \wElm
\}, \asgElm \models \trFun[\varElm]{\varthetaFrm}$.

\begin{theorem}[name =, restate = thmonestpgmc]
  \label{thm:onestpgmc}
  $\OSGMC \leq \MTL$.
\end{theorem}


%% file: SectionII-B.tex


\subsection{Alternation-Free One-Step Graded \MC}
\label{sec:contmplog;sub:altfreonestpgmc}

Alternation-Free Modal \MC (\AFMC), namely the fragment of \MC where no
alternation of fixpoint operators is allowed, is a quite expressive, still much
easier, fragment of Modal \MC that can be encoded in \WMSO, when
finitely-branching trees are considered~\cite{AN92,FVZ13,CFVZ14,CFVZ20}.
Here we analyse the alternation-free fragment of \OSGMC (\OSAFGMC) and prove
that its semantics can be encoded in \WMTL on the same class of trees.  Note
that \AFMC is known to be equivalent to \WMSO. Hence, dropping the
alternation-freeness constraint from \OSAFGMC would immediately lead us outside
\WMSO and, therefore, \WMTL.
We also look at the expressive power of \OSAFGMC in comparison with \WMPL and a
graded-on-path extension of \CTL, showing that this fragment is an interesting
logic on its own.

Let $\ThetaSet[\ZSet, \OSet][\ASym\FSym]$ and $\PhiSet[\ZSet,
\OSet][\ASym\FSym]$ denote the subsets of $\ThetaSet[\ZSet, \OSet]$ and
$\PhiSet[\ZSet, \OSet]$, respectively, containing all and only alternation-free
formulae.

It is immediate to see that the \OSAFGMC sentence $\mu \XvarElm \ldotp \aapElm
\vee (\BMod[< 1]\, \XvarElm) \in \ThetaSet[\emptyset, \emptyset][\ASym\FSym]$,
equivalent to the \CTL formula $\A \F \aapElm$, encodes the $\aapElm$-acceptance
property.

\begin{theorem}[restate = thmaltfreonestpgmcprp]
  \label{thm:altfreonestpgmcprp}
  The $\aapElm$-acceptance property is expressible in \OSAFGMC.
\end{theorem}

As an immediate corollary of Theorems~\ref{thm:altfreonestpgmcprp}
and~\ref{thm:InexpressivenessWeakMPL}, we obtain the following result.

\begin{corollary}[restate = coraltfreonestpgmcexpa]
  \label{cor:altfreonestpgmcexpa}
  $\OSAFGMC \not\leq \WMPL$.
\end{corollary}

It is well-known that \CTL is strictly subsumed by the \AFMC~\cite{Eme96,CGP02},
thanks to the one-step unfolding properties of the temporal operators $\U$ and
$\R$.
Thus, obviously, $\CTL < \OSAFGMC$ holds as well.
We can show, however, a stronger property.
In~\cite{BMM09,BMM10,BMM12}, different counting variants of \CTL and \CTLS than
those considered in the previous section have been proposed, called \emph{Graded
\CTL} (\GCTL) and \emph{Graded \CTLS} (\GCTLS), where classic path quantifiers
$\E$ and $\A$ are replaced with their graded versions $\E[][\geq k]$ and $\A[][<
k]$.
These can be read informally as ``there are at least $k$ paths'' and ``all but
less than $k$ paths'', respectively.
Now, Theorem 5.4 of~\cite{BMM12} shows that \GCTL can be encoded into \GMC via a
generalisation of the classic one-step unfolding properties.
A closer inspection of the proof, though, reveals that the translation only uses
fixpoint variables within the range of a single modal operators.
Hence, the following can be obtained.

\begin{theorem}[restate = thmaltfreonestpgmcexpb]
  \label{thm:altfreonestpgmcexpb}
  $\GCTL \leq \OSAFGMC$.
\end{theorem}

To prove that $\OSAFGMC \leq \WMTL$ on finitely-branching trees, we first need
to introduce some notation and prove some auxiliary properties that hold true
even for the full \GMC.

Given a \GMC formula $\varphiFrm$ and a variable $\XvarElm \in \VarSet[2]$, we
denote with $\varphiFrm \down[\XvarElm]$ and $\varphiFrm \up[\XvarElm]$, called
\emph{time-zero suppression}, the formulae obtained from $\varphiFrm$ by
replacing each free occurrence of $\XvarElm$ not in the scope of a modal
operator with $\Ff$ and $\Tt$, respectively.
\Eg, $(\mu \YvarElm \ldotp (\apElm \vee \XvarElm) \wedge \DMod (\XvarElm \vee
\YvarElm)) \down[\XvarElm] = \mu \YvarElm \ldotp (\apElm \vee \Ff) \wedge
\DMod[\geq 1] (\XvarElm \vee \YvarElm)$ and $(\XvarElm \wedge \mu \XvarElm
\ldotp (\apElm \vee \XvarElm) \wedge \BMod[< 2] \XvarElm) \up[\XvarElm] = \Tt
\wedge \mu \XvarElm \ldotp (\apElm \vee \XvarElm) \wedge \BMod[< 2] \XvarElm$.
For the sake of space, the formal definition of these syntactic transformations
is given in Appendix~II.\ref{app:contmplog;sub:altfreonestpgmc}.
Intuitively, the two time-zero suppressions ensure that the membership of a node
to the denotation of the resulting formula does not depend on the interpretation
of the specified variable at ``time-zero''.

\begin{proposition}
  \label{prp:gmc}
  For every \GMC formula $\varphiFrm$, variable $\XvarElm \in
  \free{\varphiFrm}$, Kripke tree $\TName = (\TSet, \Lab)$, set of nodes $\WSet
  \subseteq \TSet$, node $\wElm \in \TSet$, and assignment $\asgElm \in
  \AsgSet[\TName](\free{\varphiFrm})$, the following holds true:
  \begin{enumerate}[a)]
  \item\label{prp:gmc(mon)}
    $\denot{\varphiFrm \down[\XvarElm]}[\asgElm][\TName] \subseteq
    \denot{\varphiFrm \up[\XvarElm]}[\asgElm][\TName]$;
  \item\label{prp:gmc(shn)}
    $\denot{\varphiFrm}[\asgElm][\TName] = \denot{\varphiFrm \!\down[\XvarElm]
    \vee\, (\XvarElm \wedge \varphiFrm \up[\XvarElm]) }[\asgElm][\TName]$;
  \item\label{prp:gmc(dwn)}
    $\wElm \in \denot{\varphiFrm \down[\XvarElm]}[{{\asgElm}[\XvarElm \mapsto
    \WSet]}][\TName]$ \iff $\wElm \in \denot{\varphiFrm
    \down[\XvarElm]}[{{\asgElm}[\XvarElm \mapsto \WSet \setminus
    \{ \wElm \}]}][\TName]$;
  \item\label{prp:gmc(ind)}
    $\wElm \in \denot{\varphiFrm \down[\XvarElm]}[{{\asgElm}[\XvarElm \mapsto
    \WSet]}][\TName]$ \iff $\wElm \in \denot{\varphiFrm}[{{\asgElm}[\XvarElm
    \mapsto \WSet \setminus \{ \wElm \}]}][\TName]$.
  \end{enumerate}
\end{proposition}

Items~\ref{prp:gmc(mon)}-\ref{prp:gmc(dwn)} of the above proposition can easily
be obtained by structural induction on the formula $\varphiFrm$.
In particular, Item~\ref{prp:gmc(mon)} is used in the proof of
Item~\ref{prp:gmc(shn)} (see~\cite[Lemma 9.1.1]{AN01} for an idea of proof),
while Item~\ref{prp:gmc(ind)} is an immediate consequence of
Items~\ref{prp:gmc(shn)} and~\ref{prp:gmc(dwn)}.
It may be interesting to observe that Item~\ref{prp:gmc(shn)} is a
generalisation of Shannon's lemma for Boolean function.
Moreover, Item~\ref{prp:gmc(dwn)} formally states that a node belongs to the
denotation of a formula with a variable $\XvarElm$ suppressed regardless of its
membership in the interpretation of that variable.

\begin{table*}[t]
  \small\tabstl
  \caption{\label{tab:stl} Translation function $\trFun[\XvarElm, \varElm]{}
    \colon \STLS \to \MTL$ from \STLS to \MTL.}
  \vspace{-1.5em}
\end{table*}

Essentially, the time-zero suppressions of a formula $\varthetaFrm$ are used in
the following to make sure that the presence of a node in the denotation of
$\varthetaFrm$ (typically, the argument of some fixpoint operator) is granted
solely by the presence of its descendants and its inclusion in the assignment is
indeed redundant.
This is crucial to prove the next result, where we characterise the semantics of
alternation-free least-fixpoint formulae, by means of finite trees.
Specifically, thanks to Kleene's Theorem and the fact that the underlying tree
is finitely-branching, the witness for the membership of a node to the
denotation of these formulae, once the fixpoint variable is suppressed, is
always finite.
Moreover, Item~\ref{prp:gmc(ind)} of Proposition~\ref{prp:gmc} ensures that such
a witness is indeed a least fixpoint (and not an arbitrary one), since a node
belongs to the denotation independently of its membership to the interpretation
of the fixpoint variable.

\begin{lemma}[name =, restate = lmmaltfreonestpgmc]
  \label{lmm:altfreonestpgmc}
  For every \OSGMC formula $\varthetaFrm = \mu \XvarElm[1] \ldots \mu
  \XvarElm[k] \ldotp \varphiFrm \!\in\! \ThetaSet[\ZSet, \OSet]$, with
  $\varphiFrm \!\in\! \PhiSet[{\ZSet \cup \{ \XvarElm[1], \ldots, \XvarElm[k]
  \}, \OSet \cup \{ \XvarElm[1], \ldots, \XvarElm[k] \}}]$, finitely-branching
  Kripke tree $\TName = (\TSet, \Lab)$, node $\wElm \in \TSet$, and assignment
  $\asgElm \in \AsgSet[\TName](\ZSet \cup \OSet)$, the following properties are
  equivalent:
  \begin{itemize}
  \item
    $\wElm \in \denot{\varthetaFrm}[\asgElm][\TName]$;
  \item
    there exists a finite tree $\WSet \subseteq \TSet$ with $\wElm \in \WSet$
    such that $\vElm \in \denot{{\varphiFrm}[\XvarElm[1] / \YvarElm, \ldots,
    \XvarElm[k] / \YvarElm] \down[\YvarElm]}[{{\asgElm}[ {\YvarElm \mapsto \WSet
    \root[\vElm]} ]}][\TName]$, for all $\vElm \in \WSet$.
  \end{itemize}
\end{lemma}

Table~\ref{tab:altfreonestpgmc} reports the two cases for which the translation
function $\trFun{} \colon\!\! \VarSet[1] \!\to\! (\ThetaSet[\ZSet,
\OSet][\ASym\FSym] \!\to\! \WMTL)$ from \OSAFGMC into \WMTL differs from the one
for the general case.
As opposed to the original function, in this case it is the translation of the
greatest-fixpoint formulae that is derived from the ones for the least-fixpoint
via the duality property.
The translation for the latter is then obtained by simply encoding the property
stated in the above lemma, where we use the auxiliary formula
$\maxsubtreeMac{\YvarElm, \XvarElm, \varElm}$ to identify the maximal subtree
$\YvarElm$ fully included in the witness $\XvarElm$ and rooted at some given
node $\varElm$.
Note also that all least-fixpoint operators are merged together and transformed
as a monolithic entity, thanks to the following classic equivalence (see,
\cite[Proposition 1.3.2]{AN01}): $\mu \XvarElm[1] \ldots \mu \XvarElm[k] \ldotp
\varphiFrm \equiv \mu \YvarElm \ldotp {\varphiFrm}[\XvarElm[1] / \YvarElm,
\ldots, \XvarElm[k] / \YvarElm]$.
The expressiveness result reported below is obtained by exploiting the same line
of reasoning used for the proof of Theorem~\ref{thm:onestpgmc}.

\begin{theorem}[name =, restate = thmaltfreonestpgmc]
  \label{thm:altfreonestpgmc}
  $\OSAFGMC \leq \WMTL$ on finitely-branching trees.
\end{theorem}


%% file: SectionII-C.tex


\subsection{Substructure Temporal Logic}
\label{sec:contmplog;sub:stl}

In~\cite{BMM13,BMM15}, an extension of \CTLS, called Substructure Temporal Logic
(\STLS), is proposed, with the distinctive feature of being able to implicitly
predicate over non-blocking substructures of the underlying Kripke model.
When such models are trees, this reduces to reasoning about subtrees.
The logic is obtained by adding four (future and past) temporal-like operators
$\UU$ (until), $\RR$ (release), $\SS$ (since), and $\BB$ (before), called
\emph{semilattice operators}, whose interpretation is given relative to the join
semilattice induced by the partial order on subtrees.

For the sake of space, here we only recall the semantics of the until formulae
$\varphiFrm[1] \UU[\phiFrm] \varphiFrm[2]$, which is given for a fixed
Kripke-tree model $\TName[][*]$ and one of its non-blocking subtrees $\TName$
(see Appendix III~for the full definition).
The satisfaction relation $\TName \kmodels{\TName[][*]} \varphiFrm[1]
\UU[\phiFrm] \varphiFrm[2]$ holds if there exists a non-blocking
$\phiFrm$-preserving strict subtree $\TName'$ of $\TName$ such that $\TName'
\kmodels{\TName[][*]} \varphiFrm[2]$ and, for all non-blocking
$\phiFrm$-preserving trees $\TName''$ strictly lying between $\TName'$ and
$\TName$, it holds that $\TName'' \kmodels{\TName[][*]} \varphiFrm[1]$.
Here, $\phiFrm$-preserving means that all the children in $\TName$ of the nodes
in $\TName'$ that satisfy the formula $\phiFrm$ are kept in $\TName'$ as well.

Table~\ref{tab:stl} reports the translation function $\trFun{} \colon \VarSet[2]
\times \VarSet[1] \!\to\! (\STLS \!\to\! \MTL)$ from \STLS into \MTL that
encodes the semantics of the four semilattice operators.
The formula $\nonblockingMac{\XvarElm}$ encodes the non-blocking property for
the tree $\XvarElm$, while $\YvarElm \!\sqsubset_{\phiFrm}^{\XvarElm}\!
\ZvarElm$ encodes the $\phi$-preserving subtree relation.
The encoding of the remaining part of the logic, which corresponds to \CTLS, is
the same as the one into \MPL proposed in~\cite{HT87} and is not reported here.
The comparison between \STLS and \MPL was left open in~\cite{BMM15}. Thanks
again to Theorem~\ref{thm:inexpressivenessDensityMPL} and the fact that \STLS
can express the density property~\cite{BMM15}, the following theorem answers
that question.

\begin{theorem}[name =, restate = thmstls]
  \label{thm:stls}
  $\STLS \leq \MTL$ and $\STLS \not\leq \MPL$.
\end{theorem}

Section~5 of~\cite{BMM15} carries out an analysis of several problems involving
reasoning about games.
There it is shown that \STLS can encode \emph{\LTL reactive
synthesis}~\cite{PR89}, \emph{\CTLS module checking}~\cite{KVW01}, and the
solution of turn-based/concurrent zero/non-zero-sum games with \LTL, hence
\FO-definable, goals.
Thanks to the above theorem, the same abilities are inherited by \MTL.


%% file: Discussion.tex


\section{Discussion}

We have introduced and studied \MTL, a variant of \MSO over non-blocking trees,
where the domain of the second-order variables is restricted to subtrees of the
original structure.
An extensive comparison of the expressive power of \MTL, and its finite (\WMTL)
and co-finite (\coWMTL) variants against the corresponding variants of \MSO and
\MPL has also been provided.
Unlike \MSO, which can quantify over non-connected sets of nodes, \MTL, much
like \MPL, is designed to predicate over connected sets of nodes only.
As a consequence, while the result that \MSO (\resp, \WMSO, \coWMSO) is strictly
more expressive than \MTL (\resp, \WMTL, \coWMTL) may be not surprising and
somewhat expected, far less obvious is the result that the same relationship
holds between \MTL and \MPL.
As we have shown, this essentially follows from the ability of \MTL to express
meaningful properties of trees, such as the density property stipulating that
the set of paths of the underlying tree is uncountable or, equivalently, that
the tree contains a full binary tree as a minor, which cannot be captured by
means of path quantifications only.

These results call for a deeper investigation of the relationships between \MTL
and temporal logics.
Here we have begun this analysis by showing that \STLS, an expressive temporal
logic able to capture reasoning about strategies and games in a quite general
way, can easily be embedded into \MTL.
This suggests that \MTL is indeed enough for reasoning about games with
\FO-definable goals.
A deeper result, though, is obtained by identifying the one-step fragment of the
Graded \MC that restricts free variables to occur within the scope of at most a
single modal operator.
Such a restriction, in turn, essentially prevents the formulae of this language
from being able to predicate over non-connected sets.
We show, in fact, that this fragment is contained in \MTL but cannot be captured
by \MPL, as it can still express the density property.
This study can be viewed as a contribution of interest on its own, since the
Modal \MC, although often considered an unfriendly assembly-like specification
language, is very important from a practical viewpoint.
\emph{Symbolic model-checking tools}, indeed, exist that compute the denotation
of fixpoint expressions over the set of states of the model to verify, see, \eg,
\cite{CGP02}.

In addition to the question still open of whether \WMTL is subsumed by \MPL on
finitely-branching trees, the most relevant problem still to address is the
completeness \wrt \MTL and \WMTL of the two one-step fragments (or slight
generalisations thereof) of the \GMC.
Particular care will be required for the treatment of the alternation-free
fragment on arbitrary-branching trees, where the corresponding fragment of \MTL
would conceivably not be \WMTL but a Noetherian variant~\cite{CFVZ20} of \MTL.


%% file: AppendixA.tex


\label{app:mtl}
\begin{center}
I. Missing Proofs of Section~\ref{sec:mtl}
\end{center}

\subsection{Expressiveness under Full Quantifications}
\label{app:mtl;sub:fqn}

In this section, we provide the proofs of Proposition~\ref{prop:MTLChains} and
Lemma~\ref{lmm:InexpressivenessCCTLStarNDandD}.

\propMTLChains*
\begin{proof}
It is known that over the class of $2^{\Prop}$-labeled infinite chains,
$\FO<\MSO$ and $\MSO\equiv \WMSO \equiv \CoWMSO$~\cite{Buc66}.
Since the logic $\MTL$, $\WMTL$, and $\CoWMTL$ subsume $\FO$, it suffices to
show that over infinite $2^{\Prop}$-labeled chains, $\Logic \leq \FO$ for each
$\Logic\in \{\MTL,\WMTL,\CoWMTL\}$. We focus on the logic $\MTL$. The proof for
the weak and coWeak variants of $\MTL$ is similar.
We exploit the fact that a subtree of a chain is a path. Let $\varphi$ be a
$\MTL$ sentence and $X_1,\ldots,X_n$ be the second-order variables occurring in
$\varphi$.
Without loss of generality, we assume that distinct occurrences of second-order
quantifiers in $\varphi$ are associated to distinct second-order variables. For
each $i\in [1,n]$, let $x_i$ and $x'_i$ be fresh first-order variables.
For each subformula $\psi$ of $\varphi$, we define a $\FO$ formula
$f(\psi,t_1,\ldots,t_n)$, where $t_i\in \{\Fin,\Inf\}$ is a flag. Intuitively,
$t_i=\Inf$ means that in the second-order quantifier $\exists^{\TSym} X_i$ of
$\varphi$, an infinite subtree of the given chain is chosen.
The mapping $f$ is homomorphic w.r.t.~Boolean connectives, $\FO$-atomic
formulas, and first-order quantifiers, and is defined as follows for the other
$\MTL$ constructs:
\begin{itemize}
\item $f(z\in X_i,t_1,\ldots,t_n)\DefinedAs \left\{
  \begin{array}{ll} x_i\leq z & \text{ if } t_i=\Inf \\
    x_i \leq z \wedge z \leq x'_i & \text{ otherwise}
  \end{array}
  \right.$
\item $
  \begin{array}{ll}
    f(\exists^{\TSym} X_i \,\psi,t_1,\ldots,t_n)\DefinedAs & \exists x_i.\,
    f(\psi,t_1,\ldots,t_{i-1},\Inf,t_{i+1},\ldots,t_n) \,\vee\,\\
    & \exists x_i \, \exists x'_i.\, (x_i\leq x'_i \wedge f(\psi,t_1,\ldots,t_{i-1},\Fin,t_{i+1},\ldots,t_n)
   \end{array}$.
 \end{itemize}
 The desired $\FO$ sentence is given by $f(\varphi,\Fin,\ldots,\Fin)$.
 Correctness of the construction can be easily proved.
\end{proof}

Now, we give a proof of  Lemma~\ref{lmm:InexpressivenessCCTLStarNDandD}.
Let $T$ be a non-blocking tree, $\pi$ a path of $T$, and $0\leq i\leq |\pi|$. We use the notation 
$T,\pi,i\models \psi$,  for a $\CCTLS$  path formula $\psi$,
to mean that $\LT,\pi,i\models\psi$, where $\LT=\tpl{T,\Lab_\emptyset}$ is the Kripke tree such that 
$\Lab_\emptyset(\tau)=\emptyset$ for all $\tau\in T$.
The meaning of the notation 
 $T,\tau\models \varphi$,  for a $T$-node $\tau$ and a $\CCTLS$   state formula $\varphi$, is similar. 

\lmmInexpressivenessCCTLStarNDandD*
\begin{proof}
We assume that $T\in \D_n$ and $T'\in \ND_m$. The other cases are similar or
simpler.  The proof is given by induction on $|\varphi|$. For the base case,
$|\varphi|=1$. Hence, $\varphi$ is an atomic proposition, and the result
trivially follows (recall that we are considering unlabeled trees).
Now, assume that $|\varphi|>1$. Hence, the root operator of $\varphi$ is either
a Boolean connective, or a counting operator, or a path quantifier. The case of
Boolean connectives directly follows from the induction hypothesis. Now, we
consider the other two cases.\vspace{0.1cm}

\noindent \textbf{Case} $\varphi =\DC^{k}\theta$ for some state formula
$\theta$: assume that $ T \models \DC^{k}\theta$. We need to show that $ T'
\models \DC^{k}\theta$. If $k=0$, the result is obvious. Now, let $k>0$.  Since
$T\in \D_n$ and $T\models \DC^{k}\theta$, by construction of the class $\D_n$,
there are $k$ distinct children $\tau_1,\ldots,\tau_{k}$ of the $T$-root such
that for each $\ell\in [1,k]$:
\begin{itemize}
  \item the subtree $T_\ell$ of $T$ rooted at node $\tau_\ell$ is either in the
    class $\ND_{i}$ for some $i\in [1,n-1]$, or in the class $\D_n$,
  \item $ T_\ell \models \theta$.
\end{itemize}
Recall that $T'\in \ND_m$. Since $\min(m,n)\geq |\varphi|$ and $\varphi
=\DC^{k}\theta$, it holds that $m>k$.  We select $k$ distinct children
$\tau'_1,\ldots,\tau'_k$ of the $T'$-root such that for the subtree $T'_\ell$ of
$T'$ rooted at $\tau'_\ell$, $ T'_\ell \models \theta$. Hence, the result
follows. Assume that we have already selected the children
$\tau'_1,\ldots,\tau'_{\ell-1}$ for $\ell\in [1,k]$. We choose $\tau'_\ell$ as
follows, where $T'_\ell$ is the associated rooted subtree:
  \begin{itemize}
  \item Case $T_\ell \in \D_n$: since $m>1$, by construction, the $T'$-root has
    $m$ distinct children which are $\ND_{m-1}$-nodes. Being $k\leq m$, we can
    set $\tau'_\ell$ to one of these nodes which has not already been
    selected. Thus, being $m-1\geq |\theta|$, by the induction hypothesis,
    $T'_\ell\models \theta$.
   \item Case $T_\ell \in \ND_i$ for some $i\in [1,n-1]$ and $i<m$: by
     construction, the $T'$-root has $m$ distinct children which are
     $\ND_{i}$-nodes. Being $k\leq m$, we can set $\tau'_\ell$ to one of these
     nodes which has not already been selected.  Since the class $\ND_i$
     contains only isomorphic trees, we obtain that $T'_\ell\models \theta$.
   \item Case $T_\ell \in \ND_i$ for some $i\in [m,n-1]$: since $m>1$, by
     construction, the $T'$-root has $m$ distinct children which are
     $\ND_{m-1}$-nodes. Being $k\leq m$ and $i\geq m$, we proceed as in the
     first case.
\end{itemize}
The implication $T'\models \varphi \Rightarrow T\models \varphi$ is similar, and
we omit the details here. \vspace{0.2cm}

\noindent \textbf{Case}
$\varphi =\EQ \psi$ for some path formula $\psi$: assume that $
 T \models \EQ \psi$. We need to show that $T' \models \EQ \psi$. By hypothesis,
 there is a path $\pi$ of $T$ starting at the root such that
 $T,\pi,0\models \psi$. We assume that $\pi$ is infinite (the other case being
 similar). Note that since $T$ is in $\D_n$, by construction, for each $i\geq
 0$, $\pi(i)$ is either a $\D_n$-node or a $\ND_i$-node for some $i\in [1,n-1]$.
 We show that there exists an infinite path $\pi'$ of $T'$ starting at the
 $T'$-root such that the following \emph{invariance property} holds for all
 $\ell\geq 0$:
\begin{itemize}
  \item \emph{either} $\pi(\ell)$ is a $\D_n$-node and $\pi'(\ell)$ is an
    $\ND_m$-node;
  \item \emph{or} both $\pi(\ell)$ and $\pi'(\ell)$ are $\ND_i$-nodes for some
    $i\in [1,m-1]$ with $m\leq n$;
  \item \emph{or} $\pi(\ell)$ is an $\ND_i$ node for some $i\in [m,n-1]$ and
    $\pi'(\ell)$ is an $\ND_m$ node.
\end{itemize}
Since we are considering unlabeled trees and all trees in the class $\ND_i$ are
isomorphic, by the invariance property and the induction hypothesis, it follows
that for all $i\geq 0$ and state subformulas $\theta$ of $\psi$,
$T,\pi(i)\models \theta \Leftrightarrow T,\pi'(i)\models
\theta$. Thus,
being $T,\pi,0\models \psi$, by the $\CCTLStar$-semantics, we obtain that
$T',\pi',0\models \psi$, and the result follows. It remains to show the
existence of the infinite path $\pi'$.  Since the root of $T$ is a $\D_n$-node
and the root of $T'$ is a $\ND_m$-node, we can assume that we have already
defined the first $\ell+1$ nodes of the infinite path $\pi'$ for some $\ell\geq
0$ such that the invariance property holds for each $0\leq i\leq \ell$. Then,
the node $\pi'(\ell+1)$ of $\pi'$ is selected among the children of $\pi'(\ell)$
in $T'$ as follows:
\begin{itemize}
  \item \textbf{Case} $\pi(\ell)$ is a $\D_n$-node and $\pi'(\ell)$ is a
    $\ND_m$-node:
  \begin{itemize}
  \item if $\pi(\ell+1)$ is a $\D_n$-node, then we set $\pi'(\ell+1)$ to the
    unique $\ND_m$-child of $\pi'(\ell)$ in $T'$,
  \item if $\pi(\ell+1)$ is an $\ND_i$-node for some $i\in [1,m-1]$ (note that
    $m\leq n$), then we set $\pi'(\ell+1)$ to one of the $m$ $\ND_i$-children of
    $\pi'(\ell)$ in $T'$,
  \item if $\pi(\ell+1)$ is an $\ND_n$-node for some $i\in [m,n-1]$, then we set
    $\pi'(\ell+1)$ to the unique $\ND_m$-child of $\pi'(\ell)$ in $T'$.
\end{itemize}
  \item \textbf{Case} $\pi(\ell)$ and $\pi(\ell')$ are $\ND_i$-nodes for some
    $i\in [1,m-1]$ with $m\leq n$: hence $\pi(\ell+1)$ is a $\ND_j$-node for
    some $j\in [1,m-1]$. We set $\pi'(\ell+1)$ to some $\ND_j$-child of
    $\pi'(\ell)$ in $T'$.
    \item \textbf{Case} $\pi(\ell)$ is an $\ND_i$-node for some $i\in [m,n-1]$
      and $\pi'(\ell)$ is an $\ND_m$-node: hence, $\pi(\ell+1)$ is an
      $\ND_j$-node for some $j\in [1,i]$. If $j<m$, we set $\pi'(\ell+1)$ to one
      of the $m$ $\ND_j$-children of $\pi'(\ell)$ in $T'$. Otherwise, we set
      $\pi'(\ell+1)$ to the unique $\ND_m$-child of $\pi'(\ell)$ in $T'$.
\end{itemize}
In all the cases, the invariance property is preserved.  The implication $ T'
\models \EQ \psi \Rightarrow T \models \EQ \psi$ is similar, and we omit the
details here.  This concludes the proof of
Lemma~\ref{lmm:InexpressivenessCCTLStarNDandD}.
\end{proof}

\bigskip

\subsection{Expressiveness under Weak Quantifications: proof of Lemma~\ref{lmm:hCompatibility}}
\label{app:mtl;sub:wqn}

In this section, we provide a proof of Lemma~\ref{lmm:hCompatibility}.  We need
some technical definitions and preliminary results.  Fix $n\geq 1$ and a Kripke
tree $\LT\in \NAcc_n$. Let $\pi$ be a finite path of $\LT$.
By construction, $\pi$ is of the form $\pi'\cdot \pi''$, where $\pi'$ is either
empty or visits only nodes with empty label ($\emptyset$-nodes), and $\pi''$ is
either empty or visits only $a$-nodes. We say that $\pi'$ (resp., $\pi''$) is
the $\emptyset$-part (resp., $a$-part) of $\pi$. We denote by $\last(\pi)$ the
last node of $\pi$, and by $\pi_{\geq i}$, where $0\leq i<|\pi|$, the suffix of
$\pi$ starting from position $i$.  Let $\N_{\emptyset}(\pi)$, $\N_{a}(\pi)$, and
$\D_{a}(\pi)$ be the natural numbers defined as follows:
\begin{itemize}
  \item $\N_{\emptyset}(\pi) \DefinedAs|\pi'|$ (the length of the
    $\emptyset$-part of $\pi$);
  \item $\N_{a}(\pi) \DefinedAs|\pi''|$ (the length of the $a$-part of $\pi$);
  \item $\D_{a}(\pi) \DefinedAs 0$ if $\N_a(\pi)>0$ (i.e., $\pi$ leads to a
    $a$-node); otherwise, $\D_{a}(\pi)$ is $\ell-1$, where $\ell$ is the length
    of the smallest finite paths of $\LT$ starting from $\last(\pi)$
    and leading to a $a$-node. Note that $\D_{a}(\pi)$ is well-defined and
    $0\leq \D_{a}(\pi)\leq n$.
\end{itemize}

Next, for each $h\in [1,n]$, we introduce the notion of \emph{$h$-compatibility}
between finite paths of $\LT\in \NAcc_n$. Intuitively, this notion
provide a sufficient condition to make two finite paths of $\LT$
indistinguishable by balanced $\WCCTLStar$ path formulas having size at most
$h$.

\begin{definition}[$h$-compatibility]
Let $h\in [1,n]$. Two finite paths $\pi$ and $\pi'$ of $\LT$ are
\emph{$h$-compatible} if the following conditions hold:
\begin{itemize}
  \item $\N_a(\pi)=\N_a(\pi')$;
  \item either $\N_\emptyset(\pi)=\N_\emptyset(\pi')$, or $\N_\emptyset(\pi)\geq
    h$ and $\N_\emptyset(\pi')\geq h$;
   \item either $\D_a(\pi)=\D_a(\pi')$, or $\D_a(\pi)\geq h$ and $\D_a(\pi')\geq
     h$.
\end{itemize}
\end{definition}

We denote by $\Rel(h)$ the binary relation over the finite paths of
$\LT\in\NAcc_n$  such that $(\pi,\pi')$ if and only if $\pi$ and $\pi'$ are
$h$-compatible. Notice that $\Rel(h)$ is an equivalence relation for all $h\in
[1,n]$. Moreover, $\Rel(h)\subseteq \Rel(h-1)$, for all $h\in [2,n]$, that is,
$\Rel(h)$ is a refinement of $\Rel(h-1)$.  The following lemma establishes
useful properties of the equivalence relation $\Rel(h)$ which intuitively
capture the semantics of the temporal modalities, counting modalities, and path
quantifiers over finite paths.

\begin{lemma} 
  \label{lmm:hCompatibilityOne}
  Let $h\in [2,n]$ and $(\pi,\pi')\in\Rel(h)$. Then, the following properties
  hold:
\begin{description}
\item[(1)] If $|\pi|>1$, then $|\pi'|>1$ and $(\pi_{\geq 1},\pi'_{\geq
  1})\in\Rel(h-1)$.
\item[(2)] For each $0\leq i< |\pi|$, there is $0\leq i'< |\pi'|$ such that
  $(\pi_{\geq i},\pi'_{\geq i'})\in\Rel(\FloorL{\frac{h}{2}})$ and the
  restriction of $\Rel(\FloorL{\frac{h}{2}})$ to the pairs $(\pi_{\geq
    j},\pi'_{\geq j'})$, where $0\leq j<i$ and $0\leq j'<i'$, is
  total.\footnote{Recall that a binary relation $\Rel\subseteq S\times S'$ is
  total if for each $s\in S$ (resp., $s'\in S'$), there is $s'\in S'$ (resp.,
  $s\in S$) such that $(s,s')\in \Rel$}
\item[(3)] For all children $\tau$ of $\pi(0)$ and children $\tau'$ of
  $\pi'(0)$, $(\tau,\tau')\in\Rel(h-1)$. Moreover, either $n_0=n'_0$, or
  $n_0,n'_0\geq h+1$, where $n_0$ (resp., $n'_0$) is the number of children of
  $\pi(0)$ (resp., $\pi'(0)$).
\item[(4)] For each finite path of the form $\pi(0)\cdot \rho$, there is a
  finite path of the form $\pi'(0)\cdot \rho'$ such that $(\pi(0)\cdot
  \rho,\pi'(0)\cdot \rho')\in\Rel(\FloorL{\frac{h}{2}})$.
\end{description}
 \end{lemma} 
\begin{proof}
\emph{Proof of Property~1.} Assume that $|\pi|>1$. Since $(\pi,\pi')\in
\Rel(h)$, either $\N_\emptyset(\pi)=\N_\emptyset(\pi')$ or
$\N_\emptyset(\pi)\geq h$ and $\N_\emptyset(\pi')\geq h$. Moreover, the
$a$-parts of $\pi$ and $\pi'$ have the same length. Thus, being $h\geq 2$, the
result easily follows.\vspace{0.2cm}

\noindent \emph{Proof of Property~2.} Let $0\leq i<|\pi|$. Being
$(\pi,\pi')\in\Rel(h)$, the $a$-parts of $\pi$ and $\pi'$ have the same
length. Thus, if $\N_\emptyset(\pi)=\N_\emptyset(\pi')$, the result trivially
follows by setting $i'=i$. Otherwise, $\N_\emptyset(\pi)\geq h$ and
$\N_\emptyset(\pi')\geq h$. We distinguish three cases;
\begin{itemize}
\item $i<\FloorL{\frac{h}{2}}$. We set $i'=i$. Being $\N_\emptyset(\pi)\geq h$
  and $\N_\emptyset(\pi')\geq h$, we have that $\N_\emptyset(\pi_{\geq i})\geq
  \FloorL{\frac{h}{2}}$ and $\N_\emptyset(\pi'_{\geq i})\geq
  \FloorL{\frac{h}{2}}$. Hence the result easily follows.
\item $i\geq \FloorL{\frac{h}{2}}$ and $\N_\emptyset(\pi_{\geq i})\geq
  \FloorL{\frac{h}{2}}$.  We set $i'= \FloorL{\frac{h}{2}}$. Being
  $\N_\emptyset(\pi')\geq h$, it holds that $\N_\emptyset(\pi'_{\geq i})\geq
  \FloorL{\frac{h}{2}}$, and the result easily follows in this case as well.
\item $i\geq \FloorL{\frac{h}{2}}$ and $\N_\emptyset(\pi_{\geq i})<
  \FloorL{\frac{h}{2}}$. We set $i'$ in such a way that $\N_\emptyset(\pi'_{\geq
    i'})= \N_\emptyset(\pi_{\geq i})$. Note that being $\N_\emptyset(\pi')\geq
  h$, $i'$ is well-defined. Moreover note that $i'\geq \FloorL{\frac{h}{2}}$ and
  one can easily show that the restriction of $\Rel(\FloorL{\frac{h}{2}})$ to
  the pairs $(\pi_{\geq j},\pi'_{\geq j'})$, where $0\leq j<i$ and $0\leq
  j'<i'$, is total.
\end{itemize}

 \noindent \emph{Proof of Property~3.} Let $\tau$ be a child of $\pi(0)$ and
 $\tau'$ be a child of $\pi'(0)$.  If both $\pi(0)$ and $\pi'(0)$ are $\NAcc_n$
 nodes, or both $\pi(0)$ and $\pi'(0)$ are $a$-nodes (hence, $\Acc_1$-nodes),
 the result is trivial. Otherwise, being $(\pi,\pi')\in \Rel(h)$, one of the
 following three conditions hold:
 \begin{itemize}
   \item $\pi(0)$ is an $\Acc_\ell$-node and $\pi'(0)$ is an $\Acc_{\ell'}$-node
     for some $\ell,\ell'\in [2,n]$. By construction, it holds that
     $\D_a(\tau)=\N_\emptyset(\pi)-1 +\D_a(\pi)$ and
     $\D_a(\tau')=\N_\emptyset(\pi')-1+\D_a(\pi')$. Hence, being $(\pi,\pi')\in
     \Rel(h)$, we have that either $\D_a(\tau)=\D_a(\tau')$, or $\D_a(\tau)\geq
     h-1$ and $\D_a(\tau')\geq h-1$. In the first case, by construction, it
     follows that $\ell=\ell'$, and the result trivially follows. Otherwise,
     being $h\geq 2$, $\tau$ and $\tau'$ are both $\emptyset$-nodes. Moreover,
     by construction, $\D_a(\pi(0))=\D_a(\tau)+1$,
     $\D_a(\pi'(0))=\D_a(\tau')+1$, $\ell= \D_a(\pi(0))+1$, and $\ell'=
     \D_a(\pi'(0))+1$. Hence, being $\D_a(\tau)\geq h-1$ and $\D_a(\tau')\geq
     h-1$, we obtain that $\ell,\ell'\geq h+1$, and the result follows.
   \item $\pi(0)$ is an $\Acc_\ell$-node and $\pi'(0)$ is a $\NAcc_{n}$-node for
     some $\ell\in [2,n]$.  Hence, $\tau$ is an $\Acc_{\ell-1}$ node, and
     $\tau'$ is either a $\NAcc_{n}$-node or an $\Acc_{n}$-node. We claim that
     $\ell\geq h+1$. Hence, $\D_a(\tau)\geq h-1\geq 1$, and the result follows.
     We assume the contrary and derive a contradiction. Since
     $(\pi,\pi')\in\Rel(h)$, $\ell<h+1$, and by construction
     $\ell=\N_\emptyset(\pi)+\D_a(\pi)+1$, it holds that $\N_\emptyset(\pi)=
     \N_\emptyset(\pi')$ and $\D_a(\pi)= \D_a(\pi')$. On the other hand, being
     $\pi'(0)$ a $\NAcc_{n}$-node, it holds that
     $\N_\emptyset(\pi')+\D_a(\pi')+1> n$, which is a contradiction, and the
     result follows.
   \item $\pi(0)$ is a $\NAcc_n$-node and $\pi'(0)$ is a $\Acc_{\ell'}$-node for some $\ell'\in [2,n]$.
    This case is similar to the previous one.
 \end{itemize}\vspace{0.2cm}

  \noindent \emph{Proof of Property~4.}  If both $\pi(0)$ and $\pi'(0)$ are
  $\NAcc_n$ nodes, or both $\pi(0)$ and $\pi'(0)$ are $a$-nodes (hence,
  $\Acc_1$-nodes), the result is trivial. Otherwise, being $(\pi,\pi')\in
  \Rel(h)$, one of the following three conditions hold:
 \begin{itemize}
   \item $\pi(0)$ is an $\Acc_\ell$-node and $\pi'(0)$ is an $\Acc_{\ell'}$-node
     for some $\ell,\ell'\in [2,n]$. By reasoning as in the proof of Property~3,
     we deduce that either $\ell=\ell'$ or $\ell,\ell'\geq h+1$. Hence, the
     result easily follows.
   \item $\pi(0)$ is an $\Acc_\ell$-node and $\pi'(0)$ is a $\NAcc_{n}$-node for
     some $\ell\in [2,n]$. By reasoning as in the proof of Property~3, we deduce
     that $\ell\geq h+1$. Hence, the result easily follows in this case as well.
   \item $\pi(0)$ is a $\NAcc_n$-node and $\pi'(0)$ is a $\Acc_{\ell'}$-node for
     some $\ell'\in [2,n]$.  This case is similar to the previous one.
 \end{itemize}
 \end{proof}

By Lemma~\ref{lmm:hCompatibilityOne}, we easily deduce that every balanced
$\WCCTLStar$ path formula having size at most $h$ cannot distinguish
$h$-compatible finite paths in the fixed Kripke tree $\LT\in\NAcc_n$.

\begin{lemma} \label{lmm:hCompatibilityTwo}
Let $h\in [1,n]$ and $(\pi,\pi')\in\Rel(h)$. Then, for each balanced
$\WCCTLStar$ path formula $\psi$ such that $|\psi|\leq h$, it holds that
$\LT,\pi,0\models \psi$ if and only if $\LT,\pi',0\models
\psi$.
\end{lemma}
\begin{proof} Since $\Rel(h)$ is an equivalence relation, it suffices to show that
  $\LT,\pi,0\models \psi$ implies $\LT,\pi',0\models \psi$.  The proof is
  by induction on $|\psi|$. The cases for the Boolean connectives directly
  follow from the induction hypothesis.  As for the other cases, we proceed as
  follows:
\begin{itemize}
  \item $\psi= a$: since $(\pi,\pi')\in \Rel(h)$, it holds that either both
    $\pi(0)$ and $\pi'(0)$ are $\emptyset$-nodes, or both $\pi(0)$ and $\pi'(0)$
    are $a$-nodes. Hence, the result follows.
  \item $\psi=\Next\psi_1$: by hypothesis $|\psi_1|\leq h-1$ and $h\geq 2$. Let
    $\LT,\pi,0\models \psi$. Hence, $|\pi|>1$ and $\LT,\pi_{\geq
    1},0\models \psi_1$.  Being $(\pi,\pi')\in \Rel(h)$, by
    Lemma~\ref{lmm:hCompatibilityOne}(1), $|\pi'|>1$ and $(\pi_{\geq
      1},\pi'_{\geq 1})\in \Rel(h-1)$. Thus, by the induction hypothesis, it
    follows that $\LT,\pi'_{\geq 1},0\models \psi_1$. This means that
    $\LT,\pi',0\models \psi$, and the result follows.
  \item $\psi=\psi_1\Until \psi_2$: since $\psi$ is balanced and $|\psi|\leq h$,
    it holds that $|\psi_1|,|\psi_2|\leq \FloorL{\frac{h}{2}}$. Let
    $\LT,\pi,0\models \psi$. Hence, there is $0\leq i<|\pi|$ such that
    $\LT,\pi_{\geq i},0\models \psi_2$ and $\LT,\pi_{\geq j},0 \models
    \psi_1$ for all $0\leq j<i$. Since $(\pi,\pi')\in \Rel(h)$, by applying the
    induction hypothesis and Lemma~\ref{lmm:hCompatibilityOne}(2), it follows
    that there is $0\leq i'<|\pi'|$ such that $\LT,\pi'_{\geq i'},0\models
    \psi_2$ and $\LT,\pi'_{\geq j'},0\models \psi_1$ for all $0\leq
    j'<i'$. This means that $\LT,\pi',0\models \psi$, and the result follows.
  \item $\psi =\DC^{\ell}\psi_1$: being $|\psi|\leq h$, it holds that $h>2$,
    $\ell\leq h-1$ and $|\psi_1|\leq h-1$.  Let $\LT,\pi,0\models
    \psi$. Hence, there are $\ell$ distinct children $\tau$ of $\pi(0)$ such
    that $\LT,\tau\models \psi_1$.  We distinguish two cases:
  \begin{itemize}
   \item $\pi(0)$ is a $a$-node: hence, being $(\pi(0),\pi'(0))\in \Rel(h)$,
     $\pi'(0)$ is a $a$-node as well. Since the subtrees rooted at $a$-nodes are
     chains of $a$-nodes, we trivially deduce that $\ell=1$ and
     $\LT,\pi',0\models \psi$.
   \item $\pi(0)$ is a $\emptyset$-node: hence, being $(\pi(0),\pi'(0))\in
     \Rel(h)$, $\pi'(0)$ is a $\emptyset$-node as well. By applying the
     induction hypothesis and Lemma~\ref{lmm:hCompatibilityOne}(4), we obtain
     that for each child $\tau'$ of $\pi'(0)$, $\LT,\tau'\models \psi_1$.
     Moreover, either $n_0=n'_0$, or $n_0,n'_0\geq h+1$, where $n_0$ (resp.,
     $n'_0$) is the number of children of $\pi(0)$ (resp., $\pi'(0)$). Thus,
     since $\ell\leq h+1$, we deduce that $\LT,\pi',0\models \psi$.
 \end{itemize}
  \item $\psi=\EQ\theta$: being $|\psi|\leq h$ and $\psi$ balanced, it holds
    that $\theta$ is of the form $\theta_1\wedge \theta_2$, where
    $|\theta_1|,|\theta_2|\leq \FloorL{\frac{h}{2}}$.  Let $\LT,\pi,0\models
    \psi$. Hence, there exists a finite path of the form $\pi(0)\cdot \rho$ such
    that $\LT,\pi(0)\cdot \rho,0\models \theta_i$ for $i=1,2$. Being
    $(\pi,\pi')\in R(h)$, by Lemma~\ref{lmm:hCompatibilityOne}(4) and the
    induction hypothesis, there exists a finite path of the form $\pi'(0)\cdot
    \rho'$ such that $\LT,\pi'(0)\cdot \rho',0\models \theta_i$ for each
    $i=1,2$.  Hence, $\LT,\pi',0\models \psi$ and we are done.
\end{itemize}
\end{proof}

We now prove Lemma~\ref{lmm:hCompatibility} by exploiting
Lemma~\ref{lmm:hCompatibilityTwo}.
 
\lmmhCompatibility*
\begin{proof}
Let $\tpl{T,\Lab}\in \NAcc_n$ and $\tpl{T',\Lab'}\in \Acc_n$ where $n>
|\varphi|$, and $\Rel(n-1)$ the $(n-1)$-compatibility relation for the finite
paths of $\tpl{T,\Lab}$. By construction the root $\tau_0$ of $\tpl{T,\Lab}$ has
some child $\tau_1$ whose Kripke subtree $\tpl{T'',\Lab''}$ is in
$\Acc_n$. Moreover, by construction $(\tau_0,\tau_1)\in \Rel(n-1)$. Being
$|\varphi|\leq n-1$, by Lemma~\ref{lmm:hCompatibilityTwo}, it follows that
$\tpl{T,\Lab}\models \varphi$ if and only if $\tpl{T'',\Lab''}\models \varphi$.
Thus, being $\tpl{T',\Lab'}$ and $\tpl{T'',\Lab''}$ isomorphic, the result
follows.
\end{proof}

\bigskip
\subsection{Weak Quantifications versus coWeak Quantifications:
proof of Proposition~\ref{prop:MTLvSubumedByCoWeakMTL}}
\label{app:mtl;sub:WeakVersusCoWeak}

\propMTLvSubumedByCoWeakMTL*
\begin{proof}
We show that $\MTL\leq \CoWMTL$ (the proof of $\WMTL\leq \CoWMTL$ is similar).
Let us consider the open $\MTL$ formula $\theta(x,X)$ defined as follows:
\[
\begin{array}{ll}
\theta(x,X)\DefinedAs & x\in X\,\wedge\, \neg \exists y\in X.\, \exists^{\TSym} Y.\,[Y\subseteq X \wedge \forall z\in Y.\,(z=y\vee \Child(y,z))]\,\wedge\,\\
&\forall Y^{\TSym}.\,[(\Path(Y)\wedge Y\subseteq X)\rightarrow \exists y\in
Y.\,x<y]
\end{array}
\]
Assuming that $X$ is interpreted as an infinite tree $T$, under the $\CoWMTL$
semantics, the previous formula asserts that $T$ is finitely branching, node $x$
is in $T$, and each infinite path of $T$ visits some strict descendant of node
$x$.

Let $\varphi$ be a $\MTL$ sentence and $X_1,\ldots,X_n$ be the set variables
occurring in $\varphi$.  Without loss of generality, we assume that distinct
occurrences of second-order quantifiers in $\varphi$ are associated to distinct
set variables. For each $i\in [1,n]$, let $\overline{X}_i$ (resp.,
$\overline{x}_i$) be a fresh set (resp., fresh first-order) variable. For each
subformula $\psi$ of $\varphi$, we define a $\MTL$ formula
$f(\psi,t_1,\ldots,t_n)$, where $t_i\in \{\Fin,\Inf\}$ is a flag. Intuitively,
$t_i=\Inf$ means that in the second-order quantifier $\exists^{\TSym} X_i$ of
$\varphi$, an infinite subtree of the given not-blocking tree is chosen.
The mapping $f$ is homomorphic w.r.t.~Boolean connectives, $\FO$-atomic
formulas, and first-order quantifiers, and is defined as follows for the other
$\MTL$ constructs:
\begin{itemize}
   \item $f(z\in X_i,t_1,\ldots,t_n)\DefinedAs \left\{\begin{array}{ll} z\in X_i
        & \text{ if } t_i=\Inf \\
        z\in \overline{X}_i\wedge \neg \overline{x}_i<z & \text{
        otherwise} \end{array}\right.$

     \item $ \begin{array}{ll} f(\exists^{\TSym}
     X_i \,\psi,t_1,\ldots,t_n)\DefinedAs & \exists^{\TSym} X_i.\,
     f(\psi,t_1,\ldots,t_{i-1},\Inf,t_{i+1},\ldots,t_n) \,\vee\,\\
     &\exists^{\TSym} \overline{X}_i.\,\exists \overline{x}_i.\,(\theta(\overline{x}_i,\overline{X}_i)\wedge
     f(\psi,t_1,\ldots,t_{i-1},\Fin,t_{i+1},\ldots,t_n)) \end{array}$.  \end{itemize}
     By construction and Lemma~\ref{lemma:CharacterizationFiniteTree}, it easily
     follows that the $\MTL$ sentence $\varphi$ is equivalent to the $\MTL$
     sentence $f(\varphi,\Fin,\ldots,\Fin)$ interpreted under the $\CoWMTL$
     semantics.
\end{proof}


%% file: AppendixB.tex


\newpage
\begin{center}
II. Missing Proofs of Section~\ref{sec:contmplog}
\end{center}

\subsection{One-Step Graded \MC}
\label{app:contmplog;sub:onestpgmc}

\lmmonestpmodlog*
\begin{proof}
The proof proceeds by structural induction on the \OSGMC base formula
$\varphiFrm \in \PhiSet[\ZSet, \OSet]$, which, \wlogx, is assumed to be in
positive normal form.
\begin{itemize}
\item\textbf{[Base cases $\varphiFrm \in \{ \Ff, \Tt \} \cup \set{
  \apElm, \neg \apElm }{ \apElm \in \APSet } \cup \set{ \varthetaFrm, \neg
  \varthetaFrm }{ \varthetaFrm \in \ThetaSet[\emptyset, \emptyset]}$]:}
  Since $\free{\varphiFrm} = \emptyset$, we have that the semantics of
  $\varphiFrm$ does not depend on the valuations of the variables in $\asgElm$
  and $\asgElm'$, \ie, $\denot{\varphiFrm}[\asgElm][\TName] =
  \denot{\varphiFrm}[\asgElm'][\TName]$, from which the thesis immediately
  follows.
\item\textbf{[Base case $\varphiFrm = \XvarElm \in \ZSet$]:}
  By definition of the semantics of fixpoint variables, we have that
  $\denot{\varphiFrm}[\asgElm][\TName] = \asgElm(\XvarElm)$ and
  $\denot{\varphiFrm}[\asgElm'][\TName] = \asgElm'(\XvarElm)$.
  Now, $\asgElm(\XvarElm) \cap \WSet \subseteq \asgElm'(\XvarElm)$, since
  $\asgElm \sqsubseteq_{\WSet}^{\ZSet, \OSet} \asgElm'$ and $\XvarElm \in
  \ZSet$, so the thesis immediately follows in this case as well.
\item\textbf{[Inductive cases $\varphiFrm \in \{ \varphiFrm[1] \wedge
  \varphiFrm[2], \varphiFrm[1] \vee \varphiFrm[2] \}$]:}
  By the inductive hypothesis, it holds that
  $\denot{\varphiFrm[1]}[\asgElm][\TName] \cap \WSet \subseteq
  \denot{\varphiFrm[1]}[\asgElm'][\TName]$ and
  $\denot{\varphiFrm[2]}[\asgElm][\TName] \cap \WSet \subseteq
  \denot{\varphiFrm[2]}[\asgElm'][\TName]$.
  Hence, the thesis easily follows by analysing the semantics of the Boolean
  connectives:
  \begin{multicols}{2}
  \begin{eqnarray*}
    {\denot{\varphiFrm[1] \wedge \varphiFrm[2]}[\asgElm][\TName] \cap \WSet}
  & = &
    {(\denot{\varphiFrm[1]}[\asgElm][\TName] \cap
    \denot{\varphiFrm[2]}[\asgElm][\TName]) \cap \WSet} \\
  & = &
    {(\denot{\varphiFrm[1]}[\asgElm][\TName] \cap \WSet) \cap
  (\denot{\varphiFrm[2]}[\asgElm][\TName] \cap \WSet)} \\
  & \subseteq &
    {\denot{\varphiFrm[1]}[\asgElm'][\TName] \cap
    \denot{\varphiFrm[2]}[\asgElm'][\TName]} \\
  & = &
    {\denot{\varphiFrm[1] \wedge \varphiFrm[2]}[\asgElm'][\TName]};
  \end{eqnarray*}
  \begin{eqnarray*}
    {\denot{\varphiFrm[1] \vee \varphiFrm[2]}[\asgElm][\TName] \cap \WSet}
  & = &
    {(\denot{\varphiFrm[1]}[\asgElm][\TName] \cup
    \denot{\varphiFrm[2]}[\asgElm][\TName]) \cap \WSet} \\
  & = &
    {(\denot{\varphiFrm[1]}[\asgElm][\TName] \cap \WSet) \cup
  (\denot{\varphiFrm[2]}[\asgElm][\TName] \cap \WSet)} \\
  & \subseteq &
    {\denot{\varphiFrm[1]}[\asgElm'][\TName] \cup
    \denot{\varphiFrm[2]}[\asgElm'][\TName]} \\
  & = &
    {\denot{\varphiFrm[1] \vee \varphiFrm[2]}[\asgElm'][\TName]}.
  \end{eqnarray*}
  \end{multicols}
\item\textbf{[Inductive cases $\varphiFrm \in \{ \DMod[\geq k]\, \varphiFrm',
  \BMod[< k]\, \varphiFrm' \}$]:}
  Recall that $\varphiFrm' \in \PhiSet[\OSet, \emptyset]$ and observe that
  $\asgElm \sqsubseteq_{\WSet}^{\ZSet, \OSet} \asgElm'$ implies $\asgElm
  \sqsubseteq_{\post{\WSet}}^{\OSet, \emptyset} \asgElm'$.
  Thus, by the inductive hypothesis, we have that
  $\denot{\varphiFrm'}[\asgElm][\TName] \cap \post{\WSet} \subseteq
  \denot{\varphiFrm'}[\asgElm'][\TName]$.
  Hence, the thesis easily follows by analysing the semantics of the two modal
  operators:
  \begin{multicols}{2}
  \begin{eqnarray*}
    {\denot{\DMod[\geq k]\, \varphiFrm'}[\asgElm][\TName] \cap \WSet}
  & = &
    {\set{ \wElm \in \WSet }{ \card{\post{\wElm} \cap
    \denot{\varphiFrm'}[\asgElm][\TName]} \geq k }} \\
  & \subseteq &
    {\set{ \wElm \in \WSet }{ \card{\post{\wElm} \cap
    \denot{\varphiFrm'}[\asgElm'][\TName]} \geq k }} \\
  & \subseteq &
    {\set{ \wElm \in \TSet }{ \card{\post{\wElm} \cap
    \denot{\varphiFrm'}[\asgElm'][\TName]} \geq k }} \\
  & = &
    {\denot{\DMod[\geq k]\, \varphiFrm'}[\asgElm'][\TName]};
  \end{eqnarray*}
  \begin{eqnarray*}
    {\denot{\BMod[< k]\, \varphiFrm'}[\asgElm][\TName] \cap \WSet}
  & = &
    {\set{ \wElm \in \WSet }{ \card{\post{\wElm} \setminus
    \denot{\varphiFrm'}[\asgElm][\TName] } < k }} \\
  & \subseteq &
    {\set{ \wElm \in \WSet }{ \card{\post{\wElm} \setminus
    \denot{\varphiFrm'}[\asgElm'][\TName]} < k }} \\
  & \subseteq &
    {\set{ \wElm \in \TSet }{ \card{\post{\wElm} \setminus
    \denot{\varphiFrm'}[\asgElm'][\TName]} < k }} \\
  & = &
    {\denot{\BMod[< k]\, \varphiFrm'}[\asgElm'][\TName]}.
  \hspace{8em}\qedhere
  \end{eqnarray*}
  \end{multicols}
\end{itemize}
\end{proof}

\lmmonestpgmc*
\begin{proof}
The proof proceeds by structural induction on the \OSGMC fixpoint formula
$\varthetaFrm \in \ThetaSet[\ZSet, \OSet]$.
\begin{itemize}
\item\textbf{[Base case $\varthetaFrm \in \PhiSet[\ZSet, \OSet]$]:}
  We first apply Lemma~\ref{lmm:onestpmodlog} to $\varthetaFrm$ \wrt
  $\DeltaSet[\wElm]$, since $\asgElm \sqsubseteq_{\DeltaSet[\wElm]}^{\ZSet,
  \OSet} \asgElm'$, obtaining $\DeltaSet[\wElm] = \DeltaSet \cap
  \DeltaSet[\wElm] = \denot{\varthetaFrm}[\asgElm][\TName] \cap
  \DeltaSet[\wElm] \subseteq \denot{\varthetaFrm}[\asgElm'][\TName]$.
  At this point, by applying the $\root[\wElm]$ restriction to both sides of
  this inclusion, we derive $\DeltaSet[\wElm] = \DeltaSet[\wElm] \root[\wElm]
  \subseteq \denot{\varthetaFrm}[\asgElm'][\TName] \root[\wElm] \subseteq
  \denot{\varthetaFrm}[\asgElm][\TName] \root[\wElm] = \DeltaSet[\wElm]$, where
  the last inclusion is due to the monotonicity property of the semantics.
  Hence, $\DeltaSet[\wElm] = \denot{\varthetaFrm}[\asgElm'][\TName]
  \root[\wElm]$.
\item\textbf{[Inductive case $\varthetaFrm = \nu \XvarElm \ldotp
  \varthetaFrm'$]:}
  By definition of fixpoint, it holds that $\DeltaSet =
  \denot{\varthetaFrm}[\asgElm][\TName] =
  \denot{\varthetaFrm'}[{{\asgElm}[\XvarElm \mapsto \DeltaSet]}][\TName]$, so,
  by the inductive hypothesis, we have that $\DeltaSet[\wElm] =
  \denot{\varthetaFrm'}[\asgElm''][\TName] \root[\wElm]$, where
  \begin{eqnarray*}
    {\asgElm''}
  & \;{\defeq}\, &
    {({\asgElm}[\XvarElm \mapsto \DeltaSet]) \rst[{\DeltaSet[\wElm]}]} \\
  & = &
    {(\asgElm \rst[{\DeltaSet[\wElm]}])[\XvarElm \mapsto \DeltaSet \cap
    (\DeltaSet[\wElm] \cup \post{\DeltaSet[\wElm]}) ]} \\
  & = &
    {{\asgElm'}[\XvarElm \mapsto (\DeltaSet \cap \DeltaSet[\wElm]) \cup
    (\DeltaSet \cap \post{\DeltaSet[\wElm]}) ]} \\
  & = &
    {{\asgElm'}[\XvarElm \mapsto \DeltaSet[\wElm] ]}.
  \end{eqnarray*}
  The last equality is due to the fact that $\DeltaSet \cap
  \post{\DeltaSet[\wElm]} \subseteq \DeltaSet[\wElm] \subseteq \DeltaSet$, which
  in turn is due to the definition of the restriction operator $\root[\wElm]$,
  as all nodes of $\DeltaSet[\wElm]$ are nodes of $\DeltaSet$ and a node in
  $\DeltaSet$ that is a successor of a node in $\DeltaSet[\wElm]$ needs to
  belong to $\DeltaSet[\wElm]$ as well.
  Hence, $\DeltaSet[\wElm] = \denot{\varthetaFrm'}[{{\asgElm'}[\XvarElm \mapsto
  \DeltaSet[\wElm] ]}][\TName] \root[\wElm] \subseteq
  \denot{\varthetaFrm'}[{{\asgElm'}[\XvarElm \mapsto
  \DeltaSet[\wElm]]}][\TName]$, which, by the semantics of the greatest fixpoint
  operator, implies that $\DeltaSet[\wElm] \subseteq
  \denot{\varthetaFrm}[\asgElm'][\TName]$.
  Thus, $\DeltaSet[\wElm] = \DeltaSet[\wElm] \root[\wElm]
  \subseteq \denot{\varthetaFrm}[\asgElm'][\TName] \root[\wElm] \subseteq
  \denot{\varthetaFrm}[\asgElm][\TName] \root[\wElm] = \DeltaSet[\wElm]$ and,
  so, $\DeltaSet[\wElm] = \denot{\varthetaFrm}[\asgElm'][\TName] \root[\wElm]$.
\item\textbf{[Inductive case $\varthetaFrm = \mu \XvarElm \ldotp
  \varthetaFrm'$]:}
  By Kleene's Theorem, we have that $\DeltaSet =
  \denot{\varthetaFrm}[\asgElm][\TName] = \bigcup_{i \in \SetN} \FSet[i]$ and
  $\denot{\varthetaFrm}[\asgElm'][\TName] = \bigcup_{i \in \SetN} \FSet[i]'$,
  where $\FSet[0], \FSet[0]' \defeq \emptyset$, $\FSet[i + 1] \defeq
  \denot{\varthetaFrm'}[{{\asgElm}[\XvarElm \mapsto \FSet[i]]}][\TName]$, and
  $\FSet[i + 1]' \defeq \denot{\varthetaFrm'}[{{\asgElm'}[\XvarElm \mapsto
  \FSet[i]']}][\TName]$, for all $i \in \SetN$.
  First observe that, by inductive hypothesis, $\GammaSet[\vElm][i] \defeq
  \FSet[i] \root[\vElm] = \denot{\varthetaFrm'}[{(\asgElm
  \rst[{\GammaSet[\vElm][i]}])[\XvarElm \mapsto \XiSet[\vElm][i]]}][\TName]
  \root[\vElm]$, where $\XiSet[\vElm][i] \defeq \FSet[i - 1] \cap
  (\GammaSet[\vElm][i] \cup \post{\GammaSet[\vElm][i]})$, for all $i \in
  \SetN[+]$ and $\vElm \in \DeltaSet$.
  Let us set $\GammaSet[\vElm][0] \defeq \emptyset$, for $\vElm \in \DeltaSet$.
  Also, note that, $\GammaSet[\vElm][i] \subseteq \DeltaSet[\wElm]$, for all
  $\vElm \in \DeltaSet[\wElm]$, since $\GammaSet[\vElm][i] = \FSet[i]
  \root[\vElm] \subseteq \DeltaSet \root[\vElm] \subseteq \DeltaSet \root[\wElm]
  = \DeltaSet[\wElm]$.
  Now, via an auxiliary inductive proof on the index $i \in \SetN$, we show
  that $\FSet[i]' \subseteq \FSet[i]$ and $\GammaSet[\vElm][i] \subseteq
  \FSet[i]' \root[\vElm]$, for all $\vElm \in \DeltaSet[\wElm]$.
  \begin{itemize}
  \item\textbf{[Base case $i = 0$]:}
    The two properties trivially hold, as $\GammaSet[\vElm][0] = \FSet[0]' =
    \FSet[0] = \emptyset$.
  \item\textbf{[Inductive case $i > 0$]:}
    By inductive hypothesis, we have that $\FSet[i - 1]' \subseteq \FSet[i - 1]$
    and $\GammaSet[\vElm][i - 1] \subseteq \FSet[i - 1]' \root[\vElm]$, for all
    $\vElm \in \DeltaSet[\wElm]$.
    Obviously, $\FSet[i]' = \denot{\varthetaFrm'}[{{\asgElm'}[\XvarElm \mapsto
    \FSet[i - 1]']}][\TName] \subseteq \denot{\varthetaFrm'}[{{\asgElm}[\XvarElm
    \mapsto \FSet[i - 1]]}][\TName] = \FSet[i]$, due to the monotonicity of the
    semantics.
    Thus, the first property is verified.
    Now, let us focus on the second property.
    Nothing has to be proven when $\vElm \not\in \FSet[i]$, being
    $\GammaSet[\vElm][i] = \emptyset$, so, consider a node $\vElm \in \FSet[i]$.
    Then, we have that
    \begin{eqnarray*}
      {\GammaSet[\vElm][i]}
    & = &
      {\denot{\varthetaFrm'}[{(\asgElm \rst[{\GammaSet[\vElm][i]}])[\XvarElm
      \mapsto \XiSet[\vElm][i]]}][\TName] \root[\vElm]} \\
    & \subseteq &
      {\denot{\varthetaFrm'}[{(\asgElm \rst[{\DeltaSet[\wElm]}])[\XvarElm
      \mapsto \XiSet[\vElm][i]]}][\TName] \root[\vElm]} \\
    & \subseteq &
      {\denot{\varthetaFrm'}[{(\asgElm \rst[{\DeltaSet[\wElm]}])[\XvarElm
      \mapsto \FSet[i - 1]']}][\TName] \root[\vElm]} \\
    & = &
      {\denot{\varthetaFrm'}[{{\asgElm'}[\XvarElm \mapsto \FSet[i -
      1]']}][\TName] \root[\vElm]} \\
    & = &
      {\FSet[i]' \root[\vElm]},
    \end{eqnarray*}
    where both inclusions are again due the monotonicity of the semantics.
    Specifically, the first one is implied by the inclusion $\GammaSet[\vElm][i]
    \subseteq \DeltaSet[\wElm]$ noted above, while the second one is due to the
    inclusion $\XiSet[\vElm][i] \subseteq \FSet[i - 1]'$ easily derived from the
    inductive hypothesis as follows.
    First note that $\XiSet[\vElm][i] = \FSet[i - 1] \cap (\GammaSet[\vElm][i]
    \cup \post{\GammaSet[\vElm][i]}) \subseteq \DeltaSet \cap (\DeltaSet[\wElm]
    \cup \post{\DeltaSet[\wElm]}) = \DeltaSet[\wElm]$, so, $\XiSet[\vElm][i]
    \subseteq \FSet[i - 1]$ and $\XiSet[\vElm][i] \subseteq \DeltaSet[\wElm]$.
    Now, by the inductive hypothesis, $\uElm \in \GammaSet[\uElm][i - 1]
    \subseteq \FSet[i - 1]' \root[\uElm] \subseteq \FSet[i - 1]'$, for every
    node $\uElm \in \FSet[i - 1] \cap \DeltaSet[\wElm]$, which implies
    $\XiSet[\vElm][i] \subseteq \FSet[i - 1]'$ as needed.
    Hence, the second property is verified as well.
  \end{itemize}
  At this point, to show that $\DeltaSet[\wElm] =
  \denot{\varthetaFrm}[\asgElm'][\TName] \root[\wElm]$, we identify, for every
  node $\vElm \in \DeltaSet[\wElm]$, an index $j_{\vElm} \in \SetN$ such that
  $\vElm \in \GammaSet[\wElm][j_{\vElm}]$.
  This index necessarily exists, as we can choose for $j_{\vElm}$ any number $j
  \geq \max[\wElm \leq \uElm \leq \vElm]\, \min \set{ i \in \SetN }{ \uElm \in
  \FSet[i] }$.
  Thus, for every $\vElm \in \DeltaSet[\wElm]$, we have that $\vElm \in
  \GammaSet[\wElm][j_{\vElm}] \subseteq \FSet[j_{\vElm}]' \root[\wElm] \subseteq
  \bigcup_{i \in \SetN} (\FSet[i]' \root[\wElm]) \subseteq (\bigcup_{i \in
  \SetN} \FSet[i]') \root[\wElm] = \denot{\varthetaFrm}[\asgElm'][\TName]
  \root[\wElm]$, which implies that $\DeltaSet[\wElm] \subseteq
  \denot{\varthetaFrm}[\asgElm'][\TName] \root[\wElm] \subseteq
  \denot{\varthetaFrm}[\asgElm][\TName] \root[\wElm] = \DeltaSet[\wElm]$ and,
  so, $\DeltaSet[\wElm] = \denot{\varthetaFrm}[\asgElm'][\TName] \root[\wElm]$.
  \qedhere
\end{itemize}
\end{proof}

Before providing the proof of Theorem~\ref{thm:onestpgmc}, we report here again
the full definition of the translation function $\trFun{} \colon \VarSet[1] \to
(\ThetaSet[\ZSet, \OSet] \to \MTL)$ from the \OSGMC to \MTL:
{\small\tabonestpgmc}


\thmonestpgmc*
\begin{proof}
To show that the \OSGMC is subsumed by \MTL, we actually prove the following
stronger statement: for every \OSGMC fixpoint formula $\varthetaFrm \in
\ThetaSet[\ZSet, \OSet]$, Kripke tree $\TName$, assignment $\asgElm \in
\AsgSet[\TName](\ZSet \cup \OSet)$, and node $\wElm \in \TSet$, it holds that
\[
  \wElm \in \denot{\varthetaFrm}[\asgElm][\TName]
\text{\ \ \iff\ \ \ }
  \TName, \{ \varElm \mapsto \wElm \}, \asgElm \models
  \trFun[\varElm]{\varthetaFrm}.
\]
The proof proceeds by (extended) structural induction on $\varthetaFrm$, where
we assume the fixpoint formula $\varthetaFrm = \mu \XvarElm \ldotp
\varthetaFrm'$ to be higher in the inductive order than $\nu \XvarElm \ldotp
\pnfMac{\neg {\varthetaFrm'}[\XvarElm / \neg \XvarElm]}$.
Since the correctness of the translation $\trFun[\varElm]{\varthetaFrm}$ for all
cases but the fixpoints is straightforward, we just focus on these two
operators.
\begin{itemize}
\item\textbf{[$\varthetaFrm = \nu \XvarElm \ldotp \varthetaFrm'$]:}
  \begin{itemize}
  \item\textbf{[``if'']:}
    If $\TName, \{ \varElm \mapsto \wElm \}, \asgElm \models
    \trFun[\varElm]{\varthetaFrm}$, there exists a (tree) set $\WSet \subseteq
    \TSet$ containing $\wElm$ such that, for all $\vElm \in \WSet$, it holds
    that $\TName, \{ \varElm \mapsto \vElm \}, {\asgElm}[\XvarElm \mapsto \WSet]
    \models \trFun[\varElm]{\varthetaFrm'}$.
    By the inductive hypothesis, it holds that $\vElm \in
    \denot{\varthetaFrm'}[{{\asgElm}[\XvarElm \mapsto \WSet]}][\TName]$, for all
    $\vElm \in \WSet$, which means that $\WSet \subseteq
    \denot{\varthetaFrm'}[{{\asgElm}[\XvarElm \mapsto \WSet]}][\TName]$.
    Now, by the semantics of the greatest fixpoint operator, we have that
    $\WSet \subseteq \denot{\varthetaFrm}[\asgElm][\TName]$ and, so, $\wElm \in
    \denot{\varthetaFrm}[\asgElm][\TName]$.
  \item\textbf{[``only if'']:}
    By definition of fixpoint, it holds that $\wElm \in \DeltaSet \defeq
    \denot{\varthetaFrm}[\asgElm][\TName] =
    \denot{\varthetaFrm'}[{{\asgElm}[\XvarElm \mapsto \DeltaSet]}][\TName]$.
    Now, by Lemma~\ref{lmm:onestpgmc}, $\wElm \in \DeltaSet[\wElm] =
    \denot{\varthetaFrm'}[{\asgElm'}{[\XvarElm \mapsto
    \DeltaSet[\wElm]]}][\TName] \root[\wElm] \subseteq
    \denot{\varthetaFrm'}[{\asgElm}{[\XvarElm \mapsto
    \DeltaSet[\wElm]]}][\TName] \root[\wElm] \subseteq
    \denot{\varthetaFrm'}[{\asgElm}{[\XvarElm \mapsto
    \DeltaSet[\wElm]]}][\TName]$, where $\DeltaSet[\wElm] \defeq \DeltaSet
    \root[\wElm]$ and $\asgElm' \defeq \asgElm \rst[{\DeltaSet[\wElm]}]$.
    Note that, due to the definition of the restriction operator $\root[\wElm]$,
    the set $\DeltaSet[\wElm]$ is a subtree of $\TSet$ rooted at $\wElm$.
    Thus, to show that $\TName, \{ \varElm \mapsto \wElm \}, \asgElm \models
    \trFun[\varElm]{\varthetaFrm}$ holds true, it is enough to prove that
    $\TName, \{ \varElm \mapsto \wElm \}, {\asgElm}[\XvarElm \mapsto
    \DeltaSet[\wElm]] \models \forall \varElm \ldotp (\varElm \in \XvarElm)
    \rightarrow \trFun[\varElm]{\varthetaFrm'}$ holds too.
    At this point, the proof easily follows from the inductive hypothesis, since
    $\DeltaSet[\wElm] \subseteq \denot{\varthetaFrm'}[{\asgElm}{[\XvarElm
    \mapsto \DeltaSet[\wElm]]}][\TName]$, as observed before.
  \end{itemize}
\item\textbf{[$\varthetaFrm = \mu \XvarElm \ldotp \varthetaFrm'$]:}
  Thanks to the duality property between the least and greatest fixpoint
  operators of \MC, it is well known that $\denot{\varthetaFrm}[\asgElm][\TName]
  = \denot{\neg \nu \XvarElm \ldotp \neg {\varthetaFrm'}[\XvarElm / \neg
  \XvarElm]}[\asgElm][\TName] = \TSet \setminus \denot{\nu \XvarElm \ldotp \neg
  {\varthetaFrm'}[\XvarElm / \neg \XvarElm]}[\asgElm][\TName] = \TSet \setminus
  \denot{\nu \XvarElm \ldotp \pnfMac{\neg {\varthetaFrm'}[\XvarElm / \neg
  \XvarElm]}}[\asgElm][\TName]$.
  This means that $\wElm \in \denot{\varthetaFrm}[\asgElm][\TName]$ \iff $\wElm
  \not\in \denot{\nu \XvarElm \ldotp \pnfMac{\neg {\varthetaFrm'}[\XvarElm /
  \neg \XvarElm]}}[\asgElm][\TName]$.
  Observe that, if $\varthetaFrm' \in \ThetaSet[\ZSet \cup \{ \XvarElm \}, \OSet
  \cup \{ \XvarElm \}]$, then $\pnfMac{\neg {\varthetaFrm'}[\XvarElm / \neg
  \XvarElm]} \in \ThetaSet[\ZSet \cup \{ \XvarElm \}, \OSet \cup \{ \XvarElm
  \}]$ as well, so, $\nu \XvarElm \ldotp \pnfMac{\neg {\varthetaFrm'}[\XvarElm /
  \neg \XvarElm]} \in \ThetaSet[\ZSet, \OSet]$.
  Now, by the inductive hypothesis, $\wElm \not\in \denot{\nu \XvarElm \ldotp
  \pnfMac{\neg {\varthetaFrm'}[\XvarElm / \neg \XvarElm]}}[\asgElm][\TName]$
  \iff $\TName, \{ \varElm \mapsto \wElm \}, \asgElm \not\models
  \trFun[\varElm]{\nu \XvarElm \ldotp \pnfMac{\neg {\varthetaFrm'}[\XvarElm /
  \neg \XvarElm]}}$.
  Hence, $\wElm \in \denot{\varthetaFrm}[\asgElm][\TName]$ \iff $\TName, \{
  \varElm \mapsto \wElm \}, \asgElm \not\models \trFun[\varElm]{\nu \XvarElm
  \ldotp \pnfMac{\neg {\varthetaFrm'}[\XvarElm / \neg \XvarElm]}}$, which
  immediately implies that $\wElm \in \denot{\varthetaFrm}[\asgElm][\TName]$
  \iff $\TName, \{ \varElm \mapsto \wElm \}, \asgElm \models
  \trFun[\varElm]{\varthetaFrm}$, by definition of the translation function.
  \qedhere
\end{itemize}
\end{proof}

\subsection{Alternation-Free One-Step Graded \MC}
\label{app:contmplog;sub:altfreonestpgmc}

Here we give the full definition of the two zero-time suppression operators:
\begin{multicols}{2}
\begin{itemize}
\item
  $\Ff \down[\XvarElm] = \Ff \up[\XvarElm] \defeq\, \Ff$ and $\Tt
  \down[\XvarElm] = \Tt \up[\XvarElm] \defeq\, \Tt$;
\item
  $\apElm \down[\XvarElm] = \apElm \up[\XvarElm] \defeq\, \apElm$;
\item
  $(\DMod[\geq k]\, \varphiFrm) \down[\XvarElm] = (\DMod[\geq k]\, \varphiFrm)
  \up[\XvarElm] \defeq\, \DMod[\geq k]\, \varphiFrm$;
\item
  $(\BMod[< k]\, \varphiFrm) \down[\XvarElm] = (\BMod[< k]\, \varphiFrm)
  \up[\XvarElm] \defeq\, \BMod[< k]\, \varphiFrm$;
\item
  $\XvarElm \down[\XvarElm] \defeq\, \Ff$ and $\XvarElm \up[\XvarElm] \defeq\,
  \Tt$;
\item
  $\YvarElm \down[\XvarElm] = \YvarElm \up[\XvarElm] \defeq\, \YvarElm$, when
  $\YvarElm \neq \XvarElm$;
\item
  $(\neg \varphiFrm) \down[\XvarElm] \defeq\, \neg (\varphiFrm \down[\XvarElm])$
  and $(\neg \varphiFrm) \up[\XvarElm] \defeq\, \neg (\varphiFrm
  \up[\XvarElm])$;
\item
  $(\varphiFrm[1] \Cnt\, \varphiFrm[2]) \down[\XvarElm] \defeq\, (\varphiFrm[1]
  \down[\XvarElm]) \Cnt\, (\varphiFrm[1] \down[\XvarElm])$, $\Cnt \in \{
  \wedge,\! \vee \}$;
\item
  $(\varphiFrm[1] \Cnt\, \varphiFrm[2]) \up[\XvarElm] \defeq\, (\varphiFrm[1]
  \up[\XvarElm]) \Cnt\, (\varphiFrm[1] \up[\XvarElm])$, $\Cnt \in \{ \wedge,\!
  \vee \}$;
\item
  $(\lambda \XvarElm \ldotp \varthetaFrm) \down[\XvarElm] = (\lambda \XvarElm
  \ldotp \varthetaFrm) \up[\XvarElm] \defeq\, \lambda \XvarElm \ldotp
  \varthetaFrm$, with $\lambda \in \{ \mu, \nu \}$;
\item
  $(\lambda \YvarElm \ldotp \varthetaFrm) \down[\XvarElm] \defeq\, \lambda
  \YvarElm \ldotp (\varthetaFrm \down[\XvarElm])$, with $\lambda \in \{ \mu, \nu
  \}$, when $\YvarElm \neq \XvarElm$;
\item
  $(\lambda \YvarElm \ldotp \varthetaFrm) \up[\XvarElm] \defeq\, \lambda
  \YvarElm \ldotp (\varthetaFrm \up[\XvarElm])$, with $\lambda \in \{ \mu, \nu
  \}$, when $\YvarElm \neq \XvarElm$.
\end{itemize}
\end{multicols}

\lmmaltfreonestpgmc*
\begin{proof}
It is well known that $\DeltaSet \defeq \denot{\varthetaFrm}[\asgElm][\TName] =
\denot{\mu \XvarElm[1] \ldots \mu \XvarElm[k] \ldotp
\varphiFrm}[\asgElm][\TName] = \denot{\mu \YvarElm \ldotp
{\varphiFrm}[\XvarElm[1] / \YvarElm, \ldots, \XvarElm[k] /
\YvarElm]}[\asgElm][\TName]$ (see, \eg,~\cite[Proposition 1.3.2]{AN01}).
By Kleene's Theorem, we have that $\DeltaSet = \bigcup_{i \in \SetN} \FSet[i]$,
where $\FSet[0] \defeq \emptyset$ and $\FSet[i + 1] \defeq
\denot{{\varphiFrm}[\XvarElm[1] / \YvarElm, \ldots, \XvarElm[k] /
\YvarElm]}[{{\asgElm}[ {\YvarElm \mapsto \FSet[i]} ]}][\TName]$, for all $i \in
\SetN$.
We now proceed by showing the two directions of the equivalence separately.

\begin{itemize}
\item\textbf{[``if'']:}
  Let us consider the family $\{ \WSet[i] \}_{i \in \SetN}$ of subsets of
  $\WSet$ defined as follows: $\WSet[0] \defeq \emptyset$ and $\WSet[i + 1]
  \defeq \set{ \vElm \in \WSet }{ \post{\vElm} \cap \WSet \subseteq \WSet[i] }$,
  for all $i \in \SetN$.
  It is easy to see that, for every $i > 0$ and $\vElm \in \WSet[i]$, it holds
  that $\WSet \!\root[\vElm]\! \setminus \{ \vElm \} \subseteq \WSet[i - 1]$.
  We now show that $\WSet[i] \subseteq \FSet[i]$, for all $i \in \SetN$.
  Assume by contradiction that there exists an index $i \in \SetN$ such that
  $\WSet[i] \setminus \FSet[i] \neq \emptyset$ and consider $j \in \SetN$ as the
  minimum of such indexes.
  Obviously, $j > 0$, since $\WSet[0] = \emptyset$, and $\WSet[j - 1] \subseteq
  \FSet[j - 1]$.
  Let $\vElm \in \WSet[j] \setminus \FSet[j] \neq \emptyset$.
  By hypothesis, we have that $\vElm \in \denot{{\varphiFrm}[\XvarElm[1] /
  \YvarElm, \ldots, \XvarElm[k] / \YvarElm] \down[\YvarElm]}[{{\asgElm}[
  {\YvarElm \mapsto \WSet \root[\vElm]} ]}][\TName]$.
  Thus, $\vElm \in \denot{{\varphiFrm}[\XvarElm[1] / \YvarElm, \ldots,
  \XvarElm[k] / \YvarElm]}[{{\asgElm}[ {\YvarElm \mapsto \WSet \!\root[\vElm]\!
  \setminus \{ \vElm \}} ]}][\TName]$, thanks to Item~\ref{prp:gmc(ind)} of
  Proposition~\ref{prp:gmc}.
  Due the monotonicity of the semantics and the inclusions $\WSet
  \!\root[\vElm]\! \setminus \{ \vElm \} \subseteq \WSet[j - 1]$ and $\WSet[j -
  1] \subseteq \FSet[j - 1]$, we have that $\vElm \in
  \denot{{\varphiFrm}[\XvarElm[1] / \YvarElm, \ldots, \XvarElm[k] /
  \YvarElm]}[{{\asgElm}[ {\YvarElm \mapsto \WSet[j - 1]} ]}][\TName] \subseteq
  \denot{{\varphiFrm}[\XvarElm[1] / \YvarElm, \ldots, \XvarElm[k] /
  \YvarElm]}[{{\asgElm}[ {\YvarElm \mapsto \FSet[j - 1]} ]}][\TName] =
  \FSet[j]$, which contradicts the assumption.
  At this point, thanks to the finiteness of $\WSet$, it is immediate to see
  that $\WSet = \bigcup_{i \in \SetN} \WSet[i]$.
  Hence, $\wElm \in \WSet = \bigcup_{i \in \SetN} \WSet[i] \subseteq \bigcup_{i
  \in \SetN} \FSet[i] = \DeltaSet$, which completes the proof of this direction.
\item\textbf{[``only if'']:}
  By induction on an index $i \in \SetN$, we show that, for every node $\wElm
  \in \FSet[i]$, there exists a finite tree $\WSet \subseteq \TSet$ rooted at
  $\wElm$ such that $\vElm \in \denot{{\varphiFrm}[\XvarElm[1] / \YvarElm,
  \ldots, \XvarElm[k] / \YvarElm] \down[\YvarElm]}[{{\asgElm}[ {\YvarElm
  \mapsto \WSet \root[\vElm]} ]}][\TName]$, for all $\vElm \in \WSet$.
  \begin{itemize}
  \item\textbf{[Base case $i = 0$]:}
    The statement trivially holds true, as $\FSet[0] = \emptyset$.
  \item\textbf{[Inductive case $i > 1$]:}
    By inductive hypothesis, for every node $\uElm \in \FSet[i - 1]$, there
    exists a finite tree $\USet[\uElm]$ rooted at $\uElm$ such that $\vElm \in
    \denot{{\varphiFrm}[\XvarElm[1] / \YvarElm, \ldots, \XvarElm[k] / \YvarElm]
    \down[\YvarElm]}[{{\asgElm}[ {\YvarElm \mapsto \USet[\uElm] \root[\vElm]}
    ]}][\TName]$, for all $\vElm \in \USet[\uElm]$.
    Let us now set $\WSet \defeq \{ \wElm \} \cup\, \bigcup_{\uElm \in
    \post{\wElm} \cap \FSet[i - 1]} \USet[\uElm]$, for $\wElm \not\in \FSet[i -
    1]$.
    Obviously, $\WSet$ is a finite tree rooted at $\wElm$, since the underlying
    tree model has finite branching.
    Moreover, $\vElm \in \denot{{\varphiFrm}[\XvarElm[1] / \YvarElm, \ldots,
    \XvarElm[k] / \YvarElm] \down[\YvarElm]}[{{\asgElm}[ {\YvarElm \mapsto
    \WSet \root[\vElm]} ]}][\TName]$, for all $\vElm \in \WSet$ with $\vElm \neq
    \wElm$, since $\USet[\uElm] \root[\vElm] = \WSet \root[\vElm]$, where
    $\uElm$ is the unique child of $\wElm$ ancestor of $\vElm$.
    So, to conclude this case, we need to show that $\wElm \in
    \denot{{\varphiFrm}[\XvarElm[1] / \YvarElm, \ldots, \XvarElm[k] /
    \YvarElm] \down[\YvarElm]}[{{\asgElm}[ {\YvarElm \mapsto \WSet}
    ]}][\TName]$.
    Note that ${\asgElm}[\YvarElm \mapsto \FSet[i - 1]] \sqsubseteq_{\{\wElm
    \}}^{\ZSet \cup \{ \YvarElm \}, \OSet \cup \{ \YvarElm \}}
    {\asgElm}[\YvarElm \mapsto \WSet \setminus \{ \wElm \}]$, as $\wElm \not\in
    \FSet[i - 1]$ and $\post{\wElm} \cap \FSet[i - 1] \subseteq \WSet \setminus
    \{ \wElm \}$.
    Thus, by Lemma~\ref{lmm:onestpmodlog}, we have that $\wElm \in
    \denot{{\varphiFrm}[\XvarElm[1] / \YvarElm, \ldots, \XvarElm[k] /
    \YvarElm]}[{{\asgElm}[ {\YvarElm \mapsto \WSet \setminus \{ \wElm \}}
    ]}][\TName]$, since $\wElm \in \FSet[i] = \denot{{\varphiFrm}[\XvarElm[1] /
    \YvarElm, \ldots, \XvarElm[k] / \YvarElm]}[{{\asgElm}[ {\YvarElm \mapsto
    \FSet[i - 1]} ]}][\TName]$.
    Hence, $\wElm \in \denot{{\varphiFrm}[\XvarElm[1] / \YvarElm, \ldots,
    \XvarElm[k] / \YvarElm] \down[\YvarElm]}[{{\asgElm}[ {\YvarElm \mapsto
    \WSet} ]}][\TName]$, thanks to Item~\ref{prp:gmc(ind)} of
    Proposition~\ref{prp:gmc}.
    \qedhere
  \end{itemize}
\end{itemize}
\end{proof}

Before providing the proof of Theorem~\ref{thm:altfreonestpgmc}, we report here
the part of the definition of the translation function $\trFun{} \colon
\VarSet[1] \to (\ThetaSet[\ZSet, \OSet][\ASym\FSym] \to \WMTL)$ from the
\OSAFGMC to \WMTL that differs from the general translation of the \OSGMC to
\MTL (recall that $\YvarElm$ is an arbitrary fresh fixpoint variable not
occurring anywhere in the formula subject of the transformation):
{\small\tabaltfreonestpgmc}

\thmaltfreonestpgmc*
\begin{proof}
To show that the \OSAFGMC is subsumed by \WMTL, for the sake of presentation, we
prove the following statement for sentences: for every \OSAFGMC fixpoint
sentence $\varthetaFrm \in \ThetaSet[\emptyset, \emptyset][\ASym\FSym]$, Kripke
tree $\TName$, and node $\wElm \in \TSet$, it holds that
\[
  \wElm \in \denot{\varthetaFrm}[\emptyfun][\TName]
\text{\ \ \iff\ \ \ }
  \TName, \{ \varElm \mapsto \wElm \}, \emptyfun \models
  \trFun[\varElm]{\varthetaFrm}.
\]
The generalisation of this property to formulae is straightforward, so we leave
it to the reader.
\begin{itemize}
\item\textbf{[$\varthetaFrm = \mu \XvarElm[1] \ldots \mu \XvarElm[k] \ldotp
  \varphiFrm$, with $\varphiFrm \in \PhiSet[{\{ \XvarElm[1], \ldots, \XvarElm[k]
  \}, \{ \XvarElm[1], \ldots, \XvarElm[k] \}}][\ASym\FSym]$]:}
  First note that $\trFun[\varElm]{\varthetaFrm} = \exists^{\TSym} \XvarElm
  \ldotp (\varElm \in \XvarElm) \wedge \forall \varElm \ldotp (\varElm \in
  \XvarElm) \rightarrow \exists^{\TSym} \YvarElm \ldotp
  \maxsubtreeMac{\YvarElm, \XvarElm, \varElm} \wedge
  \trFun[\varElm]{{\varphiFrm}[\XvarElm[1] / \YvarElm, \ldots, \XvarElm[k] /
  \YvarElm] \root[\{ \YvarElm \}]}$.
  By Lemma~\ref{lmm:altfreonestpgmc}, $\wElm \in
  \denot{\varthetaFrm}[\emptyfun][\TName]$ \iff there exists a finite tree
  $\WSet \subseteq \TSet$ containing $\wElm$ such that every node $\vElm$ in
  $\WSet$ belongs to $\denot{{\varphiFrm}[\XvarElm[1] / \YvarElm, \ldots,
  \XvarElm[k] / \YvarElm] \root[\{ \YvarElm \}]}[{{\emptyfun}[ {\YvarElm \mapsto
  \WSet \root[\vElm]} ]}][\TName]$.
  At this point, the thesis easily follows by exploiting the fact that, for
  every tree $\USet \subseteq \TSet$ and node $\uElm \in \USet$, it holds that
  \emph{(i)} $\uElm \in \denot{{\varphiFrm}[\XvarElm[1] / \YvarElm, \ldots,
  \XvarElm[k] / \YvarElm] \root[\{ \YvarElm \}]}[{{\emptyfun}[ {\YvarElm \mapsto
  \USet \root[\uElm]} ]}][\TName]$ \iff $\TName, \{ \varElm \mapsto \uElm \},
  {\emptyfun}[\YvarElm \mapsto \USet \root[\uElm]] \models
  \trFun[\varElm]{{\varphiFrm}[\XvarElm[1] / \YvarElm, \ldots, \XvarElm[k] /
  \YvarElm \root[\{ \YvarElm \}]}$, thanks to Theorem~\ref{thm:onestpgmc}, and
  \emph{(ii)} $\TName, \{ \varElm \mapsto \uElm \}, {\emptyfun}[\YvarElm \mapsto
  \ZSet, \XvarElm \mapsto \USet] \models \maxsubtreeMac{\YvarElm, \XvarElm,
  \varElm}$ \iff $\ZSet = \USet \root[\uElm]$, for every tree $\ZSet \subseteq
  \TSet$.
\item\textbf{[$\varthetaFrm = \nu \XvarElm[1] \ldots \nu \XvarElm[k] \ldotp
  \varphiFrm$, with $\varphiFrm \in \PhiSet[{\{ \XvarElm[1], \ldots, \XvarElm[k]
  \}, \{ \XvarElm[1], \ldots, \XvarElm[k] \}}][\ASym\FSym]$]:}
  Thanks to the duality property between the greatest and least fixpoint
  operators of \MC, it is well known that
  \begin{eqnarray*}
    {\denot{\varthetaFrm}[\emptyfun][\TName]}
  & = &
    {\denot{\neg \mu \XvarElm[1] \ldots \mu \XvarElm[k] \ldotp \neg
    {\varphiFrm}[\XvarElm[1] / \neg \XvarElm[1], \ldots, \XvarElm[k] / \neg
    \XvarElm[k]]}[\emptyfun][\TName]} \\
  & = &
    {\TSet \setminus \denot{\mu \XvarElm[1] \ldots \mu \XvarElm[k] \ldotp \neg
    {\varphiFrm}[\XvarElm[1] / \neg \XvarElm[1], \ldots, \XvarElm[k] / \neg
    \XvarElm[k]]}[\emptyfun][\TName]} \\
  & = &
    {\TSet \setminus \denot{\mu \XvarElm[1] \ldots \mu \XvarElm[k] \ldotp
    \pnfMac{\neg {\varphiFrm}[\XvarElm[1] / \neg \XvarElm[1], \ldots,
    \XvarElm[k] / \neg \XvarElm[k]]}}[\emptyfun][\TName]}.
  \end{eqnarray*}
  This means that $\wElm \in \denot{\varthetaFrm}[\emptyfun][\TName]$ \iff
  $\wElm \not\in \denot{\mu \XvarElm[1] \ldots \mu \XvarElm[k] \ldotp
  \pnfMac{\neg {\varphiFrm}[\XvarElm[1] / \neg \XvarElm[1], \ldots, \XvarElm[k]
  / \neg \XvarElm[k]]}}[\emptyfun][\TName]$.
  Now, observe that, if $\varphiFrm \in \PhiSet[{\{ \XvarElm[1], \ldots,
  \XvarElm[k] \}, \{ \XvarElm[1], \ldots, \XvarElm[k] \}}][\ASym\FSym]$, then
  $\pnfMac{\neg  {\varphiFrm}[\XvarElm[1] / \neg \XvarElm[1], \ldots \XvarElm[k]
  / \neg \XvarElm[k]]} \in \PhiSet[{\{ \XvarElm[1], \ldots, \XvarElm[k] \}, \{
  \XvarElm[1], \ldots, \XvarElm[k] \}}][\ASym\FSym]$ as well, so, $\mu
  \XvarElm[1] \ldots \mu \XvarElm[k] \ldotp \pnfMac{\neg
  {\varphiFrm}[\XvarElm[1] / \neg \XvarElm[1], \ldots, \XvarElm[k] / \neg
  \XvarElm[k]]} \in \ThetaSet[\emptyset, \emptyset][\ASym\FSym]$.
  By the previous item, we know that
  \[
    \wElm \not\in \denot{\mu \XvarElm[1] \ldots \mu \XvarElm[k] \ldotp
    \pnfMac{\neg {\varphiFrm}[\XvarElm[1] / \neg \XvarElm[1], \ldots,
    \XvarElm[k] / \neg \XvarElm[k]]}}[\emptyfun][\TName]
  \]
  \iff
  \[
    \TName, \{ \varElm \mapsto \wElm \}, \emptyfun \not\models
    \trFun[\varElm]{\mu \XvarElm[1] \ldots \mu \XvarElm[k] \ldotp \pnfMac{\neg
    {\varphiFrm}[\XvarElm[1] / \neg \XvarElm[1], \ldots, \XvarElm[k] / \neg
    \XvarElm[k]]}}.
  \]
  Therefore, $\wElm \in \denot{\varthetaFrm}[\emptyfun][\TName]$
  \iff $\TName, \{ \varElm \mapsto \wElm \}, \emptyfun \not\models
  \trFun[\varElm]{\mu \XvarElm[1] \ldots \mu \XvarElm[k] \ldotp \pnfMac{\neg
  {\varphiFrm}[\XvarElm[1] / \neg \XvarElm[1], \ldots, \XvarElm[k] / \neg
  \XvarElm[k]]}}$.
  At this point, it is easy to see that $\trFun[\varElm]{\varthetaFrm} = \neg
  \trFun[\varElm]{\mu \XvarElm[1] \ldots \mu \XvarElm[k] \ldotp \pnfMac{\neg
  {\varphiFrm}[\XvarElm[1] / \neg \XvarElm[1], \ldots, \XvarElm[k] / \neg
  \XvarElm[k]]}}$, which immediately implies that $\wElm \in
  \denot{\varthetaFrm}[\emptyfun][\TName]$ \iff $\TName, \{ \varElm \mapsto
  \wElm \}, \emptyfun \models \trFun[\varElm]{\varthetaFrm}$.
  \qedhere
\end{itemize}
\end{proof}


%% file: AppendixC.tex


\newpage
\begin{center}
III. Syntax and Semantics of \STLS
\end{center}

\begin{definition}[\STLS Syntax]
  \label{def:stls(syntax)}
  \STLS \emph{state} ($\varphi$) and \emph{path} ($\psi$) \emph{formulas}
  are built inductively from the set of atomic propositions $\APSet$ according
  to the following grammar, where $\apElm \in \APSet$:
  \begin{enumerate}
  \item\label{def:sol(syntax:state)}
    $\varphi ::= \apElm \mid \neg \varphi \mid \varphi \wedge \varphi \mid
    \varphi \vee \varphi \mid \varphi \UU[\varphi] \varphi \mid \varphi
    \RR[\varphi] \varphi \mid \varphi \SS[\varphi] \varphi \mid \varphi
    \BB[\varphi] \varphi \mid \E \psi \mid \A \psi$;
  \item\label{def:sol(syntax:path))}
    $\psi ::= \varphi \mid \neg \psi \mid \psi \wedge \psi \mid \psi \vee \psi
    \mid \X \psi \mid \psi \U \psi \mid \psi \R \psi$.
  \end{enumerate}
\end{definition}

Given two Kripke trees $\TName$ and $\TName'$ and a set of nodes $\WSet$, we say
that $\TName'$ is a \emph{(strict) $\WSet$-subtree} of $\TName$ \wrt
$\TName[][\star]$ if $\TName'$ is a (strict) subtree of $\TName$ with the same
root as $\TName$ and such that, for all nodes $\wElm \in \WSet$ of $\TName'$,
all the children of $\wElm$ in $\TName$ also belong to $\TName'$.
We say that $\TName'$ is a \emph{(strict) $\WSet$-supertree} of $\TName$ if
$\TName$ is a (strict) $\WSet$-subtree of $\TName'$.
Moreover, given a Kripke tree $\TName$ and one of its nodes $\wElm$, we denote
with $\TName[\wElm]$ the subtree of $\TName$ rooted at $\wElm$.

\begin{definition}[\STLS Semantics]
  \label{def:stls(semantics)}
  Given two Kripke trees $\TName[][\star], \TName$ with $\TName \sqsubseteq
  \TName[][\star]$, for all \STLS state formulas $\varphi_{1}$, $\varphi_{2}$,
  and $\phi$, it holds that:
  \begin{enumerate}
  \item\label{def:stls(semantics::uuopr)}
    $\TName \kmodels{\TName[][\star]} \varphi_{1} \UU[\phi] \varphi_{2}$ if
    there exists a strict $\denot{\phi}[\TName][ {\TName[][\star]} ]$-subtree
    $\TName'$ of $\TName$ such that $\TName' \kmodels{\TName[][\star]}
    \varphi_{2}$ and, for all its strict $\denot{\phi}[\TName][
    {\TName[][\star]}]$-supertree $\TName''$ of $\TName'$ that are strict
    $\denot{\phi}[\TName][ {\TName[][\star]}]$-subtree of $\TName$, it holds
    $\TName'' \kmodels{\TName[][\star]} \varphi_{1}$;
  \item\label{def:stls(semantics::rropr)}
    $\TName \kmodels{\TName[][\star]} \varphi_{1} \RR[\phi] \varphi_{2}$ if, for
    all strict $\denot{\phi}[\TName][ {\TName[][\star]} ]$-subtrees $\TName'$
    of $\TName$, it holds $\TName' \kmodels{\TName[][\star]} \varphi_{2}$ or
    there exists a strict $\denot{\phi}[\TName][ {\TName[][\star]} ]$-supertree
    $\TName''$ of $\TName$ that is also a strict $\denot{\phi}[\TName][
    {\TName[][\star]} ]$-subtree of $\TName$ such that $\TName''
    \kmodels{\TName[][\star]} \varphi_{1}$;
  \item\label{def:stls(semantics::ssopr)}
    $\TName \kmodels{\TName[][\star]} \varphi_{1} \SS[\phi] \varphi_{2}$ if
    there exists a strict $\denot{\phi}[\TName][ {\TName[][\star]} ]$-supertree
    $\TName'$ of $\TName$ that is also a subtree of $\TName[][\star]$ such that
    $\TName' \kmodels{\TName[][\star]} \varphi_{2}$ and, for all strict
    $\denot{\phi}[\TName][ {\TName[][\star]} ]$-subtrees $\TName''$ of $\TName'$
    that are strict $\denot{\phi}[\TName][ {\TName[][\star]}]$-supertree of
    $\TName$, it holds $\TName'' \kmodels{\TName[][\star]} \varphi_{1}$;
  \item\label{def:stls(semantics::bbopr)}
    $\TName \kmodels{\TName[][\star]} \varphi_{1} \BB[\phi] \varphi_{2}$ if, for
    all strict $\denot{\phi}[\TName][ {\TName[][\star]} ]$-supertrees $\TName'$
    of $\TName$ that are also subtrees of $\TName[][\star]$, it holds $\TName'
    \kmodels{\TName[][\star]} \varphi_{2}$ or there exists a strict
    $\denot{\phi}[\TName][ {\TName[][\star]} ]$-subtree $\TName''$ of $\TName$
    that is also a strict $\denot{\phi}[\TName][ {\TName[][\star]} ]$-supertree
    of $\TName$ such that $\TName'' \kmodels{\TName[][\star]} \varphi_{1}$;
  \end{enumerate}
  where $\denot{\phi}[\TName][ {\TName[][\star]} ] \!\defeq\! \set{ \wElm \in
  \WSet[\TName] }{ \TName[\wElm] \kmodels{\TName[\wElm][\star]} \phi }$ is the
  denotation of $\phi$.
\end{definition}
